# A new concept of deformation quantization I. Normal order quantization on cotangent bundles

Markus J. Pflaum*

April 11, 1996

## Abstract

In this work we give a deformation theoretical approach to the problem of quantization. First the notion of a deformation of a noncommutative ringed space over a commutative locally ringed space is introduced within a language coming from Algebraic Geometry and Complex Analysis. Then we define what a DIRAC quantization of a commutative ringed space with a POISSON structure, the space of classical observables, is. Afterwards the normal order quantization of the POISSON space of classical polynomial observables on a cotangent bundle is constructed. By using a complete symbol calculus on manifolds we succeed in extending the normal order quantization of polynomial observables to a quantization of a POISSON space of symbols on a cotangent bundle. Furthermore we consider functorial properties of these quantizations. Altogether it is shown that a deformation theoretical approach to quantization is possible not only in a formal sense but also such that the deformation parameter $\hbar$ can attain any real value.

* Humboldt Universität zu Berlin, Mathematisches Institut, Unter den Linden 6, 10099 Berlin, email: pflaum@mathematik.hu-berlin.de

# Contents

# Introduction

One of the major problems in mathematical physics is to find an appropriate mathematical language for quantization. "In physics, the term 'quantization' refers to the process of passing from a classical-mechanical description of a physical system to a quantum-mechanical description of the same system" (cf. ALAN WEINSTEIN [61], p. 31). Classical systems like the harmonic oscillator or a free particle in EUCLIDEAN space can be quantized in a mathematical rigorous way, but a general and satisfactory scheme for quantizing all classical systems "living" on an arbitrary symplectic manifold is still lacking. So the "quantization problem" is not completely solved, though important partial results have been achieved.

## Dirac's quantization scheme

After quantum theory was developped in the twentieth of this century by HEISENBERG, SCHRÖDINGER, BOHR and others, DIRAC [12] probably was the first who tried to establish a general quantization scheme, which usually is refered to as canonical quantization. Though DIRAC'S ideas are very convincing and fruitful from the point of view of physics, they are somewhat lacking of mathematical rigor. So a lot of attempts have been undertaken to make his ideas mathematically precise. This is also the goal of this work, where we will give a proposal how to dress canonical quantization in a rigorous mathematical language by using deformation theory. Before explaining our approach in more detail, let us describe DIRAC'S idea of canonical quantization.

As already mentioned above, a classical mechanical system is modelled on a symplectic manifold, the phase space. The smooth functions on the symplectic manifold serve as observables of the system. Now the space of these functions carries two important structures, first that one of a commutative algebra and second that one of a LIE algebra, where the bracket is given by the POISSON bracket on the symplectic manifold. Passing to the quantum setting, the algebra of observables becomes a noncommutative algebra acting on a certain HILBERT space, which is interpreted as the space of pure states. It is assumed that to every classical observable there corresponds a quantized one such that selfadjoint, i. e. realvalued, classical observables are mapped to selfadjoint operators. According to DIRAC the algebra structure of the space of quantized observables should be determined by the following two conditions:

(i) The commutator of two quantized observables is given by the product of $\hbar/i$ and the POISSON bracket of its classical counterparts:

$$[\mathfrak{q}(a), \mathfrak{q}(b)] \; = \; \frac{\hbar}{i} \, \mathfrak{q}\left(\{a, b\}\right). \tag{1}$$

Here $\hbar$ is PLANCK'S constant, $a, b$ are classical observables and $\mathfrak{q}$ is the quantization map.

(*ii*) The quantization map q preserves the unit element:

$$\mathfrak{q}(1) = 1. \tag{2}$$

In other words this means there should be a LIE algebra morphism of the space of classical observables to the LIE algebra of quantum observables, where the bracket operation for quantum observables is given by the commutator times $\hbar/i$. The last assumption in DIRAC'S canonical quantization scheme is that the quantized observables have a common admissable (i.e. dense and invariant) domain in the (separable) HILBERT space of quantum states and that the action of the algebra of quantized observables is irreducible. Now GROENEWOLD [25] and VAN HOVE [59] have shown the remarkable result that there is no quantization scheme in the above sense, if one wants to quantize all observables. See also ABRAHAM, MARSDEN [1] as well as GUILLEMIN, STERNBERG [26] for a proof and discussion of this fact.

## Geometric and formal deformation quantization

By relaxing DIRAC'S original requirements, it is possible to circumvent the theorem of GROENEWOLD and VAN HOVE and set up a mathematically rigorous apparatus for the quantization of a classical physical system. Without going into details let us give two examples.

In geometric quantization by KOSTANT and SOURIAU (see for example WOODHOUSE [70]) one first constructs a certain HERMITIAN line bundle $L$ with connection $\nabla$ over $M$ for every quantizable symplectic manifold $M$, i.e. for every symplectic $M$ with symplectic form $\omega$ such that $\hbar\omega$ is integral. The line bundle $L$ gives rise to a HILBERT space $L^2(L)$ of square integrable sections of $L$ on which a classical observable $a$ acts as differential operator of the form $\mathfrak{q}(a) = \frac{\hbar}{i}\nabla_{X_a} + a$. Now the HILBERT space $L^2(L)$ is too large. By introduction of a so-called polarization on $M$ one can cut down $L^2(L)$ to the "correct" HILBERT space $\mathcal{H}$ of square integrable sections of $L$ which are covariantly constant along the leaves of the polarization. Unfortunately this implies that one also has to cut down the space of quantizable functions resp. quantizable classical observables to those whose associated differential operator $\frac{\hbar}{i}\nabla_{X_a} + a$ leaves $\mathcal{H}$ invariant. It may now happen that the space of quantizable functions on $M$ is much too small. For example it might contain only the space of smooth functions linear in momentum and thus not contain essential functions like the HAMILTONIAN which often is quadratic in momentum. But we are definitely interested in the quantization of the HAMILTONIAN. Therefore we cannot consider geometric quantization as the perfect solution to the quantization problem, though geometric quantization gives a very beautiful approach and works very well in a lot of cases. See for example RAWNSLEY [43] or BLATTNER [7] for more information on the topic of geometric quantization and its relation to other quantization schemes.

In the second approach by BAYEN ET AL. [3], every classical observable can be quantized, but here PLANCK'S constant $\hbar$ which governs quantization is considered only as a formal parameter. More precisely BAYEN ET AL. [3] regard a quantization as a formal deformation with formal parameter $\hbar$ in the sense of GERSTENHABER [19, 20]. The space to be deformed is the algebra $\mathcal{C}^\infty(M)$ (resp. the sheaf $\mathcal{C}^\infty_M$) of smooth functions on a symplectic manifold $M$. The deformed algebra is given by the linear space $\mathcal{C}^\infty(M)[[\hbar]]$ of formal power series in the variable $\hbar$ with coefficients in $\mathcal{C}^\infty(M)$ together with a noncommutative product $\star$ on $\mathcal{C}^\infty(M)[[\hbar]]$:

$$\left(\sum_{k\in\mathbb{N}} a_k \hbar^k\right) \star \left(\sum_{l\in\mathbb{N}} b_l \hbar^l\right) = \sum_{n\in\mathbb{N}} \sum_{k+l+m=n} \alpha_m(a_k, b_l)\, \hbar^n. \qquad (3)$$

Here the $a_k$, $b_l$ are elements of $\mathcal{C}^\infty(M)$ and the bilinear mappings $\alpha_m : \mathcal{C}^\infty(M) \times \mathcal{C}^\infty(M) \to \mathcal{C}^\infty(M)$ have to be appropriately chosen, such that $\alpha_0(a,b) = ab$ and such that associativity is fulfilled. Furthermore the quantization map $\mathfrak{q}$ is given by the canonical embedding $\mathcal{C}^\infty(M) \to \mathcal{C}^\infty(M)[[\hbar]]$, $a \mapsto a$, and it is assumed that DIRAC'S quantization condition holds for every $a, b \in \mathcal{C}^\infty(M)$ up to first order in $\hbar$. In particular one wants to write

$$\mathfrak{q}(a) \star \mathfrak{q}(b) = ab + \frac{\hbar}{2i}\{a,b\} + \hbar^2 \alpha_2(a,b) + ... + \hbar^m \alpha_m(a,b) + ... \qquad (4)$$

for every $a, b \in \mathcal{C}^\infty(M)$. It is highly nontrivial to see that such formal deformations exist, but by a result of DE WILDE and LECOMTE [69] one knows that every symplectic manifold can be quantized resp. deformed in the formal sense. A geometric proof of this fact is given by FEDOSOV [17]. The major set-back of formal deformation quantization now lies in the fact that PLANCK'S constant is regarded only as a formal parameter, whereas experimental physicists tell us that $\hbar$ has a certain measurable numerical value. Therefore the formal quantization scheme is inappropriate for certain aspects of quantum theory.

Though both of the two outlined approaches do not give a completely satisfactory answer to the quantization problem, they are nevertheless very useful for a better understanding of quantum theory and its mathematical description. In particular they have to be understood as important and essential steps on the tough way to the final and right mathematical language (which we all hope to exist) for quantization.

## Concrete deformation quantization

In this work we will give a new proposal for a mathematically rigorous quantization scheme and try to make DIRAC'S ideas precise. The quantization scheme we introduce is based on deformation theory but differs from the formal approach in certain important aspects. As already mentioned BAYEN ET AL. [3] use PLANCK'S constant $\hbar$ only in a formal sense to give a quantization of the algebra $\mathcal{C}^\infty(M)$ of classical observables on a symplectic manifold $M$. Now I was interested in the question whether it is possible in their approach to give $\hbar$ a real or complex value such that the formal power series

$$\mathfrak{q}(a) \star \mathfrak{q}(b) \;=\; ab + \frac{\hbar}{2i}\{a,b\} + \hbar^2\,\alpha_2(a,b) + ... + \hbar^m\,\alpha_m(a,b) + ... \tag{5}$$

for the product of two quantized observables converges for every $a,b \in \mathcal{C}^\infty(M)$ and this particular value of $\hbar$. Unfortunately even in the flat case of $M = \mathbb{R}^{2n}$ with its canonical symplectic structure one can show that the only $\hbar$ fulfilling the above requirement is $\hbar = 0$. It even seems in general to be impossible to construct a nontrivial subalgebra of observables $a,b \in \mathcal{C}^\infty(M)$ for which the series in Eq. (5) does not have a vanishing convergence radius. The reason of this is that the bilinear mappings $\alpha_m : \mathcal{C}^\infty(M) \times \mathcal{C}^\infty(M) \to \mathcal{C}^\infty(M)$ governing the product expansion in Eq. (5) are constructed by a complicated cohomological argument, which does not allow any considerations about the growth of the $\alpha_m$. This suggests that one should consider another path for deformation quantization and leads us to our main question:

> Is it possible to find a quantization scheme using concrete deformations over real or complex parameter spaces (and not formal ones) such that the algebras of classical observables which can be quantized are large enough for purposes of theoretical physics and such that calculations can be carried out?

It is the main result of this work that we can give a positive answer to the above question. In particular we prove that a concrete deformation quantization of the algebra of classical polynomial observables on a cotangent bundle exists and also one of the algebra of classical observables coming from a symbol space. Hereby we use a formalism which is closer to the deformation theory in complex analysis resp. algebraic geometry than to the deformation theory of GERSTENHABER [19, 20].

Before explaining our approach in more detail let us mention that the problem to find a concrete resp. converging deformation quantization is also mentioned several times in the mathematical physics literature, for example in BLATTNER [7], RIEFFEL [47], VAISMAN [58] or WEINSTEIN [62]. Whereas BLATTNER, VAISMAN and WEINSTEIN do not further pursue this problem, RIEFFEL [47] tackles it by using a different approach to ours, namely the language of $C^*$-algebras.

Several difficulties arise when one tries to establish a quantization scheme built upon concrete deformations. As I intended to include local aspects in our quantization scheme, it was necessary to use the language of sheaves resp. ringed spaces for spaces of classical and quantized observables as well as for parameter spaces. With the help of this language it is no problem anymore to formulate what a family of algebras continuously varying over a parameter space i.e. a deformation of an algebra should be; it is just a certain sheaf over the parameter space. But, as the regarded family of algebras in general does not comprise a bundle of finite dimensional vector spaces, it is not yet immediately clear how to express the requirement that the family of algebras is in a certain sense a "locally trivial" one. The notion of a flat morphism of algebras now gives the right tool for this. More precisely we require that the algebra morphisms between the stalks of the parameter space and those of the family of algebras are flat. In concrete models of our quantization scheme we therefore always have to prove that the flatness condition or a slightly weaker condition is fulfilled.

A second problem in a concrete deformation theory is that unlike in the case of formal deformations one does not have a natural quantization map which associates to any classical observable a quantized one. This suggests that a quantization is more than just a deformation and that one needs an appropriate definition of what a deformation quantization should be. In particular we had to clarify in the definition where the quantization map comes in and which properties it should have. In the context of $C^*$-algebras such a definition was given by MARC RIEFFEL in [44, 45, 46, 47] with his concept of a "strict deformation quantization". But the $C^*$-algebra setting turned out not to be applicable for our purposes, as our approach first uses sheaves and second at the best algebras with a FRÉCHET-$*$-structure but not with a $C^*$-structure. So a new notion of a deformation quantization appropriate for our needs had to be found. The outcome of these attempts is a definition of a deformation quantization which on a first glance looks quite complicated, but is capable to comprise smooth, holomorphic, real analytic or even formal deformations and quantizations in one language. Another point is that though the definition of a quantization is complicated, calculations in concrete models of our approach are a lot easier than for example in geometric quantization. Furthermore they are closer to the methods used in theoretical physics. Compare for example the quantization of the HAMILTONIAN of a free particle moving on a RIEMANNIAN manifold in section 2.4 with the one in WOODHOUSE, *Geometric Quantization* [70]. Finally we succeed in giving a definition of what a category of quantizations should be, which opens up a new way to study functorial properties of quantizations.

## Acknowledgements

This article arose from my PhD-thesis at the Department of Mathematics at the University of Munich. Therefore I first would like to thank my supervisor Martin Schottenloher for his continuous interest and valuable advice during the past three years. I am indepted to Julius Wess, who helped with advice and supported the plan of a research trip to Berkeley.

A grant by the "Graduiertenkolleg *Mathematik im Bereich ihrer Wechselwirkung mit der Physik*" and a travel grant by the German Academic Exchange Service (DAAD) are greatly acknowledged.

Nicolai Reshetikhin invited me to the Mathematics Department of the University of California at Berkeley, where many ideas for this work were born. I am grateful to Alan Weinstein for pointing out to me the articles of Harold Widom and to Robert Blattner, Claudio Emmrich, Dale Husemoller, Marc Rieffel and Pierre Schapira for valuable discussions.

# 1 Foundations

## 1.1 Deformation theory

In this paragraph we introduce the notion of a deformation. We keep the language close to algebraic geometry (see HARTSHORNE [27] plus references therein) and complex analysis (see GINDIKIN, KHENKIN [22]). That means we use sheaves of algebras respectively ringed spaces as the spaces to be deformed and as parameter spaces. The new aspect is that sheaves of noncommutative algebras are considered for deformation.

First we recall the notion of a ringed space. A **ringed space** is a pair $(X, \mathcal{A})$ where $X$ is a topological space and $\mathcal{A}$ a sheaf of rings on $X$. $\mathcal{A}$ is called the **structure sheaf** of the ringed space. It gives the topological space $X$ an additional structure like that of a smooth manifold, complex space, scheme or supermanifold and can be regarded as a space of germs of "admissible generalized functions" on $X$. If the stalks $\mathcal{A}_x$ with $x$ in $X$ are all commutative (resp. noncommutative, local) we call $(X, \mathcal{A})$ a commutative (resp. noncommutative, locally) ringed space. If all the sections $\mathcal{A}(U)$ with $U$ open in $X$ are $k$-algebras, where $k$ is a field, one says $(X, \mathcal{A})$ to be a $k$-ringed space or a ringed space over $k$. More generally if $\underline{A}$ is a subcategory of the category of rings and all the stalks $\mathcal{A}_x$ take values in $\underline{A}$, $(X, \mathcal{A})$ is said to be a ringed space with values in $\underline{A}$.

A **morphism of ringed spaces** from $(X, \mathcal{A})$ to $(Y, \mathcal{B})$ is a pair $(f, \phi)$, where $f : X \to Y$ is a continuous map and $\phi : \mathcal{B} \to f_*(\mathcal{A})$ a morphism of sheaves on $Y$ with values in the category of rings. Here $f_*(\mathcal{A})$ is the direct image of $\mathcal{A}$ via $f$, that means $f_*(\mathcal{A}(U)) = \mathcal{A}(f^{-1}(U))$ for all open $U \subset Y$. If the ringed spaces $(X, \mathcal{A})$ and $(Y, \mathcal{B})$ are locally ringed (resp. take values in $\underline{A}$), we require $\phi_x : \mathcal{B}_{f(x)} \to \mathcal{A}_x$ to be local (resp. to be a morphism in $\underline{A}$) for all $x$ in $X$. In case $(X, \mathcal{A})$ and $(Y, \mathcal{B})$ are ringed spaces over $k$, it is assumed that $\phi(U) : \mathcal{B}(U) \to \mathcal{A}(f^{-1}(U))$ is a homomorphism of $k$-algebras for all $U \subset X$ open.

It is obvious that ringed spaces (resp. ringed spaces with values in $\underline{A}$) and their morphisms form a subcategory of the category of sheaves on topological spaces.

Let us now write down some special commutative ringed spaces:

| | | | |
|---|---|---|---|
| $(\mathbb{R}^n, \mathcal{C})$, | where | $\mathcal{C}$ | - sheaf of continuous functions on $\mathbb{R}^n$, $n \in \mathbb{N}$, |
| $(\mathbb{R}^n, \mathcal{C}^r)$, | where | $\mathcal{C}^r$ | - sheaf of $r$-times continuously differentiable functions on $\mathbb{R}^n$, $n \in \mathbb{N}$, $r \in \mathbb{N}^* \cup \{\infty\}$, |
| $(\mathbb{R}^n, \mathcal{C}^\omega)$, | where | $\mathcal{C}^\omega$ | - sheaf of real analytic functions on $\mathbb{R}^n$, $n \in \mathbb{N}$, |
| $(\mathbb{C}^n, \mathcal{O})$, | where | $\mathcal{O}$ | - sheaf of holomorphic functions on $\mathbb{C}^n$, $n \in \mathbb{N}$. |

These examples of commutative ringed spaces can be generalized.

**Example 1.1** Let $n \in \mathbb{N}$ and $r \in \mathbb{N}^* \cup \{\infty\}$. A ringed space $(X, \mathcal{A})$ is called a

(i) topological manifold of dimension $n$, if $(X, \mathcal{A})$ is locally isomorphic to $(\mathbb{R}^n, \mathcal{C})$,

(ii) differentiable $r$-manifold of dimension $n$, if $(X, \mathcal{A})$ is locally isomorphic to $(\mathbb{R}^n, \mathcal{C}^r)$,

(iii) real analytic manifold of dimension $n$, if $(X, \mathcal{A})$ is locally isomorphic to $(U, \mathcal{C}^\omega)$, where $U \subset \mathbb{R}^n$ is open,

(iv) complex manifold of dimension $n$, if $(X, \mathcal{A})$ is locally isomorphic to $(U, \mathcal{O}_n|_U)$, where $U \subset \mathbb{C}^n$ is open,

(v) scheme, if for every $x \in X$ there exists an open neighbourhood $U$ of $x$, such that $(U, \mathcal{A}|_U)$ is isomorphic to an affine scheme.

The structure sheaf of a topological (resp. $r$-times differentiable, real analytic or complex) manifold $X$ is denoted by $\mathcal{C}_X$ (resp. $\mathcal{C}_X^r$, $\mathcal{C}_X^\omega$ or $\mathcal{O}_X$).

Supersymmetric structures (see WESS, BAGGER [64]) are our first examples of noncommutative ringed spaces. Most easily this can be seen with the definition of superspaces according to MANIN [34].

**Definition 1.2** *A* **superspace** *consists of a pair* $(M, \mathcal{O}_M)$, *where $M$ is a topological space and $\mathcal{O}_M$ a sheaf of supercommutative rings, such that all stalks $\mathcal{O}_{M,x}$, $x \in M$ are local.*

**Supermanifolds** are superspaces which locally split into an even and odd part such that the splitting is differentiable and the odd part is a locally free module sheaf over the even part.

**Example 1.3** Superspaces and supermanifolds are noncommutative ringed spaces.

The notion of a fibered morphism of ringed spaces is crucial for the definition of a deformation.

**Definition 1.4** *A morphism $(f, \phi) : (X, \mathcal{A}) \to (P, \mathcal{S})$ of ringed spaces is called* **fibered**, *if the following conditions are fulfilled:*

(i) $(P, \mathcal{S})$ *is a commutative locally ringed space.*

(ii) $f : X \to P$ *is surjective.*

(iii) $\phi_x : \mathcal{S}_{f(x)} \to \mathcal{A}_x$ *maps $\mathcal{S}_{f(x)}$ into the center of $\mathcal{A}_x$ for all $x \in X$.*

*The* **fiber** *of $F$ over a point $p$ of $P$ then is the ringed space $(X_p, \mathcal{A}_p)$ defined by*

$$X_p = f^{-1}(p), \quad \mathcal{A}_p = \mathcal{A}|_{f^{-1}(p)} / \mathfrak{m}_p \mathcal{A}|_{f^{-1}(p)},$$

*where $\mathfrak{m}_p$ is the maximal ideal in $\mathcal{S}_p$ which acts on $\mathcal{A}|_{f^{-1}(p)}$ via $\phi$.*

A fibered morphism of ringed spaces can be pictured in the following way:

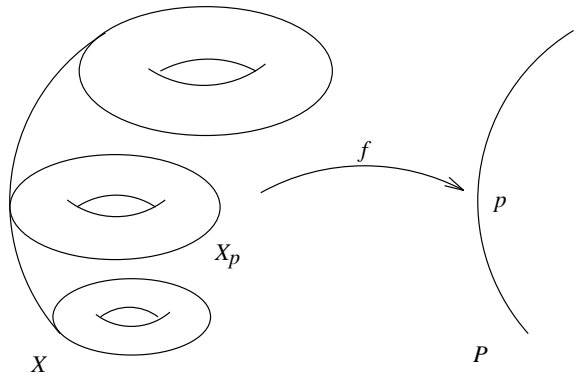

Additionally to this intuitive picture on the "space level", the sheaf morphism $\phi$ and condition $(iii)$ in 1.4 guarantee that the structure sheaves on $X$, $P$ and the $X_p$ fit to the above picture. More precisely does condition $(iii)$ imply that the stalks $\mathcal{A}_x$ are central extensions of $\mathcal{A}_x/\mathfrak{m}_{f(x)}\mathcal{A}_x$ by $\mathcal{S}_{f(x)}$.

**Remark 1.5** By using the language of sheaves resp. ringed spaces we allow the considered spaces to have singularities.

**Example 1.6** $(i)$ Let $P$ and $Y$ be smooth manifolds, $X = Y \times P$ the product of the two and $f = pr_2 : X \to P$ the projection on the second cooordinate. Then $(f, f_*) : (X, \mathcal{C}_X^\infty) \to (P, \mathcal{C}_P^\infty)$ gives rise to a fibered morphism, which we call **trivial**.

$(ii)$ More generally let $f : X \to P$ be a smooth fiber bundle. Then it induces a fibered morphism $(f, f_*) : (X, \mathcal{C}_X^\infty) \to (P, \mathcal{C}_P^\infty)$.

A deformation of a ringed space $(X, \mathcal{A})$ now consists of a fibered morphism $(d, \Delta) : (Y, \mathcal{B}) \to (P, \mathcal{S})$ such that one of its fibers is isomorphic to $(X, \mathcal{A})$ and such that the fibered morphism $(d, \Delta)$ is "locally trivial" in a certain sense. As the fibers of $(d, \Delta)$ in general neither comprise a (smooth) fiber bundle nor are finite dimensional, we have to express the condition of local triviality by another notion, namely the one of (algebraic) flatness (see Appendix A for the definition of flat modules and morphisms). It is a well-known fact in algebraic geometry and complex analysis (see HARTSHORNE [27]) that under certain conditions flatness is a good algebraic substitute for the geometric notion of local triviality. Therefore we will require that the morphism $\Delta : \mathcal{S} \to \mathcal{B}$ is flat resp. flatly filtered. Here we use flatly filtered morphisms to include the case, where the parameter spaces are neither algebraic nor complex but come from smooth manifolds. Flatness then is not anymore an appropriate substitute for "local triviality", but flat filteredness is. In abstract terms we use flat filteredness for the case when the stalks of the parameter space are not HAUSDORFF with respect to the KRULL topology.

**Definition 1.7** Let $(P, \mathcal{S})$ be a commutative locally ringed space over a field $k$, $*$ a distinguished point in $P$, and $(X, \mathcal{A})$ a $k$-ringed space.

A **deformation** of $(X, \mathcal{A})$ over the **parameter space** $(P, \mathcal{S})$ *(with **distinguished point** $*$)* is given by a fibered morphism $(d, \Delta) : (Y, \mathcal{B}) \to (P, \mathcal{S})$ over $k$ together with an isomorphism $(i, \iota) : (X, \mathcal{A}) \to (Y_*, \mathcal{B}_*)$ such that for every $p \in P$ and $y \in d^{-1}(p)$ the following condition holds:

$\Delta_y : \mathcal{S}_p \to \mathcal{B}_y$ is flatly filtered, that is it induces for every $m \in \mathbb{N}^* = \{1, 2, 3, ...\}$ a flat morphism $\Delta_y^m : \mathcal{S}_p/\mathfrak{m}_p^m \to \mathcal{B}_y/\mathfrak{m}_p^m \mathcal{B}_y$, where $\mathfrak{m}_p$ is the maximal ideal of $\mathcal{S}_p$, and $\mathcal{B}_y$ is filtered by $\mathcal{B}_y \subset \mathfrak{m}_p \mathcal{B}_y \subset ... \subset \mathfrak{m}_p^m \mathcal{B}_y \subset ...$ .

A deformation of $(X, \mathcal{A})$ is called **flat**, if for every $p \in P$ and $x \in d^{-1}(p)$ the homomorphism $\Delta_x : \mathcal{S}_p \to \mathcal{B}_x$ is flat.

Now let $(\tilde{d}, \tilde{\Delta}) : (\tilde{Y}, \tilde{\mathcal{B}}) \to (P, \mathcal{S})$, $(\tilde{i}, \tilde{\iota}) : (X, \mathcal{A}) \to (\tilde{Y}_*, \tilde{\mathcal{B}}_*)$ be a deformation over $(P, \mathcal{S})$ of a second $k$-ringed space $(\tilde{X}, \tilde{\mathcal{A}})$. Furthermore assume that we have a morphism $(f, \phi) : (X, \mathcal{A}) \to (\tilde{X}, \tilde{\mathcal{A}})$ of $k$-ringed spaces. Then a **morphism of deformations** from $(d, \Delta)$ to $(\tilde{d}, \tilde{\Delta})$ over $(f, \phi)$ is given by a morphism of $k$-ringed spaces $(b, \beta) : (Y, \mathcal{B}) \to (\tilde{Y}, \tilde{\mathcal{B}})$ such that the diagrams

$$\begin{array}{ccc}
(Y, \mathcal{B}) & \xrightarrow{(b, \beta)} & (\tilde{Y}, \tilde{\mathcal{B}}) \\
& \searrow (d,\Delta) \quad (\tilde{d},\tilde{\Delta}) \swarrow & \\
& (P, \mathcal{S}) &
\end{array} \qquad \begin{array}{ccc}
(X, \mathcal{A}) & \xrightarrow{(f, \phi)} & (\tilde{X}, \tilde{\mathcal{A}}) \\
(i,\iota) \downarrow & & \downarrow (\tilde{i},\tilde{\iota}) \\
(Y_*, \mathcal{B}_*) & \xrightarrow{(b|_{Y_*}, \beta|_{Y_*})} & (\tilde{Y}_*, \tilde{\mathcal{B}}_*)
\end{array}$$

commute.

It is obvious by definition that deformations resp. flat deformations over a commutative locally $k$-ringed space $(P, \mathcal{S})$ with distinguished point $*$ form a category $\underline{\text{Def}}_{*,(P,\mathcal{S})}$ resp. $\underline{\text{Def}}^{\text{flat}}_{*,(P,\mathcal{S})}$.

In case all the ringed spaces $(X, \mathcal{A})$, $(Y, \mathcal{B})$ and $(P, \mathcal{S})$ of a deformation $(d, \Delta) : (Y, \mathcal{B}) \to (P, \mathcal{S})$ have the form $(\{*\}, A)$, $(\{*\}, B)$ resp. $(\{*\}, S)$, where $*$ is a point and $A$, $B$, and $S$ are $k$-algebras, we sometimes say by abuse of language that the algebra $B$ or the algebra morphism $\Delta : S \to B$ is a deformation of $A$ over $S$.

Before giving some examples of deformations let us state a proposition which provides a useful criterion to decide whether a deformation is flat.

**Proposition 1.8** Let $(d, \Delta) : (Y, \mathcal{B}) \to (P, \mathcal{S})$ be a fibered morphism, where $(P, \mathcal{S})$ is commutative and locally ringed. Then the condition

(i) For every $p \in P$ and $y \in d^{-1}(p)$ the homomorphism $\Delta_y : \mathcal{S}_p \to \mathcal{B}_y$ is flatly filtered.

*implies condition*

(*ii*) *For every $p \in P$ and $y \in d^{-1}(p)$ the homomorphism $\Delta_y : \mathcal{S}_p \to \mathcal{B}_y$ is flat.*

*If for every $p \in P$ and $y \in d^{-1}(p)$ the stalk $\mathcal{S}_p$ is* NOETHERIAN *and the* KRULL *topology defined on $\mathcal{B}_y$ by the maximal ideal $\mathfrak{m}_p$ of $\mathcal{S}_p$ is* HAUSDORFF, (*i*) *and* (*ii*) *are equivalent.*

PROOF: This is a direct consequence from Theorem 1 in Chapter III, §5.2 of BOURBAKI [9], *Commutative Algebra*. □

**Example 1.9** (*i*) (Products of *k*-ringed spaces) Let $(X, \mathcal{A})$ be any *k*-ringed space and $(P, \mathcal{S})$ a *k*-scheme. Then the product $(X \times P, \mathcal{B}) = (X, \mathcal{A}) \times_k (P, \mathcal{S})$ is a flat deformation of $(X, \mathcal{A})$ with distinguished point $*$ for any closed point $* \in P$. This can be seen easily from the fact that $\mathcal{B}_{(x,p)} = \mathcal{A}_x \otimes_k \mathcal{S}_p$ for every $x \in X$ and $p \in P$.

(*ii*) (Deformation of a scheme) Consider the *k*-scheme $Y = \operatorname{Spec} k[x, y, t]/(xy - t)$. It gives rise to a fibration $Y \to \operatorname{Spec} k[t]$, whose fibers $Y_a$ with $a \in k$ are hyperbola $xy = a$, when $a \neq 0$, and consist of the two axes $x = 0$ and $y = 0$, when $a = 0$. Let us illustrate this deformation in the following picture.

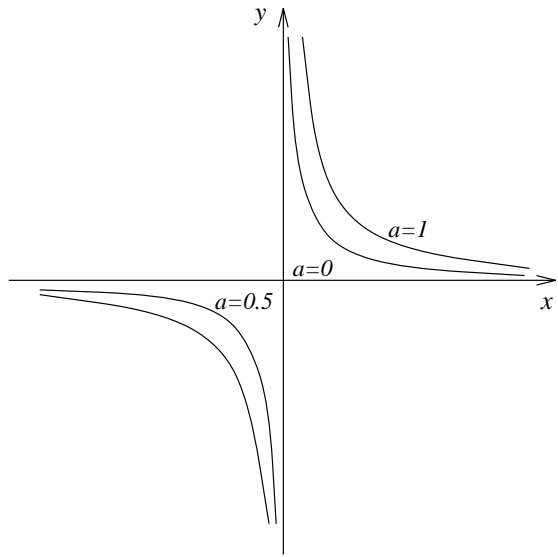

(Cf. HARTSHORNE [27], Example 3.3.2.)

(*iii*) (First order deformation of algebras) Consider a *k*-algebra $A$ and the algebra $S = k[\varepsilon]/\varepsilon^2 k[\varepsilon]$. Furthermore let $\alpha : A \times A \to A$ be a HOCHSCHILD 2-cocycle of $A$, in other words let the relation

$$a\,\alpha(b, c) - \alpha(ab, c) + \alpha(a, bc) - \alpha(a, b)\,c = 0 \tag{6}$$

hold for all $a, b, c \in A$. Then we can define a new $k$-algebra $B$, whose underlying linear structure is isomorphic to $A \otimes_k S$ and whose product is given by the following construction: Any element $\in B$ can be written uniquely in the form $= a + b\,\varepsilon = a \otimes 1 + b \otimes \varepsilon$ with $a, b \in A$. Then the product of $= a + b\,\varepsilon \in B$ and $= c + d\,\varepsilon \in B$ is given by

$$= a\,c + [\alpha(a, c) + ad + bc]\,\varepsilon. \tag{7}$$

By Eq. (6) this product is associative. We thus receive a flat deformation $\Delta : S \to B$ of the algebra $A$ and call it the **first order deformation** of $A$ through the HOCHSCHILD cocycle $\alpha$. See GERSTENHABER, SCHACK [21] for further information on this and the connection between deformation theory and HOCHSCHILD cohomology.

(*iv*) (Formal deformation of an algebra) Let us generalize the preceeding example. Assume again $A$ to be an arbitrary $k$-algebra and choose bilinear maps $\alpha_n : A \times A \to A$ for $n \in \mathbb{N}$ such that $\alpha_0$ is the product on $A$ and $\alpha_1$ a HOCHSCHILD cocycle. Furthermore let $S$ be the algebra $k[[t]]$ of formal power series in one variable over $k$, and $B = A[[t]]$ the space of formal power series in one variable with coefficients in $A$. Denoting by $\mathfrak{m}$ the maximal ideal in $k[[t]]$ the spaces $B/\mathfrak{m}^m \cong A \otimes_k k^m$ are for every integer $m \geq 1$ flat modules over the ring $k[[t]]/\mathfrak{m}^m \cong k^m$. As $k[[t]]$ is NOETHERIAN and the KRULL topology on $B$ defined by $\mathfrak{m}$ obviously is HAUSDORFF, Proposition 1.8 entails that $B$ is a flat $k[[t]]$-module. Now let us consider the following bilinear map:

$$\star : B \times B \to B, \quad \left( \sum_{n \in \mathbb{N}} a_n\, t^n, \sum_{n \in \mathbb{N}} b_n\, t^n \right) \mapsto \sum_{n \in \mathbb{N}} \sum_{\substack{k, l, m \in \mathbb{N} \\ k + l + m = n}} \alpha_m(a_k, b_l)\, t^n. \tag{8}$$

If $B$ together with $\star$ becomes a $k$-algebra or in other words if $\star$ is associative, we receive a flat deformation of $A$ over $S = k[[t]]$. In that case we say that $B$ is a **formal deformation** of $A$ through the family $(\alpha_n)_{n \in \mathbb{N}^*}$. Contrarily to the preceeding example there might not exist for every HOCHSCHILD cocycle $\alpha$ on $A$ a formal deformation $B$ of $A$ defined by a family $(\alpha_n)_{n \in \mathbb{N}^*}$ such that $\alpha_1 = \alpha$. In case it exists, we call the deformation $B$ of $A$ **induced** by $\alpha$. If the third HOCHSCHILD cohomology group $\mathrm{HH}^3(A, A)$ vanishes, there exists for every HOCHSCHILD cocycle $\alpha$ on $A$ a deformation $B$ of $A$ induced by $\alpha$. See again GERSTENHABER, SCHACK [21] for further information.

(*v*) (Formal deformation on symplectic manifolds) Let us consider the last two examples for the case $A = \mathcal{C}^\infty(M)$ where $M$ is a symplectic manifold. Then the POISSON bracket $\{\,,\,\}$ gives a HOCHSCHILD cocycle on $\mathcal{C}^\infty(M)$. There now exists a first order deformation of $C^\infty(M)$ through $\frac{1}{2i}\{\,,\,\}$ and, though $\mathrm{HH}^3(A,A)$ might not always vanish, even a formal deformation of $C^\infty(M)$ induced by $\frac{1}{2i}\{\,,\,\}$. This fact was first proven by DEWILDE and LECOMTE [69]. A second and more geometric proof was given by FEDOSOV [17].

(vi) (Quantized universal enveloping algebras according to DRINFELD) A **quantized universal enveloping algebra** for a complex LIE algebra $\mathfrak{g}$ is a HOPF algebra $A$ over $\mathbb{C}[[\hbar]]$ such that $A$ is a formal deformation (in the sense of Example (iv)) of the universal enveloping algebra $U\mathfrak{g}$ of $\mathfrak{g}$. In particular $(A/\hbar A)$ is isomorphic to $U\mathfrak{g}$, and $A$ is a flat $\mathbb{C}[[\hbar]]$-module. See DRINFELD [14] or KASSEL [31] for details and examples of quantized universal enveloping algebras.

(vii) (Quantum plane) Consider the tensor algebra $T = \bigoplus_{n \in \mathbb{N}} (\mathbb{R}^2)^{\otimes n}$ of the 2-dimensional real vector space $\mathbb{R}^2$, and let $(x,y)$ be the canonical basis of $\mathbb{R}^2$. Now form the tensor product sheaf $T_{\mathbb{C}^*} = T \otimes_{\mathbb{R}} \mathcal{O}_{\mathbb{C}^*}$ of $T$ with the sheaf of holomorphic functions on $\mathbb{C}^*$ and let $I_{\mathbb{C}^*}$ be the ideal sheaf in $T_{\mathbb{C}^*}$ generated by the relations

$$x \otimes y - z\,y \otimes x = 0, \qquad (9)$$

where $z : \mathbb{C}^* \to \mathbb{C}^*$ is the identity function. The quotient sheaf $B = B_{\mathbb{C}^*} = T_{\mathbb{C}^*}/I_{\mathbb{C}^*}$ then is a sheaf of $\mathbb{C}$-algebras and an $\mathcal{O}_{\mathbb{C}^*}$-module. Now using Eq. (9) move all occurances of $x$ in an element of $B_{\mathbb{C}^*}$ to the right of all $y$'s. By using the fact that $\frac{1}{z}$ is well-defined over $\mathbb{C}^*$ one can thus show that $B_{\mathbb{C}^*}$ is a free $\mathcal{O}_{\mathbb{C}^*}$-module. Hence $B_{\mathbb{C}^*}$ is flat over $\mathcal{O}_{\mathbb{C}^*}$. Further it is easy to see that for every $q \in \mathbb{C}^*$ the $\mathbb{C}$-algebra $A_q = B_q/\mathfrak{m}_q A_q$ is freely generated by elements $x, y$ with relations

$$x \otimes y - q\,y \otimes x = 0. \qquad (10)$$

We call $A_q$ the $q$**-deformed quantum plane** and $B$ the **over $\mathbb{C}^*$ universally deformed quantum plane**. Altogether we can now interpret $B$ as a flat deformation of $A_q$ over $\mathbb{C}^*$, in particular as a flat deformation of $A_1 = T \otimes_{\mathbb{R}} \mathbb{C} = \mathbb{C}[x,y]$, the algebra of complex polynomials in two generators.

In the same way one can deform function algebras on higher dimensional vector spaces as well as function algebras on certain LIE groups. We thus receive for example the quantum group $SU_q(2)$ as a deformation of a HOPF algebra of functions on $SU(2)$. See for example FADDEEV, RESHETIKHIN, TAKHTAJAN [16], MANIN [35] and WESS, ZUMINO [65] for more information on $q$-deformations of vector spaces, LIE groups, differential calculi and all that.

(viii) (Noncommutative Fiber Bundles) Let $SU_q(2)$ with $q \in \mathbb{C}^*$ be the $q$-deformed $SU(2)$. Then it is possible to construct a sheaf $\mathcal{M}$ over $S^4$ and for every $k \in \mathbb{Z}$ a sheaf $\mathcal{P}^k$ over $S^4$ together with a morphism $\rho_k : \mathcal{M} \to \mathcal{P}^k$ such that $\rho_k : \mathcal{M} \to \mathcal{P}^k$ can be regarded as a deformation of the classical $SU(2)$-bundle over $S^4$ with index $k$ by the parameter $q$. Therefore $\rho_k : \mathcal{M} \to \mathcal{P}^k$ is called the **noncommutative $SU_q(2)$-fiber bundle over $\mathcal{M}$** with index $k \in \mathbb{Z}$. Note that both sheaves $\mathcal{M}$ and $\mathcal{P}^k$ are not commutative for $q \neq 1$. Here we cannot go into the details of the construction of noncommutative fiber bundles but refer the interested reader to PFLAUM [40, 37] for exact definitions and statements.

Every deformation gives rise to an exact sequence

$$0 \longrightarrow \mathfrak{m}_* \mathcal{B}|_{Y_*} \longrightarrow \mathcal{B}|_{Y_*} \xrightarrow{\pi} \mathcal{A} \longrightarrow 0 \tag{11}$$

of sheaves on $Y_*$. A **trivialization** of $(d, \Delta)$ then is a multiplicative right inverse $\rho : \mathcal{A} \to \mathcal{B}|_{Y_*}$ of $\pi$. In case such a trivialization exists, we call the deformation $(d, \Delta)$ **trivial**.

**Example 1.10**  (*i*) Consider Example 1.9 (i). Obviously it has a trivialization induced by all maps $\mathcal{A}_x \to \mathcal{B}_{(x,p)}$, $a \mapsto a \otimes_k 1$ with $x \in X$ and $p \in P$.

(*ii*) If the HOCHSCHILD cocycle $\alpha$ in Example 1.9 (iii) is zero, the deformation $B$ of $A$ through $\alpha$ is obviously trivial. More generally any first order deformation of $A$ has a trivialization, if $\mathrm{HH}^2(A, A) = 0$. In the same way any formal deformation of $A$ is trivial, if $\mathrm{HH}^2(A, A)$ vanishes. See GERSTENHABER, SCHACK [21] and references therein for details.

## 1.2 Quantization

In this paragraph we introduce quantizations as deformations together with an appropriate family of isomorphisms which linearly map the ringed space to be deformed to the fibers of the deformation. In more physical terms do these isomorphisms associate to any classical observable one and only one quantized observable for any value of the deformation or quantization parameter. The quantization is called DIRAC, if the space to be quantized is a POISSON space and the commutator of two quantized variables is equal to the quantization of the POISSON bracket of the two observables up to higher orders of PLANCK'S constant which serves as deformation parameter. See Appendix A for some algebraic geometry notions used in the following.

**Definition 1.11** *A **deformation quantization** of a k-ringed space $(X, \mathcal{A})$ consists of a deformation $(d, \Delta) : (Y, \mathcal{B}) \to (P, \mathcal{S})$ of $(X, \mathcal{A})$ with distinguished point $*$ and a linear morphism $(q, \mathfrak{q}) : (Y, \mathcal{B}) \to (X, \mathcal{A})$ of k-ringed spaces, such that:*

*(i) $q : Y \to X$ is onto and transversal to $d$ in the sense that every fiber $q^{-1}(x)$ with $x \in X$ intersects $Y_p$ for any $p \in P$ in exactly one point.*

*(ii) (Correspondence principle) $(q, \mathfrak{q})$ is a linear left inverse to the canonical inclusion $(X, \mathcal{A}) \to (Y_*, \mathcal{B}_*) \to (Y, \mathcal{B})$.*

*If $(d, \Delta)$ is flat, the deformation quantization is called bf flat.*

Now let $(\tilde{q}, \tilde{\mathfrak{q}})$ with $(\tilde{d}, \tilde{\Delta}) : (\tilde{Y}, \tilde{\mathcal{B}}) \to (P, \mathcal{S})$, $(\tilde{\imath}, \tilde{\iota}) : (\tilde{X}, \tilde{\mathcal{A}}) \to (\tilde{Y}_*, \tilde{\mathcal{B}}_*)$ and $(\tilde{q}, \tilde{\mathfrak{q}}) : (\tilde{Y}, \tilde{\mathcal{B}}) \to (\tilde{X}, \tilde{\mathcal{A}})$ be a deformation quantization of a second k-ringed space $(\tilde{X}, \tilde{\mathcal{A}})$. Furthermore assume that we have a morphism $(f, \phi) : (X, \mathcal{A}) \to (\tilde{X}, \tilde{\mathcal{A}})$ of k-ringed spaces. Then a **morphism of deformation quantizations** from $(q, \mathfrak{q})$ to $(\tilde{q}, \tilde{\mathfrak{q}})$ over $(f, \phi)$ is given by a morphism $(b, \beta) : (Y, \mathcal{B}) \to (\tilde{Y}, \tilde{\mathcal{B}})$ such that $(b, \beta)$ gives rise to a morphism of deformations from $(d, \Delta)$ to $(\tilde{d}, \tilde{\Delta})$ over $(f, \phi)$, and such that the diagram

$$\begin{array}{ccc} (Y, \mathcal{B}) & \xrightarrow{(b, \beta)} & (\tilde{Y}, \tilde{\mathcal{B}}) \\ (q, \mathfrak{q}) \downarrow & & \downarrow (\tilde{q}, \tilde{\mathfrak{q}}) \\ (X, \mathcal{A}) & \xrightarrow[(f, \phi)]{} & (\tilde{X}, \tilde{\mathcal{A}}) \end{array}$$

*commutes.*

Note that, as all quantizations regarded in this work are deformation quantizations anyway, we usually just say quantization instead of deformation quantization. So we call $(q, \mathfrak{q})$ a (flat) quantization of $(X, \mathcal{A})$ over the deformation $(d, \Delta)$, or $\big((q, \mathfrak{q}), (d, \Delta)\big)$ a (flat) quantization of $(X, \mathcal{A})$.

It is obvious by definition that deformation quantizations (resp. flat deformation quantizations) over a commutative locally $k$-ringed space $(P, \mathcal{S})$ with distinguished point $*$ form a category $\underline{\mathsf{Qu}}_{*,(P,\mathcal{S})}$ (resp. $\underline{\mathsf{Qu}}^{\mathrm{flat}}_{*,(P,\mathcal{S})}$).

Let us now illustrate the "spacial part" of the definition in a little picture.

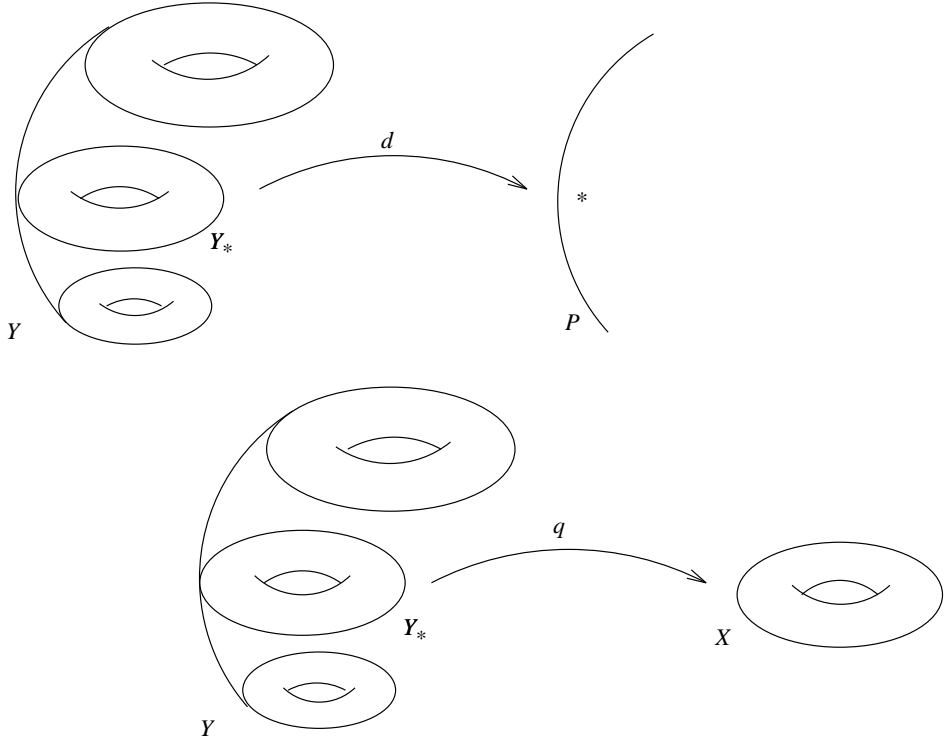

By condition $(i)$ above $Y$ almost looks like the product $X \times P$, and $q$ and $d$ can be regarded as the projections on the first and second coordinates respectively. But more important than this technical observation is condition $(ii)$. It namely expresses the fact that the morphism $\mathfrak{q} : \mathcal{A} \to \mathcal{B}$ is a quantization map, i.e. associates to any observable $f \in \mathcal{A}(U)$ with $U \subset X$ open a universally quantized observable $\mathfrak{q}(f) \in \mathcal{B}(q^{-1}(U))$ such that

$$\mathfrak{q}(f)|_U = f. \tag{12}$$

In other words this means that the classical limit of the universal quantization of a classical observable is the originally given observable. So the correspondence principle is fulfilled. If one now wants to quantize a classical observable $f \in \mathcal{A}(U)$ for a specific quantization parameter $p \in P$, one can form the universal quantization $\mathfrak{q}(f)$ of $f$ and then project this object down to the fiber lying over $p$. The resulting element $\mathfrak{q}_p(f) \in \mathcal{B}_p\left(q^{-1}(U) \cap Y_p\right)$ is the quantization of $f$ for the specific parameter $p$. The quantization $\mathfrak{q}(f)$ thus gives all possible quantizations of $f$, and $\mathfrak{q}_p(f)$ only the one for a specific point $p$ in $P$.

**Remark 1.12** The reason why we allow very general spaces as parameter spaces for a quantization lies in the fact that one for example would like to include some coupling constants besides the parameter $\hbar$ in the space $(P, \mathcal{S})$.

**Example 1.13** (i) Consider the trivial deformation $(d, \Delta) : (X \times P, \mathcal{B}) \to (P, \mathcal{S})$ of a $k$-ringed space $(X, \mathcal{A})$ in Example 1.9 (i). Now let $(q, \mathfrak{q}) : (X \times P, \mathcal{B}) \to (X, \mathcal{A})$ be the canonical morphism given by

$$q : X \times P \to X, \quad (x, p) \mapsto x \tag{13}$$

$$\mathfrak{q}_{(x,p)} : \mathcal{A}_x \to \mathcal{B}_{(x,p)} = \mathcal{A}_x \otimes_k \mathcal{S}_p, \quad f \mapsto f \otimes 1. \tag{14}$$

Then $(q, \mathfrak{q})$ obviously is a quantization of $(X, \mathcal{A})$ which we call a **trivial** or an **unphysical** one.

(ii) Consider the formal deformation $B$ of a $k$-algebra $A$ in Example 1.9 (iv). The natural injection $\mathfrak{q} : A \to B$ together with the trivial map $q : \{*\} \to \{*\}$ then give rise to a quantization of the ringed space $(\{*\}, A)$ over the (formal) parameter space $(\{*\}, k[[\hbar]])$. In that case we usually call $B$ a **formal deformation quantization** of $A$ over $k[[\hbar]]$.

In mathematical physics we are mainly interested in the quantization of a certain class of ringed spaces, namely those which can be interpreted as a classical phase space plus a sheaf of observables on that phase space. These ringed spaces are called POISSON spaces and will be defined in the following.

**Definition 1.14** *A **Poisson $k$-space** is a commutative $k$-ringed space $(X, \mathcal{A})$ together with a bilinear map $\{ \, , \, \} : \mathcal{A} \times \mathcal{A} \to \mathcal{A}$, called **Poisson bracket**, such that the following conditions hold:*

(i) *(anti-symmetry)*
$\{f, g\} = -\{g, f\}$ *for $f, g \in \mathcal{A}(U)$, $U \subset X$ open.*

(ii) *(JACOBI identity)*
$\{\{f, g\}, h\} + \{\{g, h\}, f\} + \{\{h, f\}, g\} = 0$ *for $f, g, h \in \mathcal{A}(U)$.*

(iii) *(LEIBNIZ rule)*
$\{f, gh\} = g\{f, h\} + \{f, g\}h$ *for $f, g, h \in \mathcal{A}(U)$.*

*A **morphism of Poisson $k$-spaces** is a morphism $(f, \phi) : (X, \mathcal{A}) \to (Y, \mathcal{B})$ of the underlying $k$-ringed spaces such that the equation*

$$\phi\bigl(\{a, b\}_Y\bigr) = \{\phi(a), \phi(b)\}_X \tag{15}$$

*holds for every $a, b \in \mathcal{B}(U)$, $U \subset Y$ open.*

Note that by definition the POISSON $k$-spaces form a category $\underline{\mathrm{Pois}}_k$. As in the following we will mainly consider the case $k = \mathbb{C}$, we just say POISSON space instead of POISSON $\mathbb{C}$-space.

**Example 1.15** (*i*) Consider the even dimensional EUCLIDEAN space $\mathbb{R}^{2n}$ and the sheaf $\mathcal{C}^{\infty}_{\mathbb{R}^{2n}}$ of smooth functions on it. Denote any element of $\mathbb{R}^{2n}$ by a $2n$-tupel $(x_1, ..., x_n, \xi_1, ..., \xi_n)$. Then the equation

$$\{f, g\} = \sum_{1 \leq k \leq n} \frac{\partial f}{\partial x_k} \frac{\partial g}{\partial \xi_k} - \frac{\partial f}{\partial \xi_k} \frac{\partial g}{\partial x_k} \tag{16}$$

with $f, g \in \mathcal{C}^{\infty}_{\mathbb{R}^{2n}}(U)$ and $U \subset \mathbb{R}^{2n}$ open defines on the ringed space $\left(\mathbb{R}^{2n}, \mathcal{C}^{\infty}_{\mathbb{R}^{2n}}\right)$ the structure of a POISSON space.

(*ii*) Let $Y$ be a symplectic manifold with symplectic form $\omega$. If $\mathcal{C}^{\infty}$ denotes the sheaf of differentiable functions on $Y$, define $\{\,,\,\} : \mathcal{C}^{\infty} \times \mathcal{C}^{\infty} \to \mathcal{C}^{\infty}$ by $\{f, g\} = \omega(X_f, X_g)$ with $f, g \in \mathcal{C}^{\infty}(U)$, $U \in X$ open and $X_f, X_g$ the Hamiltonian vector fields over $U$ associated to $f$ resp. $g$. Then $(Y, \mathcal{C}^{\infty})$ together with $\{\,,\,\}$ forms a POISSON space. By setting $Y = \mathbb{R}^{2n}$ (*i*) is a special case of this example. Now let $Y'$ be another symplectic manifold and $\Omega : Y \to Y'$ a symplectomorphism. Then the pullback $\Omega^* : \mathcal{C}^{\infty}_{Y'} \to \mathcal{C}^{\infty}_{Y}$ gives rise to a POISSON morphism from $(Y, \mathcal{C}^{\infty}_Y)$ to $(Y', \mathcal{C}^{\infty}_{Y'})$. We thus receive a functor from the category <u>Symp</u> of symplectic manifolds and symplectomorphisms to the category <u>Pois</u>$_{\mathbb{C}}$ of $\mathbb{C}$-POISSON spaces.

(*iii*) Let $X$ be any manifold, $\omega$ the canonical symplectic form on $T^*X$ and $\mathcal{D}$ be the direct image sheaf $\pi_* \mathcal{C}^{\infty}_{T^*X}$, where $\pi : T^*X \to X$ is the canonical projection and $\mathcal{C}^{\infty}_{T^*X}$ the sheaf of differentiable functions on $T^*X$. In more down to earth terms we then have $\mathcal{D}(U) = \mathcal{C}^{\infty}\left(\pi^{-1}(U)\right)$ for every $U \subset X$ open. Now define a POISSON bracket on $\mathcal{D}$ similar to (*i*) by $\{f, g\} = \omega(X_f, X_g)$ for $f, g \in \mathcal{D}(U)$. This bracket gives rise to a POISSON structure on the ringed space $(X, \mathcal{D})$.

(*iv*) A **Poisson manifold** is a manifold $Y$ together with a POISSON bracket $\{\,,\,\}$ on the sheaf of smooth functions on $Y$. Obviously a POISSON manifold is a POISSON space. See VAISMAN [58] as a reference for POISSON manifolds.

(*v*) More generally take a symplectic manifold $Y$ with (real) polarization $P$. Define $X$ to be the space of leaves of $P$ equipped with the quotient topology defined by the canonical projection $\pi : Y \to X$. Let $\mathcal{P}$ be the direct image $\pi_* \mathcal{C}^{\infty}_Y$ and pull down the POISSON bracket on $\mathcal{C}^{\infty}_Y$ to a bracket $\{\,,\,\}$ on $\mathcal{P}$ via $\pi_*$. We thus receive a POISSON space $(X, \mathcal{P})$ with bracket $\{\,,\,\}$ and call it associated to the symplectic manifold $(Y, \omega)$ and polarization $P$.

(*vi*) A **Poisson Lie group** is a LIE group $G$ together with a POISSON bracket $\{,\}$ on the sheaf $\mathcal{C}^{\infty}_G$ of smooth functions on $G$ such that the multiplication on $G$ gives rise to a morphism of POISSON spaces. Thus by definition every POISSON LIE group forms a POISSON space. See CHARI, PRESSLEY [10] for further information on POISSON LIE groups.

(*vii*) Let $G$ be a symmetry LIE group of a HAMILTONIAN system $(M, \omega, H)$, where $M$ is a smooth manifold with a symplectic form $\omega$, and $H$ a smooth function on $M$. Further assume that $G$ has an equivariant moment map $m : M \to \mathfrak{g}^*$. For any regular value $\xi \in \operatorname{im} m \subset \mathfrak{g}^*$ the set $M_\xi = m^{-1}(\xi)$ is a submanifold of $M$. Now there exists a foliation on $M_\xi$ generated by the "kernel distribution" $\omega^\perp$ of $\omega$. This gives rise to the quotient space $Q_\xi$ of all leaves of the foliation and to the sheaf $\mathcal{M}_\xi$ of all smooth functions on $M_\xi$ which are constant on the leaves. Note that $Q_\xi$ in general neither needs to be a smooth manifold nor even a HAUSDORFF topological space. Now push-down $\mathcal{M}_\xi$ via the canonical projection $\pi : M_\xi \to Q_\xi$ to the sheaf $\mathcal{Q} = \pi_*(\mathcal{M}_\xi)$. Then $\mathcal{Q}$ is a sheaf on $Q_\xi$, which canonically inherits a POISSON bracket from the one on $\mathcal{M}_\xi$. We thus receive a new POISSON space $(Q_\xi, \mathcal{Q})$, the so-called **Marsden-Weinstein quotient**.

**Definition 1.16** *Let $(X, \mathcal{A})$ be a POISSON space and $(P, \mathcal{S})$ be one of the following parameter spaces:*

$$
\begin{aligned}
(P, \mathcal{S}) &= (\mathbb{C}, \mathcal{O}) & & \textit{(holomorphic case)} \\
&= (\mathbb{R}, \mathcal{C}^\omega) & & \textit{(real analytic case)} \\
&= (\mathbb{R}, \mathcal{C}^\infty_\mathbb{R}), (\mathbb{C}, \mathcal{C}^\infty_\mathbb{C}) & & \textit{(differentiable case)} \\
&= (*, \mathcal{F}_1) = (*, \mathbb{C}[[\hbar]]) & & \textit{(formal case)}.
\end{aligned}
$$

*A holomorphic (resp. real analytic, differentiable or formal)* **Dirac quantization** *of $(X, \mathcal{A})$ is then given by a quantization $(q, \mathfrak{q}) : (Y, \mathcal{B}) \to (X, \mathcal{A})$ of $(X, \mathcal{A})$ over $(d, \Delta)$ with distinguished point $0$ (resp. $*$ in the formal case) such that the following condition holds: For every $f, g \in \mathcal{A}(U)$, $U \subset X$ open the commutator $[\mathfrak{q}(f), \mathfrak{q}(g)]$ can be expanded in the form*

$$[\mathfrak{q}(f), \mathfrak{q}(g)] = -i\, d\, \mathfrak{q}(\{f, g\}) + d^2\, r_{f,g}, \tag{17}$$

*where $r_{f,g} \in \mathcal{A}(U \times \mathbb{C})$, and $d$ denotes the function $d = \Delta(id_\mathbb{C}) : U \times \mathbb{C} \to \mathbb{C}, (x, \hbar) \to \hbar$ in the holomorphic, real analytic and differentiable case and the element $\Delta(\hbar) = \hbar$ in the formal case.*

*A morphism of holomorphic (resp. real analytic, differentiable or formal)* DIRAC *quantizations is given by a morphism of quantizations with parameter space $(P, \mathcal{S})$ over a morphism of* POISSON *spaces.*

It is obvious by definition that holomorphic (resp. real analytic, differentiable or formal) DIRAC quantizations form a category $\underline{\mathsf{Qu}}^{\text{Dirac}}_{(\mathbb{C}, \mathcal{O})}$ (resp. $\underline{\mathsf{Qu}}^{\text{Dirac}}_{(\mathbb{R}, \mathcal{C}^\omega)}$, $\underline{\mathsf{Qu}}^{\text{Dirac}}_{(\mathbb{R}, \mathcal{C}^\infty_\mathbb{R})}$, $\underline{\mathsf{Qu}}^{\text{Dirac}}_{(\mathbb{C}, \mathcal{C}^\infty_\mathbb{C})}$ or $\underline{\mathsf{Qu}}^{\text{Dirac}}_{(*, \mathcal{F}_1)}$).

**Remark 1.17** (*i*) Suppose we are given a holomorphic, real analytic or formal DIRAC quantization whose underlying deformation $(d, \Delta) : (Y, \mathcal{B}) \to (P, \mathcal{S})$ is such that the stalks of $\mathcal{B}$ are HAUSDORFF with respect to the KRULL topology generated by the maximal ideals of the stalks of $\mathcal{S}$. By Proposition 1.8 the deformation $(d, \Delta)$ then is automatically flat, as $\mathcal{O}, \mathcal{C}^\omega$ and $\mathcal{F}_1$ are NOETHERIAN.

(ii) In case the quantization is holomorphic, real analytic or differentiable, formula (17) gives rise to the equation

$$[\mathfrak{q}(f), \mathfrak{q}(g)]_\hbar = -i\hbar\, \mathfrak{q}(\{f,g\})_\hbar + \hbar^2\, (r_{f,g})_\hbar \tag{18}$$

over the fiber $Y_\hbar = d^{-1}(\hbar)$ of $\hbar \in \mathbb{C}$. This is the more familiar form of the DIRAC quantization (modulo higher orders of $\hbar$).

Let us now give some examples for quantizations of certain POISSON spaces in order to clarify and justify the definition of a DIRAC quantization.

**Example 1.18** (WEYL calculus and MOYAL product) Consider the symplectic manifold $T^*\mathbb{R}^n \cong \mathbb{R}^{2n}$ with its canonical symplectic structure $\omega = \sum_{k=1}^{n} dx_k \wedge d\xi_k$, where $(x_1, ..., x_n, \xi_1, ..., \xi_n)$ are canonical coordinates. We want to show that the MOYAL product, which will be defined in the following with the help of the WEYL calculus, gives rise to a quantization of a certain POISSON space of so-called symbols on $\mathbb{R}^n$. For the required definitions and theorems about symbol spaces, asymptotic expansions and all that we refer the reader to appendix B, in particular Definition B.1, or to HÖRMANDER [29], GRIGIS, SJÖSTRAND [24] and SHUBIN [49].

The POISSON space to be quantized in this example is $(\mathbb{R}^n, S_{1,0}^\infty)$, where the symbol sheaf $S_{1,0}^\infty$ is defined by $S_{1,0}^\infty(U) = \bigcup_{\mu \in \mathbb{R}} S_{1,0}^\mu(U)$ for every $U \subset \mathbb{R}^n$ open and by

$$\begin{aligned} S_{1,0}^\mu(U) = \{ b \in \mathcal{C}^\infty(U \times \mathbb{R}^n) : \forall K \subset U \text{ compact}, \alpha, \beta \in \mathbb{N}^n \; \exists C_{K,\alpha,\beta} > 0 \text{ s.t.} \\ \forall x \in K \; \forall \xi \in \mathbb{R}^n \sup_{x \in K} \left| \partial_x^\alpha \partial_\xi^\beta a(x,\xi) \right| \leq C_{K,\alpha,\beta} (1+|\xi|)^{\mu-\alpha} \} \end{aligned} \tag{19}$$

for every $\mu \in \mathbb{R}$. The POISSON bracket on $S_{1,0}^\infty(U)$ is the one induced by $\omega$ (cf. Examples 1.15 (i) and (iii)). Furthermore let $\Psi_{1,0}^\infty$ be the sheaf of pseudo-differential operators of type $(1,0)$ on $\mathbb{R}^n$.

Now we can define for any symbol $a \in S_{1,0}^\mu(U)$ with $U$ convex and any $\hbar \in \mathbb{R}^*$ the pseudo-differential operator $\Psi_\hbar^W(a) \in \Psi_{1,0}^\mu(U)$ by the following oscillatory integral:

$$\left[\Psi_\hbar^W(a) u\right](x) = \frac{1}{(2\pi)^n} \int\int e^{i\hbar \langle \xi, (x-y) \rangle} a\left(\frac{x+y}{2}, \xi\right) u(y) \, dy \, d\xi, \tag{20}$$

where $u \in \mathcal{D}_0^\infty(U)$. The operator $\Psi_\hbar^W(a)$ is called the **Weyl quantization** of $a$ for $\hbar$. Note that in physics $\Psi_\hbar^W(a)$ is usually formally written as

$$\Psi_\hbar^W(a) = \frac{1}{(2\pi)^n} \int\int \hat{a}(y,\zeta)\, e^{i \sum_{k=1}^{n} y_k \cdot \mathfrak{q}_\hbar(\xi_k) + \zeta_k \cdot \mathfrak{q}_\hbar(x_k)} dy\, d\zeta, \tag{21}$$

where $\hat{a}$ is the FOURIER transform $\hat{a}(y,\zeta) = \frac{1}{(2\pi)^n} \int\int e^{-i(\langle \xi, y \rangle + \langle \zeta, x \rangle)} a(x, \xi)\, dx\, d\xi$, $\mathfrak{q}_\hbar(\xi_k) = -i\hbar \frac{\partial}{\partial x_k}$ and $\mathfrak{q}_\hbar(x_k)$ is the multplication operator with $x_k$. As the set of open convex subsets of $\mathbb{R}^n$ is a basis of the topology of $\mathbb{R}^n$, the intersection of two convex sets is again convex and $\Psi_{1,0}^\infty$ is a sheaf, we can construct for every $U \subset \mathbb{R}^n$ open and $a \in S_{1,0}^\mu(U)$ a section $\Psi_\hbar^W(a) \in \Psi_{1,0}^\infty(U)$ by patching together the WEYL quantizations on open convex subsets of $U$.

Let us consider the product of $\Psi_\hbar^W(a)$ and $\Psi_\hbar^W(b)$ with $a,b \in S_{1,0}^\infty(U)$. By methods of the theory of pseudo-differential operators given for example in SHUBIN [49] it is a well-known fact that $\Psi_\hbar^W$ has an inverse $\sigma_\hbar^W : \Psi_{1,0}^\mu(U) \to S_{1,0}^\mu(U)$, which asssociates to every pseudo-differential operator its so-called **Weyl symbol**. Furthermore one knows that for symbols $a,b$ and $\hbar \neq 0$ the relation

$$\Psi_\hbar^W(a) \cdot \Psi_\hbar^W(b) = \Psi_\hbar^W(c) \tag{22}$$

holds, where the symbol $c$ has the following asymptotic expansion:

$$c(x,\xi) \sim \sum_{\alpha,\beta \in \mathbb{N}^n} (-1)^{|\beta|} \left(\frac{-i\hbar}{2}\right)^{|\alpha+\beta|} \left(\frac{\partial^{|\alpha|}}{\partial \xi^\alpha} \frac{\partial^{|\beta|}}{\partial x^\beta} a(x,\xi)\right) \left(\frac{\partial^{|\beta|}}{\partial \xi^\beta} \frac{\partial^{|\alpha|}}{\partial x^\alpha} b(x,\xi)\right). \tag{23}$$

In particular we have

$$\left[\Psi_\hbar^W(a), \Psi_\hbar^W(b)\right] = -i\hbar \, \Psi_\hbar^W(\{a,b\}) + \hbar^2 R, \quad R \in \Psi_{1,0}^\infty(U), \tag{24}$$

which justifies to call $\Psi_\hbar^W$ a quantization map. MOYAL now used Eq. (24) and the above symbol map to define an $\hbar$-dependent family of noncommutative products on a space of smooth functions on $\mathbb{R}^{2n}$. In particular MOYAL thus constructed a deformation respectively a quantization of a space of classsical observables on the symplectic manifold $\mathbb{R}^{2n}$.

Let us translate in the following MOYAL'S ideas to our deformation theoretical language. The arguments will be rather sketchy, as similar ones will be used later on in section 3 in a more general context. Define on the sheaf $S_{1,0}^\infty$ for $\hbar \neq 0$ a noncommutative associative product by

$$S_{1,0}^\infty(U) \times S_{1,0}^\infty(U) \ni (a,b) \mapsto \sigma_\hbar^W\left(\Psi_\hbar^W(a)\Psi_\hbar^W(b)\right) \in S_{1,0}^\infty(U). \tag{25}$$

Denote the resulting sheaf of algebras by $S_{1,0;\hbar}^\infty$, and by $S_{1,0;0}^\infty$ or just $S_{1,0}^\infty$ the sheaf $S_{1,0}^\infty$ with its canonical (commutative) algebra structure of pointwise multiplication of functions. Next construct the sheaf $S_{1,0;d}^\infty$ on $\mathbb{R}^{n+1}$, such that $S_{1,0;d}^\infty(U \times O)$ for $U \subset \mathbb{R}^n$ and $O \subset \mathbb{R}$ open consists of all smooth functions $U \times \mathbb{R}^n \times O \to \mathbb{C}$ whose restriction to $U \times \mathbb{R}^n$ (uniformly) belongs to $S_{1,0}^\infty(U)$. Embed $\mathcal{C}_\mathbb{R}^\infty$ into $S_{1,0;d}^\infty$ by $f \overset{\Delta}{\mapsto} f \circ pr_{2n+1}$ and check that one can give $S_{1,0;d}^\infty$ a unique structure of a sheaf of algebras such that the fibers of the morphism $\left(pr_{n+1}, \Delta\right) : \left(\mathbb{R}^{n+1}, S_{1,0;d}^\infty\right) \to (\mathbb{R}, \mathcal{C}^\infty)$ are isomorphic to $S_{1,0;\hbar}^\infty$. Thus we receive a deformation of $S_{1,0}^\infty$ and by Eq. (24) even a differentiable DIRAC quantization of $\left(\mathbb{R}^n, S_{1,0}^\infty\right)$.

**Example 1.19** (DEWILDE, LECOMTE; FEDOSOV) There exists a formal DIRAC quantization of the sheaf of smooth functions on a symplectic manifold $M$.

PROOF: By DEWILDE, LECOMTE [69] or FEDOSOV [17] there exists a noncommutative associative product $m_\hbar$ (cf. Example 1.9 $(iv)$) on the sheaf $\mathcal{A} = \mathcal{C}_M^\infty[[\hbar]]$ of formal power series in $\hbar$ with coefficients in $\mathcal{C}_M^\infty$ such that for $U \subset M$ open the following conditions hold:

(i) $m_\hbar(f, g) = fg + \frac{\hbar}{2i}\{f, g\} + ...$ for every $f, g \in \mathcal{C}_M^\infty(U)$.

(ii) $m_\hbar(\hbar^k f, \hbar^l g) = \hbar^{k+l} m_\hbar(f, g)$ for every $k, l \in \mathbb{N}$ and $f, g \in \mathcal{C}_M^\infty(U)$.

(iii) $m_\hbar(z, f) = m_\hbar(f, z) = zf$ for every constant function $z \in \mathbb{C}$ and every $f \in \mathcal{C}_M^\infty(U)$.

Now the canonical embeddings $\phi_U : \mathbb{C}[[\hbar]] \to \mathcal{A}(U)$ with $U \subset M$ open induces a fibered morphism $(f, \phi) : (M, \mathcal{A}) \to \left(\{*\}, \mathbb{C}[[\hbar]]\right)$, where $f : M \to \{*\}$ is the constant function. The stalk $\mathcal{A}_m$ with $m \in M$ is a $\mathbb{C}[[\hbar]]$-module. The maximal ideal $\mathfrak{m} = \hbar\mathbb{C}[[\hbar]]$ induces on $\mathcal{A}_m$ a topological structure, namely the KRULL-topology with respect to the filtration

$$\mathcal{A}_m \supset \mathfrak{m}\mathcal{A}_m \supset \mathfrak{m}^2\mathcal{A}_m \supset ... \supset \mathfrak{m}^n\mathcal{A}_m \supset ... . \tag{26}$$

Now the KRULL-topology on $\mathcal{A}_m$ is HAUSDORFF, and $\mathcal{A}_m/\mathfrak{m}^n\mathcal{A}_m \cong \mathcal{C}_m^\infty \otimes_\mathbb{C} \mathbb{C}[[\hbar]]/\mathfrak{m}^n$ for every $n \in \mathbb{N}^*$. So $\mathcal{A}_m/\mathfrak{m}^n\mathcal{A}_m$ is a flat $\mathbb{C}[[\hbar]]/\mathfrak{m}^n$-module. As $\mathbb{C}[[\hbar]]$ is NOETHERIAN, Proposition 1.8 implies that $\mathcal{A}_m$ is a flat $\mathbb{C}[[\hbar]]$-module. Furthermore we have $\mathcal{A}_m\big/\mathfrak{m}\mathcal{A}_m \cong \mathcal{C}_m^\infty$, so the fiber $(M, \mathcal{A}_*)$ is isomorphic to $(M, \mathcal{C}_M^\infty)$. The quantization morphism $\mathfrak{q} : \mathcal{C}_M^\infty \to \mathcal{A}$ is canonically given by $\mathfrak{q}(f) = f$ for any $f \in C_M^\infty(U)$. Hence by the above condition (i) DIRAC'S quantization condition holds, and the claim follows. □

**Remark 1.20** (i) Further results on the formal quantization of POISSON spaces are given in DONIN [13].

(ii) Formal quantizations lack of physical applicability, as $\hbar$ cannot be given a nonvanishing numerical value. We therefore try to circumvent this obstacle by establishing concrete resp. convergent deformation quantizations. This will be the main object of discussion from section 2 on.

## 1.3 Product and monoidal structures

It is a well-known principle in quantum physics that the state space of two coupled systems is the tensor product of the state spaces of both systems. In classical mechanics a similar situation holds: the phase space resp. configuration space of two systems is the CARTESIAN product of both phase spaces resp. configuartion spaces. A quantization theory now should obey these principles or, to put it differently, the category of deformation quantizations resp. appropriate subcategories of it should have a natural monoidal structure. We will now show in this section that this is the case indeed.

Let us first define the category $\underline{\mathsf{Esp}}^{\mathcal{C}^\infty}$ of smooth ringed spaces. The class of objects consists of all ringed spaces $(X, \mathcal{A})$, such that $X$ is a smooth manifold and such that the structure sheaf $\mathcal{A}$ carries besides its algebra structure that one of a locally free (left) $\mathcal{C}_X^\infty$-module. The morphisms $(f, \varphi)$ in $\underline{\mathsf{Esp}}^{\mathcal{C}^\infty}$ all have the property that $f$ is a smooth map between manifolds. In $\underline{\mathsf{Esp}}^{\mathcal{C}^\infty}$ there now exists a natural monoidal structure given by the following. Let $(X, \mathcal{A})$ and $(Y, \mathcal{B})$ be objects in $\underline{\mathsf{Esp}}^{\mathcal{C}^\infty}$. Then regard the projections $pr_1 : X \times Y \to X$ and $pr_2 : X \times Y \to Y$ on the first resp. second coordinate and define the external tensor product $\mathcal{A} \boxtimes \mathcal{B}$ by

$$\mathcal{A} \boxtimes \mathcal{B} = pr_1^{-1}\mathcal{A} \otimes pr_2^{-1}\mathcal{B}. \tag{27}$$

Defining
$$(X, \mathcal{A}) \times (Y, \mathcal{B}) = (X \times Y, \mathcal{A} \boxtimes \mathcal{B}) \tag{28}$$
we get a product $\times$ on $\underline{\mathsf{Esp}}^{\mathcal{C}^\infty}$ with projections $(pr_1, \mathfrak{p}_1) : (X \times Y, \mathcal{A} \boxtimes \mathcal{B}) \to (X, \mathcal{A})$ and $(pr_2, \mathfrak{p}_2) : (X \times Y, \mathcal{A} \boxtimes \mathcal{B}) \to (Y, \mathcal{B})$, where $\mathfrak{p}_1(f) = f \otimes 1$ and $\mathfrak{p}_2(g) = 1 \otimes g$ for $f \in \mathcal{A}(U)$, $g \in \mathcal{B}(V)$ and $U \subset X$, $V \subset Y$ open. The product $\times$ also gives rise to a tensor product functor on $\underline{\mathsf{Esp}}^{\mathcal{C}^\infty}$, where the unit is obviously given by the one point space $(\{*\}, \mathbb{C})$ with its canonical structure of a zero dimensional manifold.

**Note 1.21** By using the inverse image sheaves $pr_1^{-1}\mathcal{A}$ and $pr_2^{-1}\mathcal{B}$ and the external tensor product $\mathcal{A} \boxtimes \mathcal{B}$ we bypass the problem of tackling with topologically completed tensor products. In particular we get advantage of the fact that $pr_1^{-1}\mathcal{A}$ and $pr_2^{-1}\mathcal{B}$ are $\mathcal{C}_{X \times Y}^\infty$-modules by definition, so $\mathcal{A} \boxtimes \mathcal{B}$ is one, too.

Next let $(P, \mathcal{S}) = (P, \mathcal{C}_P^\infty)$ be a smooth manifold and $(f, \phi) : (X, \mathcal{A}) \to (P, \mathcal{S})$ a fibered morphism. We then call $(f, \phi)$ a **smooth fibered morphism**, if it is a morphism in $\underline{\mathsf{Esp}}^{\mathcal{C}^\infty}$ and if $f$ is a submersion. Note that $(P, \mathcal{S})$ is a commutative locally ringed space and an object in $\underline{\mathsf{Esp}}^{\mathcal{C}^\infty}$. Now consider the subcategory $\underline{\mathsf{Def}}^{\mathcal{C}^\infty}_{*,(P,\mathcal{S})}$ of $\underline{\mathsf{Def}}_{*,(P,\mathcal{S})}$ consisting of all deformations such that the underlying fibered morphism is a smooth one. If then $(d, \Delta) : (Y, \mathcal{B}) \to (P, \mathcal{S})$ and $(\tilde{d}, \tilde{\Delta}) : (\tilde{Y}, \tilde{\mathcal{B}}) \to (P, \mathcal{S})$ denote deformations in $\underline{\mathsf{Def}}^{\mathcal{C}^\infty}_{*,(P,\mathcal{S})}$ of $(X, \mathcal{A})$ resp. $(\tilde{X}, \tilde{\mathcal{A}})$, the fiber product $(Y, \mathcal{B}) \times_{(P,\mathcal{S})} (\tilde{Y}, \tilde{\mathcal{B}}) = (Y \times_P \tilde{Y}, \mathcal{B} \boxtimes_\mathcal{S} \tilde{\mathcal{B}})$ in $\underline{\mathsf{Esp}}^{\mathcal{C}^\infty}$ is well-defined, as $\mathcal{S}$ lies in the center of $\mathcal{B}$ and $\tilde{\mathcal{B}}$. Note that hereby

$$\mathcal{B} \boxtimes_\mathcal{S} \tilde{\mathcal{B}} = pr_1^{-1}\mathcal{B} \otimes_\mathcal{S} pr_2^{-1}\tilde{\mathcal{B}}. \tag{29}$$

Furthermore $(Y, \mathcal{B}) \times_{(P,\mathcal{S})} (\tilde{Y}, \tilde{\mathcal{B}}) \to (P, \mathcal{S})$ is a smooth fibered morphism such that the fiber over $*$ is isomorphic to the product $(X, \mathcal{A}) \times (\tilde{X}, \tilde{\mathcal{A}})$. If we can now show that the condition of flat filteredness is fulfilled for $(Y, \mathcal{B}) \times_{(P,\mathcal{S})} (\tilde{Y}, \tilde{\mathcal{B}}) \to (P, \mathcal{S})$, this morphism even comprises a deformation of $(X, \mathcal{A}) \times (\tilde{X}, \tilde{\mathcal{A}})$ and can be interpreted as the product of $(d, \Delta)$ and $(\tilde{d}, \tilde{\Delta})$ in $\underline{\mathsf{Def}}^{\mathcal{C}^\infty}_{*,(P,\mathcal{S})}$. So let us show the flat filteredness and first prove a little lemma.

**Lemma 1.22** *Let $(f, \phi) : (Y, \mathcal{B}) \to (P, \mathcal{S})$ be a smooth fibered morphism. Then the sheaf $\mathcal{A}$ is flatly filtered over $\mathcal{S}$.*

PROOF: As $\mathcal{B}$ is a locally free $\mathcal{C}_Y^\infty$-module sheaf, it suffices to show that $\mathcal{C}_Y^\infty$ is flatly filtered over $\mathcal{S} = \mathcal{C}_P^\infty$. So assume $p \in P$, $y \in f^{-1}(p)$, $m \in \mathbb{N}$ and let $\mathfrak{m}_p$ be the maximal ideal in $\mathcal{S}_p = \mathcal{C}_p^\infty$. Then as $f$ is a submersion, there exists an open neighborhood $U$ of $y$ in $Y$ such that $V = f^{-1}(p) \cap U$ is a submanifold of $Y$. So the stalk $\mathcal{C}_{V,y}^\infty$ of smooth functions on $V$ at $y$ is well-defined, and we have

$$\mathcal{C}_y^\infty \big/ \mathfrak{m}_p^m \mathcal{C}_y^\infty \;\cong\; \mathcal{C}_{V,y}^\infty \otimes_\mathbb{C} \left( \mathcal{S}_p \big/ \mathfrak{m}_p^m \right). \tag{30}$$

But the latter $\mathcal{S}_p \big/ \mathfrak{m}_p^m$-module is flat, hence the claim follows. $\square$

Now $\mathcal{B} \boxtimes_{(P,\mathcal{S})} \tilde{\mathcal{B}}$ is a locally free $\mathcal{C}_{Y \times_P \tilde{Y}}^\infty$-module sheaf, as $\mathcal{B}$ and $\tilde{\mathcal{B}}$ are locally free over $\mathcal{C}_Y^\infty$ resp. $\mathcal{C}_{\tilde{Y}}^\infty$. But then the above lemma entails that $\mathcal{B} \boxtimes_{(P,\mathcal{S})} \tilde{\mathcal{B}}$ is flatly filtered over $\mathcal{S}$. Thus we can construct for two deformations $(d, \Delta)$ and $(\tilde{d}, \tilde{\Delta})$ their product $(d, \Delta) \times (\tilde{d}, \tilde{\Delta})$ in $\underline{\mathsf{Def}}_{*,(P,\mathcal{S})}^{\mathcal{C}^\infty}$. Together with the **null deformation** $D_\mathrm{n} = 1_{(P,\mathcal{S})} : (P, \mathcal{S}) \to (P, \mathcal{S})$ as unit we receive a tensor product $\times$ on $\underline{\mathsf{Def}}_{*,(P,\mathcal{S})}^{\mathcal{C}^\infty}$.

The same idea now leads to a product and a tensor product in the subcategory $\underline{\mathsf{Qu}}_{*,(P,\mathcal{S})}^{\mathcal{C}^\infty}$ of $\underline{\mathsf{Qu}}_{*,(P,\mathcal{S})}$ consisting of quantizations $(q, \mathfrak{q})$ over $(d, \delta)$ such that $(d, \delta)$ lies in $\underline{\mathsf{Def}}_{*,(P,\mathcal{S})}^{\mathcal{C}^\infty}$ and $(q, \mathfrak{q})$ is a morphism in $\underline{\mathsf{Esp}}^{\mathcal{C}^\infty}$. From quantizations $(q, \mathfrak{q}) : (Y, \mathcal{B}) \to (X, \mathcal{A})$ and $(\tilde{q}, \tilde{\mathfrak{q}}) : (\tilde{Y}, \tilde{\mathcal{B}}) \to (\tilde{X}, \tilde{\mathcal{A}})$ in $\underline{\mathsf{Qu}}_{*,(P,\mathcal{S})}^{\mathcal{C}^\infty}$ over $(d, \Delta)$ resp. $(\tilde{d}, \tilde{\Delta})$ construct $((q, \mathfrak{q}) \times (\tilde{q}, \tilde{\mathfrak{q}})) : (Y, \mathcal{B}) \times_{(P,\mathcal{S})} (\tilde{Y}, \tilde{\mathcal{B}}) \to (X, \mathcal{A}) \times (\tilde{X}, \tilde{\mathcal{A}})$ and check that together with $(d, \Delta) \times (\tilde{d}, \tilde{\Delta})$ this forms the claimed product. The product functor $\times$ in $\underline{\mathsf{Qu}}_{*,(P,\mathcal{S})}^{\mathcal{C}^\infty}$ is also a tensor product with unit given by the **null quantization** $Q_\mathrm{n} : (P, \mathcal{S}) \to (*, \mathbb{C})$ over the null deformation.

In the same way like above we can construct the category $\underline{\mathsf{Esp}}^\mathcal{O}$ of holomorphic ringed spaces and define the categories $\underline{\mathsf{Def}}_{*,(P,\mathcal{S})}^\mathcal{O}$ and $\underline{\mathsf{Qu}}_{*,(P,\mathcal{S})}^\mathcal{O}$ of flat deformations and quantizations in $\underline{\mathsf{Esp}}^\mathcal{O}$. Like for the smooth case there exist canonical products respectively tensor products on $\underline{\mathsf{Esp}}^\mathcal{O}$, $\underline{\mathsf{Def}}_{*,(P,\mathcal{S})}^\mathcal{O}$ and $\underline{\mathsf{Qu}}_{*,(P,\mathcal{S})}^\mathcal{O}$.

**Remark 1.23** We have introduced the categories $\underline{\mathsf{Esp}}^{\mathcal{C}^\infty}$ and $\underline{\mathsf{Esp}}^\mathcal{O}$ to have a notion of noncommutatively ringed spaces such that the structure sheaves have certain common features, namely are locally free modules of smooth resp. holomorphic functions. There are more and easier to carry out constructions - like for example the construction of a tensor product - in $\underline{\mathsf{Esp}}^{\mathcal{C}^\infty}$ and $\underline{\mathsf{Esp}}^\mathcal{O}$ than in the huge category of arbitrary ringed spaces.

# 2 Quantization of a Cotangent Bundle

Our goal in this section is to define a DIRAC quantization of the POISSON space of polynomial observables associated to the cotangent bundle $T^*X$ of a RIEMANNIAN manifold $X$. The manifold $X$ is regarded as the **configuration space** of a classical physical system and the symplectic manifold $T^*X$ with its canonical symplectic structure as the **phase space** or **space of states** of the considered system. By quantizing polynomial observables on $T^*X$ we want to construct a model for the abstract definition of a quantization given in the first section. The hereby considered space of polynomial observables gives rise to a POISSON space and is large enough to contain all physically interesting observables. It is not so much a problem to find a deformation of this space, though the construction of the deformation is somewhat technical, but more to find an appropriate quantization map. Using the fact that $X$ carries a natural connection which gives rise to normal coordinates on $X$ we succeed in constructing such a quantization map. Then, roughly speaking, a classical polynomial observable is quantized by first expanding the observable over a point of $X$ in normal coordinates of that point. Afterwards its momentum and spacial components are quantized separately in such a way that momentum observables - expressed in normal coordinates - are written to the right of the spacial observables. Therefore this quantization corresponds to what physicists would call a normal order quantization. After having constructed a DIRAC quantization for polynomial observables we show that it even gives rise to a functor going from the category of RIEMANNIAN manifolds to differentiable DIRAC quantizations such that the natural monoidal structures on these categories are preserved.

In case $X$ is a complex manifold we also define a deformation of the sheaf of polynomial observables over $T^*X$ and then extend it in section 2.5 to a deformation of the ringed space $(X, \mathcal{D}_0^\infty)$ of holomorphic functions on $T^*X$ which are of bounded growth.

## 2.1 Deformed polynomial observables

In this first paragraph of section 2 we will show, how the sheaf of polynomial observables on the cotangent bundle $T^*X$ can be deformed. The parameter space $(P, \mathcal{S})$ will be given by $P = \mathbb{R}$ or $\mathbb{C}$ and by the sheaf $\mathcal{S}$ of smooth (resp. holomorphic) functions on $P$.

So let $X$ be a smooth (resp. complex) manifold, $\mathcal{E}_X$ the sheaf of smooth complex valued (resp. holomorphic) functions on $X$ and for every $\nu \in \mathbb{N}$ let $\mathcal{E}_{T^*X}(\nu)$ the sheaf $\mathcal{C}^\infty_{T^*X}(\nu)$ (resp. $\mathcal{O}_{T^*X}(\nu)$) on $X$ consisting of complex valued smooth (resp. holomorphic) functions on $T^*X$ which are $\nu$-homogenous in the fibers of $T^*X$. In more detail, if $\pi : T^*X \to X$ denotes the canonical projection, then $\mathcal{E}(\nu)(U)$ is given by

$$\mathcal{E}_{T^*X}(\nu)(U) = \{a \in \mathcal{E}(\pi^{-1}(U)) : a \text{ is } \nu\text{-homogeneous on every fiber}\}. \tag{31}$$

**Definition 2.1** *Let $U \subset X$ be open and define $\mathcal{D}_0(U)$ to be the $\mathbb{C}$-algebra $\bigoplus\limits_{\nu \in \mathbb{N}} \mathcal{E}(\nu)(U)$ of polynomial functions on $\pi^{-1}(U)$. It is called the **algebra of classical polynomial ob-*** **servables** *over $U$. Variation of $U$ over all open sets in $X$ defines a sheaf $\mathcal{D}_0$ of $\mathbb{C}$-algebras on $X$. The ringed space $(X, \mathcal{D}_0)$ is the space of* **classical polynomial observables** *or of* **polynomial symbols** *on $X$.*

It is our goal to quantize $\mathcal{D}_0$ by a deformation through a complex parameter $\hbar$. To achieve that we first choose a coordinate patch $U$ in $X$ with coordinates $z = (z_1, ..., z_n) : U \to k^n$, where $k = \mathbb{R}, \mathbb{C}$. They induce a chart $(z, \xi) = (z_1, ..., z_n, \xi_1, ..., \xi_n) : T^*U \to k^{2n}$. In these local coordinates $\mathcal{D}_0(U)$ is isomorphic to the polynomial ring $\mathcal{E}(U)[\xi_1, ..., \xi_n]$ over $\mathcal{E}(U)$ in $n$ variables.

Now one can define a chart dependent algebra $\mathcal{D}_\hbar(z)$ as the algebra of all sums $a = \sum_{\nu \in \mathbb{N}} a_\nu$, $a_\nu \in \mathcal{E}(\nu)(U)$ with the following properties:

$(i)$ There exists an integer $\nu_0$ such that $a_\nu = 0$ for all $\nu \geq \nu_0$.

$(ii)$ The multiplication is given by

$$\left(\sum_\nu a_\nu\right) \cdot_\hbar \left(\sum_\mu b_\mu\right) = \left(\sum_\rho c_\rho\right) \text{ with } c_\rho = \sum_{\nu+\mu-|\alpha|=\rho} \frac{(-i\hbar)^{|\alpha|}}{\alpha!} \frac{\partial^{|\alpha|}}{\partial \xi^\alpha} a_\nu \frac{\partial^{|\alpha|}}{\partial z^\alpha} b_\mu. \tag{32}$$

Then $\mathcal{D}_\hbar(z)$ carries the structure of a $\mathbb{C}$-algebra such that the diagram

$$\begin{array}{ccc} \mathcal{E}(U) & \longrightarrow & \mathcal{D}_\hbar(z) \\ & \searrow \quad \nearrow & \\ & \mathbb{C} & \end{array}$$

commutes. Additionally the following lemma holds.

**Lemma 2.2** *Fix $\hbar \in \mathbb{C}$. Then $\mathcal{D}_\hbar(z)$ is isomorphic as $\mathbb{C}$-algebra to the $\mathbb{C}$-algebra generated by $\mathcal{E}(U)$ and the elements $\xi_1, ..., \xi_n$ modulo the relations*

$$\xi_k \cdot \xi_l = \xi_l \cdot \xi_k \quad \text{and} \quad \xi_k \cdot f = \left(-i\hbar \frac{\partial}{\partial z_k} f\right) + f \cdot \xi_k, \tag{33}$$

*where $f \in \mathcal{E}(U)$ and $1 \leq k, l \leq n$.*

PROOF: It is easy to see that $\mathcal{E}(U)$ and $\xi_1, ..., \xi_n$ generate $\mathcal{D}_\hbar(z)$ and that the relations (33) denoted in the following by $(R)$ are true. Therefore we have a surjective morphism $\mathbb{C} < \mathcal{E}(U), \xi_1, ..., \xi_n / R > \to \mathcal{D}_\hbar(z)$. On the other hand define $\mathcal{D}_\hbar(z) \to \mathbb{C} < \mathcal{E}(U), \xi_1, ..., \xi_n / R >$ by $\sum_{\nu \in \mathbb{N}} a_\nu \mapsto \sum_{\alpha \in \mathbb{N}^n} a_\alpha \xi^\alpha$ where $a_\nu = \sum_{\alpha \in \mathbb{N}^n, |\alpha|=\nu} (a_\alpha \circ \pi) \xi^\alpha$ with $a_\alpha \in \mathcal{E}(U)$. By using the Leibniz rule one realizes that this mapping is in fact a morphism of $\mathbb{C}$-algebras and inverse to $\mathbb{C} < \mathcal{E}(U), \xi_1, ..., \xi_n / R > \longrightarrow \mathcal{D}_\hbar(U, z)$. □

Now assume $\tilde{z} : U \to k^n$ to be another coordinate system on $U$. Then we have in $\mathcal{D}_\hbar(\tilde{z})$

$$\begin{aligned}\left(\sum_l \frac{\partial \tilde{z}_l}{\partial z_k}\tilde{\xi}_l\right) \cdot_\hbar f &= \sum_l \frac{\partial \tilde{z}_l}{\partial z_k}\left(-i\hbar \frac{\partial}{\partial \tilde{z}_l}f\right) + f\sum_l \frac{\partial \tilde{z}_l}{\partial z_k}\tilde{\xi}_l \\ &= -i\hbar \frac{\partial f}{\partial z_k} + f\sum_l \frac{\partial \tilde{z}_l}{\partial z_k}\tilde{\xi}_l \end{aligned} \quad (34)$$

and

$$\begin{aligned}\left(\sum_r \frac{\partial \tilde{z}_r}{\partial z_k}\tilde{\xi}_r\right) \cdot_\hbar \left(\sum_s \frac{\partial \tilde{z}_s}{\partial z_l}\tilde{\xi}_s\right) &= \sum_{r,s}\left(\frac{\partial \tilde{z}_r}{\partial z_k}\frac{\partial \tilde{z}_s}{\partial z_l}\right)\tilde{\xi}_r\tilde{\xi}_s + \sum_{r,s} \frac{\partial \tilde{z}_r}{\partial z_k}\left(-i\hbar \frac{\partial}{\partial \tilde{z}_r}\frac{\partial \tilde{z}_s}{\partial z_l}\right)\tilde{\xi}_s \\ &= \sum_{r,s}\left(\frac{\partial \tilde{z}_r}{\partial z_k}\frac{\partial \tilde{z}_s}{\partial z_l}\right)\tilde{\xi}_r\tilde{\xi}_s + \sum_s\left(-i\hbar \frac{\partial^2 \tilde{z}_s}{\partial z_k \partial z_l}\right)\tilde{\xi}_s \\ &= \left(\sum_s \frac{\partial \tilde{z}_s}{\partial z_l}\tilde{\xi}_s\right) \cdot_\hbar \left(\sum_r \frac{\partial \tilde{z}_r}{\partial z_k}\tilde{\xi}_r\right). \end{aligned} \quad (35)$$

Hence the next lemma follows easily.

**Lemma 2.3** *There exists a unique homomorphism*

$$\mathcal{D}_\hbar(\tilde{z}, z) : \mathcal{D}_\hbar(z) \to \mathcal{D}_\hbar(\tilde{z}) \quad (36)$$

*satisfying $\mathcal{D}_\hbar(\tilde{z}, z)(f) = f$ for $f \in \mathcal{E}(U)$ and $\mathcal{D}_\hbar(\tilde{z}, z)(\xi_k) = \sum_l \frac{\partial \tilde{z}_l}{\partial z_k}\tilde{\xi}_l$ for $k = 1, ..., n$. The $\mathcal{D}_\hbar(\tilde{z}, z)$ are isomorphisms such that*

$$\mathcal{D}_\hbar(\hat{z}, z) = \mathcal{D}_\hbar(\hat{z}, \tilde{z}) \circ \mathcal{D}_\hbar(\tilde{z}, z) \quad \text{and} \quad \mathcal{D}_\hbar(z, z) = id \quad (37)$$

*hold for any additional coordinate map $\hat{z} : U \to k^n$.*

PROOF: Equations (34) and (2.1) and Lemma 2.2 entail that $\mathcal{D}_\hbar(\tilde{z}, z)$ respects the relations defining $\mathcal{D}_\hbar(z)$, so $\mathcal{D}_\hbar(\tilde{z}, z)$ is well-defined. The rest is obvious. □

**Note 2.4** Evaluation $\mathcal{D}_\hbar(z) \ni a \overset{e_\zeta}{\mapsto} a(\zeta) \in \mathbb{C}$ at $\zeta \in T^*U$ gives a linear functional on $\mathcal{D}_\hbar(z)$ which is a character iff $\hbar = 0$.

The proposition shows that one can introduce an equivalence relation between elements $\sum_\nu a_\nu \in \mathcal{D}_\hbar(z)$ and $\sum_\nu b_\nu \in \mathcal{D}_\hbar(\tilde{z})$ by $\sum_\nu a_\nu \sim \sum_\nu b_\nu$ iff $\mathcal{D}_\hbar(\tilde{z}, z)\left(\sum_\nu a_\nu\right) = \sum_\nu b_\nu$. As $\mathcal{D}_\hbar(\tilde{z}, z)$ is linear and multiplicative, the set $\mathcal{D}_\hbar(U)$ of equivalence classes $[\sum_\nu a_\nu]$ carries in a natural way the structure of a $\mathbb{C}$-algebra. It becomes an $\mathcal{E}_X(U)$-module by the algebra map $\mathcal{E}_X(U) \ni f \to [f] \in \mathcal{D}_\hbar(U)$. The $\mathcal{D}_\hbar(U)$ can be interpreted as spaces of sections of a sheaf $\mathcal{D}_\hbar$ of $\mathbb{C}$-algebras on $X$. The restriction morphisms are given by $\mathcal{D}_\hbar(U) \ni [\sum_\nu a_\nu] \to [\sum_\nu a_\nu|_V] \in \mathcal{D}_\hbar(V)$ for $V \subset U$ open. $\mathcal{D}_\hbar$ is unique up to isomorphism and carries the structure of an $\mathcal{E}_X$-module because the coordinate patches $U$ form a basis of the topology of $X$. Furthermore $\mathcal{D}_0$ is isomorphic to the sheaf of classical polynomial observables defined in Definition 2.1.

**Definition 2.5** *The ringed space $(X, \mathcal{D}_\hbar)$ is called the* **space of $\hbar$-deformed polynomial observables** *on $X$.*

If $\mathcal{V}$ is a smooth (resp. holomorphic) vector field over $U$, we want to assign to $\mathcal{V}$ an element $\mathcal{V}^\hbar \in \mathcal{D}_\hbar(U)$ such that for any additional smooth (resp. holomorphic) vector field $\mathcal{W}$ over $U$ the equation

$$\mathcal{V}^\hbar \cdot \mathcal{W}^\hbar - \mathcal{W}^\hbar \cdot \mathcal{V}^\hbar = -i\hbar\, [\mathcal{V},\, \mathcal{W}]^\hbar \tag{38}$$

is valid in $\mathcal{D}_\hbar(U)$. Writing $\mathcal{V} = \sum v_k \frac{\partial}{\partial z_k}$ define $\mathcal{V}^\hbar$ to be the equivalence class of $\sum i\, v_k\, \xi_k \in \mathcal{D}_\hbar(U)$. It is obvious by the definition of $\mathcal{D}_\hbar(U)$ that $\mathcal{V}^\hbar$ is independent of the specially chosen coordinate system $z$. If $\mathcal{W} = \sum w_k \frac{\partial}{\partial z_k}$, we have in the Lie algebra of vector fields

$$[\mathcal{V},\, \mathcal{W}] = \sum_{k,l} \left( v_k \frac{\partial w_l}{\partial z_k} - w_k \frac{\partial v_l}{\partial z_k} \right) \frac{\partial}{\partial z_l} \tag{39}$$

and in $\mathcal{D}_\hbar(z)$

$$\left(\sum_k i\, v_k\, \xi_k\right) \cdot \left(\sum_l i\, w_l\, \xi_l\right) - \left(\sum_k i\, w_k\, \xi_k\right) \cdot \left(\sum_l i\, v_l\, \xi_l\right) = -i\hbar \sum_{k,l} i \left(v_k \frac{\partial w_l}{\partial z_k} - w_k \frac{\partial v_l}{\partial z_k}\right) \xi_l. \tag{40}$$

Hence equation (38) follows.

The generators $\xi_1, ..., \xi_n$ of $\mathcal{D}_\hbar(z)$ correspond to the elements $-i\frac{\partial^\hbar}{\partial^\hbar z_1}, ..., -i\frac{\partial^\hbar}{\partial^\hbar z_n} \in \mathcal{D}_\hbar(U)$. Now if $a = \sum_\nu a_\nu$ is a sum of fiberwise $\nu$-homogeneous smooth (resp. holomorphic) functions $a_\nu$ over $\pi^{-1}(U)$, we therefore sometimes write $a(-i\frac{\partial^\hbar}{\partial^\hbar z}) = \sum_\nu a_\nu(-i\frac{\partial^\hbar}{\partial^\hbar z})$ for the equivalence class in $\mathcal{D}_\hbar(U)$ defined by $a$. Furthermore we denote by $\frac{\partial^{\hbar|\alpha|}}{\partial^\hbar z^\alpha}$, $\alpha \in \mathbb{N}^n$ the element $\left(\frac{\partial^\hbar}{\partial^\hbar z_1}\right)^{\alpha_1} \cdot ... \cdot \left(\frac{\partial^\hbar}{\partial^\hbar z_n}\right)^{\alpha_n} \in \mathcal{D}_\hbar(U)$.

Let $(e_1, ..., e_n)$ be a frame of $TX$ over $U \subset X$. Then the vector fields $\frac{\partial}{\partial z_k}$, $k = 1, ..., n$ can be expanded as linear combinations of $(e_1, ..., e_n)$. Hence the elements $(e_1^\hbar, ..., e_n^\hbar)$ together with $\mathcal{E}(U)$ generate the algebra $\mathcal{D}_\hbar(U)$. Note that because of equation (38) the commutator $e_k^\hbar \cdot e_l^\hbar - e_l^\hbar \cdot e_k^\hbar$ is in general not zero for $k \neq l$.

Varying the parameter $\hbar$ over $\mathbb{C}$, the $\mathcal{D}_\hbar$ shall be interpreted as fibers of a certain deformation $(X \times \mathbb{C}, \mathcal{D}_d) \to (\mathbb{C}, \mathcal{E}_\mathbb{C})$. The sheaf $\mathcal{D}_d$ is defined locally in coordinates. First let $d: X \times \mathbb{C} \to \mathbb{C}$ be the canonical projection and consider the pull-back bundle $d^*T^*X = T^*X \times \mathbb{C}$. Then we have the sheaf $\mathcal{E}_{d^*T^*X}$ of smooth (resp. holomorphic) functions on $d^*T^*X$ and the sheaves $\mathcal{E}_{d^*T^*X}(\nu)$ on $X \times \mathbb{C}$ of in the fibers $\nu$-homogeneous functions on $d^*T^*X$. For open $O$ in $\mathbb{C}$ and coordinate patches $U$ in $X$, $\mathcal{D}_{\tilde{d}}(z, O)$ shall now be the $\mathbb{C}$-algebra of sums $p = \sum_\nu p_\nu$, $p_\nu \in \mathcal{E}_{d^*T^*X}(U \times O)$ such that the following conditions hold:

(i) There exists an integer $\nu_0$ such that $p_\nu = 0$ for all $\nu \geq \nu_0$.

(ii) The multiplication is given by

$$\left(\sum_\nu p_\nu\right) \cdot \left(\sum_\mu q_\mu\right) = \left(\sum_\rho r_\rho\right) \tag{41}$$

with

$$r_\rho(-,\hbar) = \sum_{\nu+\mu-|\alpha|=\rho} \frac{(-i\hbar)^{|\alpha|}}{\alpha!} \frac{\partial^{|\alpha|}}{\partial \xi^\alpha} p_\nu(-,\hbar) \frac{\partial^{|\alpha|}}{\partial z^\alpha} q_\mu(-,\hbar), \quad \hbar \in O. \tag{42}$$

The associativity of the product in (ii) follows by an easy calculation. Reasoning like for $\mathcal{D}_\hbar$ we can construct transition homomorphisms $\mathcal{D}_{\tilde{d}}(\tilde{z}, z, O) : \mathcal{D}_{\tilde{d}}(z, O) \to \mathcal{D}_{\tilde{d}}(\tilde{z}, O)$ which give rise to a congruence relation between elements of $\mathcal{D}_{\tilde{d}}(z, O)$ and $\mathcal{D}_{\tilde{d}}(\tilde{z}, O)$. The algebra $\mathcal{D}_{\tilde{d}}(U \times O)$ of equivalence classes is isomorphic to $\mathcal{D}_{\tilde{d}}(z, O)$. As $\mathcal{D}_{\tilde{d}}(\tilde{z}, z, O) = \mathcal{D}_{\tilde{d}}(\tilde{z}, \hat{z}, O) \circ \mathcal{D}_{\tilde{d}}(\hat{z}, z, O)$ for a third coordinate system $\hat{z}$ we finally receive a sheaf $\mathcal{D}_{\tilde{d}}$ of noncommutative $\mathbb{C}$-algebras on $X \times \mathbb{C}$, called the sheaf of **universally deformed polynomial observables** on $X$.

Denoting by $d$ the canonical projection $d : X \times \mathbb{C} \to \mathbb{C}$ we want to associate to any smooth (resp. holomorphic) vector field $\mathcal{V}$ over $U \subset X$ an element $\mathcal{V}^d$ in $\mathcal{D}_{\tilde{d}}(U \times \mathbb{C})$ such that for any second vector field $\mathcal{W}$ over $U$ the equation

$$\mathcal{V}^d \cdot \mathcal{W}^d - \mathcal{W}^d \cdot \mathcal{V}^d = -id[\mathcal{V}, \mathcal{W}]^d \tag{43}$$

holds in $\mathcal{D}_{\tilde{d}}(U \times \mathbb{C})$. But this is accomplished exactly like in the definition of $\mathcal{V}^\hbar \in \mathcal{D}_\hbar(U)$. Express $\mathcal{V} = \sum v_k \frac{\partial}{\partial z_k}$ in terms of local coordinates and let $\mathcal{V}^d$ be the equivalence class of $\sum i v_k (\xi_k \circ q) \in \mathcal{D}_{\tilde{d}}(z, \mathbb{C})$ in $\mathcal{D}_{\tilde{d}}(U \times \mathbb{C})$, where $q : X \times \mathbb{C} \to X$ is the projection on the first coordinate. This $\mathcal{V}^d$ fulfills (43).

As $-i\frac{\partial^d}{\partial^d z_k}$ is the equivalence class of $\xi_k$ in $\mathcal{D}_{\tilde{d}}(U \times O)$, we denote by $p\left(-i\frac{\partial^d}{\partial^d z}\right)$ the equivalence class of any finite sum $p = \sum p_\nu \in \mathcal{D}_{\tilde{d}}(z, O)$ of fiberwise $\nu$-homogeneous holomorphic functions $p_\nu$.

The commutative ringed space $(\mathbb{C}, \mathcal{E})$ can be embedded into the center of $\mathcal{D}_{\tilde{d}}$ by $\mathcal{E}(O) \ni b \mapsto \Delta(b) = b \circ d \in \mathcal{D}_{\tilde{d}}(X \times O)$, $O \subset \mathbb{C}$. Together with the canonical projection $d : X \times \mathbb{C} \to \mathbb{C}$ this defines a morphism $(d, \Delta) : (X \times \mathbb{C}, \mathcal{D}_{\tilde{d}}) \to (\mathbb{C}, \mathcal{E}_\mathbb{C})$.

**Theorem 2.6** *Let $X$ be a smooth (resp. complex) manifold and $\mathcal{E}_\mathbb{C}$ the sheaf of smooth (resp. holomorphic) complex valued functions on $\mathbb{C}$. Then the morphism $(d, \Delta) : (X \times \mathbb{C}, \mathcal{D}_{\bar{d}}) \to (\mathbb{C}, \mathcal{E}_\mathbb{C})$ defined on the space of universally deformed polynomial observables on $X$ is fibered and for any $\hbar \in \mathbb{C}$ a deformation of $(X, \mathcal{D}_\hbar)$ over $(\mathbb{C}, \mathcal{E}_\mathbb{C})$ with distinguished point $\hbar$. In the complex case $(d, \Delta)$ even gives rise to a flat deformation. Altogether we receive a functor $\mathcal{D}$ going from the category of smooth (resp. complex) manifolds to the category $\underline{\mathrm{Def}}_{0,(\mathbb{C},\mathcal{C}^\infty)}$ (resp. $\underline{\mathrm{Def}}_{0,(\mathbb{C},\mathcal{O})}$) of differentiable (resp. holomorphic) deformations over $\mathbb{C}$.*

PROOF: By definition it is clear that $(d, \Delta)$ is fibered. To show that it is a deformation, let us first regard the complex case. By the definition of the product in $(\mathcal{D}_{\bar{d}})_{(x,\hbar)}$ and the product in $(\mathcal{D}_\hbar)_x$ it is obvious that $(\mathcal{D}_{\bar{d}})_{(x,\hbar)} / \mathfrak{m}_\hbar (\mathcal{D}_{\bar{d}})_{(x,\hbar)} \cong (\mathcal{D}_\hbar)_x$ for all $x \in X$. It remains to prove the flatness of $\mathcal{E}_\hbar \to (\mathcal{D}_{\bar{d}})_{(x,\hbar)}$. Now as $\mathcal{E}_\hbar$-module, $(\mathcal{D}_{\bar{d}})_{(x,\hbar)}$ is isomorphic to $\mathcal{E}_{n+1} \otimes \mathbb{C}[\xi_1, ..., \xi_n]$, where the last variable of $\mathcal{E}_{n+1}$ is $\hbar$. But $\mathcal{E}_{n+1} \otimes \mathbb{C}[\xi_1, ..., \xi_n]$ is flat as an $\mathcal{E}_\hbar$-module, hence also $(\mathcal{D}_{\bar{d}})_{(x,\hbar)}$. This proves $(d, \Delta)$ being a flat deformation.

Next we have to show for smooth $X$ that $(d, \Delta)$ is a flatly filtered deformation. So let $(\mathcal{D}_{\bar{d}})^m_{(x,\hbar)}$, $m \in \mathbb{N}$ be the right ideal in $(\mathcal{D}_{\bar{d}})_{(x,\hbar)}$ consisting of all germs of sums $\sum_{|\alpha|\leq\nu} p_\alpha \xi^\alpha$, $\nu \in \mathbb{N}$ such that $p_\alpha$ is a smooth function defined on a neighborhood of $(x, \hbar)$ and $\left[\frac{\partial^\mu \partial^{|\beta|} p_\alpha}{\partial \hbar^\mu \partial z^\beta}\right](x, \hbar) = 0$ for every $\alpha, \beta \in \mathbb{N}^n$ and $\mu \in \mathbb{N}$ with $\mu + |\beta| \leq m$. Then $\left((\mathcal{D}_{\bar{d}})^m_{(x,\hbar)}\right)_{m\in\mathbb{N}}$ gives a filtration of $(\mathcal{D}_{\bar{d}})_{(x,\hbar)}$. Furthermore $\Delta(\mathfrak{m}_\hbar^m) \subset (\mathcal{D}_{\bar{d}})^m_{(x,\hbar)}$, where $\mathfrak{m}_\hbar$ is the maximal ideal in $\mathcal{E}_\hbar$. The right ideal $(\mathcal{D}_{\bar{d}})^\infty_{(x,\hbar)} = \bigcap_{m\in\mathbb{N}} (\mathcal{D}_{\bar{d}})^m_{(x,\hbar)}$ is a left ideal in $(\mathcal{D}_{\bar{d}})_{(x,\hbar)}$, too and consists of all germs given by sums $\sum_{\alpha\leq\nu} p_\alpha \xi^\alpha$ such that $\left[\frac{\partial^\mu \partial^{|\beta|} p_\alpha}{\partial \hbar^\mu \partial z^\beta}\right](x, \hbar) = 0$ for every $\alpha, \mu \in \mathbb{N}$ and $\beta \in \mathbb{N}^n$. By BOREL's Theorem $(\mathcal{D}_{\bar{d}})_{(x,\hbar)} / (\mathcal{D}_{\bar{d}})^\infty_{(x,\hbar)}$ is isomorphic to $\mathbb{C}[[x_1, ..., x_n, \hbar]][\xi_1, ..., \xi_n]$ which is obviously flat over $\mathcal{E}_\hbar / \mathfrak{m}_\hbar^\infty \cong \mathbb{C}[[\hbar]]$. Hence Theorem 1. in BOURBAKI [9], *Commutative Algebra*, Chapter II §5.2 implies that $(\mathcal{D}_{\bar{d}})_{(x,\hbar)} / \mathfrak{m}_\hbar^m (\mathcal{D}_{\bar{d}})_{(x,\hbar)} \cong \left((\mathcal{D}_{\bar{d}})_{(x,\hbar)} / (\mathcal{D}_{\bar{d}})^\infty_{(x,\hbar)}\right) / \left(\mathfrak{m}_\hbar^m (\mathcal{D}_{\bar{d}})_{(x,\hbar)} / (\mathcal{D}_{\bar{d}})^\infty_{(x,\hbar)}\right)$ is a flat $\mathcal{E}_\hbar / \mathfrak{m}_\hbar^m \cong (\mathcal{E}_\hbar / \mathfrak{m}_\hbar^\infty) / (\mathfrak{m}_\hbar^m / \mathfrak{m}_\hbar^\infty)$ module. The claim follows.

The functorial property of $\mathcal{D}$ is obvious. $\square$

## 2.2 Quantized polynomial observables

After having defined a deformation of $(X, \mathcal{D}_0)$ we will now construct a quantization of $(X, \mathcal{D}_0)$, the so called normal order quantization. This method works for every smooth paracompact $X$ and for every complex manifold for which a connection with holomorphic CHRISTOFFEL symbols exists. It will be obvious by construction that the quantization is dependent on the choice of a connection on $X$.

Choose a smooth connection $\nabla$ (resp. a connection $\nabla$ with holomorphic CHRISTOFFEL symbols) on $X$ and around any $x \in X$ normal coordinates $\exp_x : K_x \to B_x$, where $K_x$ is an open neighborhood of 0 in $T_x X$ and $B_x$ one of $x$ in $X$. The exponential map induces charts $T\exp_x : K_x \times T_x X \cong TK_x \to TB_x \subset TX$ and $T^*\exp_x : K_x \times T_x^* X \cong T^* K_x \to T^* B_x \subset T^* X$.

Now let $a$ be any sum $a = \sum_\nu a_\nu \in \mathcal{D}_0(U)$, where $U \subset X$ is open. Let $\tilde{x} \in U$ and $z : V_{\tilde{x}} \to k^n$ be a coordinate system on an open neighborhood $V_{\tilde{x}}$ of $\tilde{x}$ in $U$. After possibly shrinking $V_{\tilde{x}}$ we can assume that the mappings $\exp_x^{-1} : V_{\tilde{x}} \to T_x^* X$ are well-defined for every $x \in V_{\tilde{x}}$. Furthermore we can assume that on $V_{\tilde{x}}$ there exists a smooth (resp. holomorphic) covariant frame $e_1, ..., e_n$ on $TX$, i.e. $\nabla e_k = 0$ for $k = 1, ..., n$. Then we can define the normal coordinates $z_x : V_{\tilde{x}} \to k^n$ for $x \in V_{\tilde{x}}$ by $\exp_x^{-1} = \sum_{k=1}^{n} z_{x,k} \cdot e_k(x)$. Now we have the means to define the function $a^{\mathrm{n},z} : (V_{\tilde{x}} \times T^* V_{\tilde{x}} \times \mathbb{C}) \to \mathbb{C}$ by

$$a^{\mathrm{n},z}(x, \xi, \hbar) = \mathcal{D}_\hbar(z, z_x)(a) \tag{44}$$

for $x \in V_{\tilde{x}}$, $\xi \in T^* V_{\tilde{x}}$ and $\hbar \in \mathbb{C}$. $a^{\mathrm{n},z}$ obviously does not depend on the specially chosen covariant frame $e_1, ..., e_n$ over $V_{\tilde{x}}$. Furthermore by Lemma 2.3, LEIBNIZ rule and the chain rule $a^{\mathrm{n},z}$ is a smooth (resp. holomorphic) function, because $V_{\tilde{x}} \times V_{\tilde{x}} \ni (x,y) \mapsto \exp_x^{-1}(y) \in T^* X$ is smooth (resp. holomorphic) both in $x$ and $y$. Finally set

$$a^z : T^* V_{\tilde{x}} \times \mathbb{C} \to \mathbb{C}, \quad (\xi, \hbar) \mapsto a^{\mathrm{n},z}(\pi(\xi), \xi, \hbar). \tag{45}$$

Then by running $\tilde{x}$ through $U$ and choosing appropriate coordinate systems $z : V_{\tilde{x}} \to k^n$ in the above sense, we can now define the **operator of normal order quantization** $\mathfrak{q}_U(a) \in \mathcal{D}_d(U)$ of $a$ by

$$\mathfrak{q}_U(a)\Big|_{T^* V_{\tilde{x}}} = a^z \left( -i \frac{\partial^d}{\partial^d z} \right). \tag{46}$$

To show that $\mathfrak{q}_U(a)$ is well-defined, we have to prove that $\mathcal{D}_d(\tilde{z}, z, O) a^z = a^{\tilde{z}}$ for any second coordinate system $\tilde{z} : V_{\tilde{x}} \to k^n$. But by Eq. (45) we have

$$a^z|_{T_x^* U \times \mathbb{C}} = a^{\mathrm{n},z}(x, -, \cdot)|_{T_x^* U \times \mathbb{C}} = \mathcal{D}_d(z, z_x, \mathbb{C})(a \circ q)|_{T_x^* U \times \mathbb{C}}, \tag{47}$$

where $q : T^* X \times \mathbb{C} \to T^* X$ is the projection on the first coordinate. Therefore the equation

$$\begin{aligned}
\mathcal{D}_d(\tilde{z}, z, \mathbb{C}) a^z|_{T_x^* U \times \mathbb{C}} \\
= \mathcal{D}_d(\tilde{z}, z, \mathbb{C}) \mathcal{D}_d(z, z_x, \mathbb{C})(a \circ q)|_{T_x^* U \times \mathbb{C}} = \mathcal{D}_d(\tilde{z}, z_x, \mathbb{C})(a \circ q)|_{T_x^* U \times \mathbb{C}} \\
= a^{\tilde{z}}|_{T_x^* U \times \mathbb{C}}
\end{aligned} \tag{48}$$

holds. This proves the claim.

We need a further lemma for later considerations about the quantization map $\mathfrak{q}_U$.

**Lemma 2.7** *Let $a \in \mathcal{D}_0(U)$ be a monomial in the $\xi$-s of the form $a = (f \circ \pi) \cdot \xi_1^{\alpha_1} \cdot ... \cdot \xi_n^{\alpha_n}$, $\alpha \in \mathbb{N}^n$. Then $a^z$ can be written as a polynomial*

$$a^z(-,\hbar) = \sum_{\substack{\beta \in \mathbb{N}^n \\ \beta \leq \alpha}} f_\beta(\pi(-)) \hbar^{|\alpha-\beta|} \cdot \xi^\beta(-), \tag{49}$$

*where the $f_\beta : U \to \mathbb{C}$ are smooth (resp. holomorphic) and $f_\alpha(\cdot) = f(\cdot)$.*

PROOF: Use induction on the order $m = |\alpha| = \alpha_1 + ... + \alpha_n$.

If $a = f \circ \pi$ it is clear that $a^z(-,\hbar) = f \circ \pi(-)$, so $a^z$ is smooth (resp. holomorphic) in all variables and a polynomial of order 0 in every fiber. Now assume that any monomial $b = (f \circ \pi) \cdot \xi_1^{\alpha_1} \cdot ... \cdot \xi_n^{\alpha_n} \in \mathcal{D}_0$ of order $m$ is mapped to a $b^z$ of the form $b^z(-,\hbar) = \sum_{\beta \in \mathbb{N}^n, \beta \leq \alpha} g_\beta(\pi(-)) \hbar^{|\alpha-\beta|} \cdot \xi_1^{\beta_1} \cdot ... \cdot \xi_n^{\beta_n}$ with smooth (resp. holomorphic) $g_\beta$ and $g_\alpha(\cdot) = f(\cdot)$. Let $a = (f \circ \pi) \cdot \xi_1^{\alpha_1} \cdot ... \cdot \xi_n^{\alpha_n} \cdot \xi_i$ be of order $m+1$. Then for all $x \in U$

$$a = \sum_{1 \leq k \leq n} \frac{\partial z_{x,k}}{\partial z_i} b \cdot \xi_{z_{x,k}} \quad . \tag{50}$$

Applying $\mathcal{D}_{\tilde{d}}(z, z_x, \mathbb{C})$ and evaluating at $(-,\hbar) \in T_x^*X \times \mathbb{C}$ gives

$$\begin{aligned}
a^z(-,\hbar) &= \sum_{1 \leq k \leq n} \frac{\partial z_{\pi(-),k}}{\partial z_i}(\pi(-)) \sum_{\substack{\beta \in \mathbb{N}^n \\ \beta \leq \alpha}} \sum_{\substack{\gamma \in \mathbb{N}^n \\ \gamma \leq \beta}} \sum_{1 \leq l \leq n} \frac{(-i\hbar)^{|\gamma|}}{\gamma!} g_\beta(\pi(-)) \hbar^{|\alpha-\beta|} \cdot \\
&\quad \cdot \left(\frac{\partial^{|\gamma|}}{\partial z^\gamma} \frac{\partial z_l}{\partial z_{\pi(-),k}}\right)(\pi(-)) \cdot \xi^{\beta-\gamma}(-) \cdot \xi_l(-) \\
&= f(\pi(-)) \xi^{(\alpha_1,...,\alpha_i+1,...,\alpha_n)}(-) + \sum_{|\delta| \leq |\alpha|} f_\delta(\pi(-)) \hbar^{|\alpha-\delta|+1} \cdot \xi^\delta(-) \tag{51}
\end{aligned}$$

So $a^z$ is also smooth (resp. holomorphic) and a polynomial of order $\leq m+1$ with respect to the $\xi$-s. Furthermore equation (51) shows that the polynomial expansion of $a^z$ has the required form. This proves the lemma. $\square$

Now we have enough tools to formulate and prove the following theorem.

**Theorem 2.8** *Let $X$ be a smooth (resp. complex) manifold, $\nabla$ a connection on $X$ (with holomorphic CHRISTOFFEL symbols in the complex case) and $\mathcal{E}_\mathbb{C}$ the sheaf of smooth (resp. holomorphic) functions on $\mathbb{C}$. Associate to $X$ the sheaves $\mathcal{D}_0$ and $\mathcal{D}_{\tilde{d}}$ of classical and universally deformed polynomial observables on $X$. Then the operator $\mathsf{q}_U : \mathcal{D}_0 \to \mathcal{D}_{\tilde{d}}(U \times \mathbb{C})$ of normal order quantization is linear and gives rise to a linear sheaf morphism $\mathsf{q} : \mathcal{D}_0 \to q_* \mathcal{D}_{\tilde{d}}$ over $q = pr_1 : X \times \mathbb{C} \to X$. Furthermore $(q, \mathsf{q}) : (X \times \mathbb{C}, \mathcal{D}_{\tilde{d}}) \to (X, \mathcal{D}_0)$ is a quantization of $(X, \mathcal{D}_0)$ over $(\tilde{d}, \Delta) : (X \times \mathbb{C}, \mathcal{D}_{\tilde{d}}) \to (\mathbb{C}, \mathcal{E}_\mathbb{C})$ and is called the **normal order quantization** of $(X, \mathcal{D}_0)$ with respect to $\nabla$.*

PROOF: The $\mathfrak{q}_U$ are linear by definition and coordinate independent according to our above considerations. As they obviously commute with restriction morphisms, the $\mathfrak{q}_U$ uniquely define a sheaf morphism $\mathfrak{q} : \mathcal{D}_0 \to q_* \mathcal{D}_{\bar{d}}$.

To verify that $Q = (q, \mathfrak{q})$ indeed is a quantization over $D$, we have to show $\mathfrak{j} \circ \mathfrak{q} = id_{\mathcal{D}_0}$. Here $\mathfrak{j} : \mathcal{D}_{\bar{d}} \to j_* \mathcal{D}$ is the sheaf morphism over $j : X \to X \times \mathbb{C}$, $x \mapsto (x, 0)$ mapping $(\mathcal{D}_{\bar{d}})_{(x,0)}$ to the fiber $(\mathcal{D}_{\bar{d}})_{(x,0)} / \mathfrak{m}_0 (\mathcal{D}_{\bar{d}})_{(x,0)} \cong (\mathcal{D}_0)_x$ for every $x \in X$. We prove the relation $\mathfrak{j} \circ \mathfrak{q} = id_{\mathcal{D}_0}$ by inductively showing $\mathfrak{j} \circ \mathfrak{q}(a) = a$ for every monomial in the $\xi$-s. If $a = f \circ \pi$, we obviously have $\mathfrak{j} \circ \mathfrak{q}(f \circ \pi) = \mathfrak{j}(f \circ q) = f \circ q \circ j = f \circ \pi$. In case $\mathfrak{j} \circ \mathfrak{q}(b) = b$ for $b = (f \circ \pi) \cdot \xi_1^{\alpha_1} \cdot \ldots \cdot \xi_n^{\alpha_n} \in \mathcal{D}_0$ of order $m$, equation (51) entails for $a = b \cdot \xi_i$ on $T_x^* X$

$$
\begin{aligned}
a^z(-, 0) &= \sum_{1 \le k \le n} \frac{\partial z_{x,k}}{\partial z_i}(\pi(-)) \cdot \sum_{\substack{\beta \in \mathbb{N}^n \\ |\beta| \le m}} \sum_{1 \le l \le n} g_\beta(\pi(-), 0) \cdot \frac{\partial z_l}{\partial z_{x,k}}(\pi(-)) \cdot \xi^\beta(-) \cdot \xi_l(-) \\
&= b(-) \cdot \xi_i(-) = F(-),
\end{aligned}
\tag{52}
$$

so the relation $\mathfrak{j} \circ \mathfrak{q}(a) = a$ is shown for all monomials of order $m + 1$. □

**Corollary 2.9** *Associate to the cotangent bundle $T^* X$ with its canonical symplectic structure the POISSON space $(X, \mathcal{D}_0)$ with POISSON bracket $\{\,,\,\}$ like in example 1.15 ii). Then the normal order quantization $(q, \mathfrak{q})$ is even a differentiable (resp. holomorphic) DIRAC quantization, that is for any pair $a, b \in \mathcal{D}_0(U)$, $U \subset X$ open there exists an element $r_{a,b} \in \mathcal{D}_{\bar{d}}(U \times \mathbb{C})$ such that*

$$[\mathfrak{q}(a), \mathfrak{q}(b)] = -i\,d\,\mathfrak{q}(\{a, b\}) + \bar{d}^2\, r_{a,b}. \tag{53}$$

PROOF: Let $a = (f \circ \pi) \xi^\alpha$, $b = (g \circ \pi) \xi^\beta$ be monomials of order $l$ respectively $m$. Then

$$
a^z(-, \hbar) = (f \circ \pi)(-) \xi^\alpha(-) + \sum_{\substack{\tilde{\alpha} \le \alpha \\ \tilde{\alpha} \ne \alpha}} f_{\tilde{\alpha}}(\pi(-)) \, \hbar^{|\alpha - \tilde{\alpha}|} \xi^{\tilde{\alpha}}(-), \tag{54}
$$

$$
b^z(-, \hbar) = (g \circ \pi)(-) \xi^\beta(-) + \sum_{\substack{\tilde{\beta} \le \beta \\ \tilde{\beta} \ne \beta}} g_{\tilde{\beta}}(\pi(-)) \, \hbar^{|\beta - \tilde{\beta}|} \xi^{\tilde{\beta}}(-). \tag{55}
$$

Therefore

$$
\mathfrak{q}(a) = (f \circ \pi)(-i)^{|\alpha|} \frac{\partial^{\bar{d}|\alpha|}}{\partial^{\bar{d}} z^\alpha} + \bar{d} \cdot P(-i \frac{\partial^{\bar{d}}}{\partial^{\bar{d}} z}), \tag{56}
$$

$$
\mathfrak{q}(b) = (g \circ \pi)(-i)^{|\beta|} \frac{\partial^{\bar{d}|\beta|}}{\partial^{\bar{d}} z^\beta} + \bar{d} \cdot Q(-i \frac{\partial^{\bar{d}}}{\partial^{\bar{d}} z}) \tag{57}
$$

with $P, Q$ fiberwise polynomial functions on $\pi^{-1}(U)$. But now in $\mathcal{D}_{\bar{d}}(U \times \mathbb{C})$

$$
\left[ (f \circ \pi)(-i)^{|\alpha|} \frac{\partial^{\bar{d}|\alpha|}}{\partial^{\bar{d}} z^\alpha}, (g \circ \pi)(-i)^{|\beta|} \frac{\partial^{\bar{d}|\beta|}}{\partial^{\bar{d}} z^\beta} \right] = -i\,\bar{d} \left\{ (f \circ \pi) \xi^\alpha, (g \circ \pi) \xi^\beta \right\} \left( -i \frac{\partial^{\bar{d}}}{\partial^{\bar{d}} z} \right)
$$

$$+ \tilde{d}^2 \cdot r_{f,g}, \tag{58}$$

$$\left[\tilde{d} \cdot P\left(-i\frac{\partial^{\tilde{d}}}{\partial^{\tilde{d}}z}\right), (g \circ \pi)(-i)^{|\beta|}\frac{\partial^{\tilde{d}|\beta|}}{\partial^{\tilde{d}}z^{\beta}}\right] = \tilde{d}^2 \cdot r_{P,g}, \tag{59}$$

$$\left[(f \circ \pi)(-i)^{|\alpha|}\frac{\partial^{\tilde{d}|\alpha|}}{\partial^{\tilde{d}}z^{\alpha}}, \tilde{d} \cdot Q\left(-i\frac{\partial^{\tilde{d}}}{\partial^{\tilde{d}}z}\right)\right] = \tilde{d}^2 \cdot r_{f,Q}, \tag{60}$$

$$\left[\tilde{d} \cdot P\left(-i\frac{\partial^{\tilde{d}}}{\partial^{\tilde{d}}z}\right), \tilde{d} \cdot Q\left(-i\frac{\partial^{\tilde{d}}}{\partial^{\tilde{d}}z}\right)\right] = \tilde{d}^2 \cdot r_{P,Q}, \tag{61}$$

where $r_{f,g}$, $r_{P,g}$, $r_{f,Q}$, $r_{P,Q}$ are appropriate elements of $\mathcal{D}_{\tilde{d}}(U \times \mathbb{C})$. As furthermore

$$\mathfrak{q}(\{a,b\}) = \left\{(f \circ \pi)\xi^{\alpha}, (g \circ \pi)\xi^{\beta}\right\}\left(-i\frac{\partial^{\tilde{d}}}{\partial^{\tilde{d}}z}\right) + \tilde{d} \cdot s_{a,b} \tag{62}$$

with $s_{a,b} \in \mathcal{D}_{\tilde{d}}(U \times \mathbb{C})$, equations (58) to (61) entail the expansion in (53). □

The normal order quantization has the nice feature of being functorial. This is expressed in the following proposition.

**Theorem 2.10** *Associate to any* RIEMANNIAN *manifold $X$ its normal order quantization $(q, \mathfrak{q}) : (X \times \mathbb{C}, \mathcal{D}_{\tilde{d}}) \to (X, \mathcal{D}_0)$ of the* POISSON *space $(X, \mathcal{D}_0)$ with respect to the* LEVI-CIVITA *connection $\nabla$. This gives rise to a functor $\mathcal{Q}^{no}$ from the category* <u>Riem</u> *of* RIEMANNIAN *manifolds with isometric embeddings as morphisms to the category* $\underline{\mathrm{Qu}}^{\mathrm{Dirac}}_{(\mathbb{C},\mathcal{C}^{\infty}_{\mathbb{C}})}$ *of differentiable* DIRAC *quantizations. $\mathcal{Q}^{no}$ has range in $\underline{\mathrm{Qu}}^{\mathcal{C}^{\infty}}_{0,(\mathbb{R},\mathcal{C}^{\infty})}$ and preserves the monoidal structures by mapping products in* <u>Riem</u> *to (tensor) products in* $\underline{\mathrm{Qu}}^{\mathcal{C}^{\infty}}_{0,(\mathbb{R},\mathcal{C}^{\infty})}$.

PROOF: We postpone the proof till the end of the next section, where we will provide for some mathematical tools giving a short proof of the theorem. □

**Note 2.11** An element $P = p\left(-i\frac{\partial^{\tilde{d}}}{\partial^{\tilde{d}}z}\right) \in \mathcal{D}_{\tilde{d}}(U \times O)$ is already characterized by all its projections

$$P_{\hbar} := p_{\hbar}\left(-i\frac{\partial^{\hbar}}{\partial^{\hbar}z}\right) := p\left(-i\frac{\partial^{\hbar}}{\partial^{\hbar}z}, \hbar\right) \in \mathcal{D}_{\tilde{d}}(U \times \{\hbar\})/\mathfrak{m}_{\hbar}\mathcal{D}_{\tilde{d}}(U \times \{\hbar\}) \cong \mathcal{D}_{\hbar}(U) \tag{63}$$

with $\hbar \in O$ and $p_{\hbar}(-) = p(-, \hbar) : T^*U \to \mathbb{C}$. We will write $\mathfrak{q}_{\hbar}$ for the composition of $\mathfrak{q}$ with the projection of $\mathcal{D}_{\tilde{d}}$ on $\mathcal{D}_{\hbar}$. In that notation equation (53) then reads

$$[\mathfrak{q}_{\hbar}(a), \mathfrak{q}_{\hbar}(b)] = -i\hbar\,\mathfrak{q}_{\hbar}(\{a,b\}) + \hbar^2\,r_{a,b,\hbar} \tag{64}$$

for every $\hbar \in \mathbb{C}$. Up to first order in $\hbar$ this is the standard DIRAC quantization condition.

**Note 2.12** Let us briefly motivate that the morphism q corresponds to what physicists call a normal order quantization. In case $X \cong \mathbb{R}^n$, we have a global and flat coordinate system $(z, \xi) : T^*X \to \mathbb{R}^{2n}$ of $T^*X$ and

$$\mathfrak{q}\left((f \circ \pi) \cdot \xi^\alpha\right) = (-i\,d)^{|\alpha|} f \frac{\partial^{d|\alpha|}}{\partial^d z^\alpha} \tag{65}$$

for $f \in \mathcal{E}(U)$, $U \in X$ open. In particular $\mathfrak{q}(z_k \cdot \xi_l) = (-id)\, z_k \frac{\partial^d}{\partial^d z_l} = \mathfrak{q}(z_k) \cdot \mathfrak{q}(\xi_l)$, which shows that the quantization map q normally orders the coordinate functions $z_k$ and $\xi_l$. In other words this means that quantized momentum variables are written to the right of quantized space variables.

In case $X$ is not flat, let us describe the operator $(\ )^z$ which was used to define the quantization map in terms of local coordinates. First $(\ )^z$ expands any $a \in \mathcal{D}(U)$ over a point $x \in U$ as a polynomial in the $\xi_{x,k}$, which correspond to a normal coordinate system $z_x : U \to \mathbb{C}^n$. This expansion induces the normal ordering. Afterwards $(\ )^z$ transforms it back over $x$ to the given coordinate system $z$ by means of the operator $\mathcal{D}_{\bar{d}}(z, z_x, \mathbb{C})$.

## 2.3 Action on smooth, holomorphic and integrable functions

Let $X$ be an arbitrary smooth or complex manifold and denote by $\mathcal{E}_X$ the sheaf of smooth (resp. holomorphic) functions on $X$. In this section we will now first introduce the natural action of the sheaf $\mathcal{D}_\hbar$, $\hbar \in \mathbb{C}$ of $\hbar$-deformed polynomial observables on $X$ on the sheaf $\mathcal{E}_X$. Likewise we will define a natural action of the sheaf $\mathcal{D}_{\bar{d}}$ of universally deformed polynomial observables on $X$ on the sheaf $\mathcal{E}_{X \times \mathbb{C}}$. Then later on in this section we will show how one can use this for calculational purposes in the case, where $X$ is a RIEMANNIAN manifold. Additionally we give some facts concerning a HILBERT space representation of quantized observables and show that we can thus construct a functor acting on a category of RIEMANNIAN manifolds and with range in the category of HILBERT space representations.

Now let us define the first sheaf morphism $\mathcal{D}_\hbar \times \mathcal{E}_X \to \mathcal{E}_X$ in local coordinates $z : U \to k^n$ by

$$\left(a\left(-i\frac{\partial^\hbar}{\partial^\hbar z}\right), f\right) \longmapsto a\left(-i\frac{\partial^\hbar}{\partial^\hbar z}\right) \cdot f = \sum_{\alpha \in \mathbb{N}^n} (-i\hbar)^{|\alpha|} a_\alpha \frac{\partial^{|\alpha|} f}{\partial z^\alpha}, \tag{66}$$

where $a = \sum (a_\alpha \circ \pi) \xi^\alpha \in \mathcal{E}(T^*U)$, $a_\alpha, f \in \mathcal{E}(U)$ and $a_\alpha = 0$ for all but finitely many $\alpha \in \mathbb{N}^n$. Define a second sheaf morphism $\mathcal{D}_{\bar{d}} \times \mathcal{E}_{X \times \mathbb{C}} \to \mathcal{E}_{X \times \mathbb{C}}$ in the same way by

$$\left(p\left(-i\frac{\partial^d}{\partial^d z}\right), r\right) \longmapsto p\left(-i\frac{\partial^d}{\partial^d z}\right) \cdot r = \sum_{\alpha \in \mathbb{N}^n} (-i\bar{d})^{|\alpha|} p_\alpha \frac{\partial^{|\alpha|} r}{\partial z^\alpha}, \tag{67}$$

where $p = \sum (p_\alpha \circ \pi \circ pr_1) (\xi^\alpha \circ pr_1) \in \mathcal{E}(T^*U \times O)$, $O \subset \mathbb{C}$ open, $pr_1 : T^*X \times \mathbb{C} \to T^*X$ is the projection on the first coordinate, $p_\alpha, r \in \mathcal{E}(U \times O)$, and $p_\alpha = 0$ for all but finitely many $\alpha \in \mathbb{N}^n$. The reader will easily verify that these morphisms are well-defined and give rise to actions $\mathcal{D}_\hbar \mapsto \mathcal{H}om(\mathcal{E}_X, \mathcal{E}_X)$ and $\mathcal{D}_{\tilde{d}} \mapsto \mathcal{H}om(\mathcal{E}_{X \times \mathbb{C}}, \mathcal{E}_{X \times \mathbb{C}})$. Because the sheaf $\mathcal{D}_Y$ of differential operators on a smooth or complex manifold $Y$ acts faithfully on $\mathcal{E}_Y$, the above definitions show that the action of $\mathcal{D}_\hbar$ on $\mathcal{E}_X$ is faithfull in the case $\hbar \neq 0$ and has kernel $\mathcal{K}$ with $\mathcal{K}(U) = \{a = \sum_{\nu \in \mathbb{N}} a_\nu \in \mathcal{D}_0(U) : a_\nu \in \mathcal{E}(\nu)(U), a_0 = 0\}$ for $\hbar = 0$. The sheaf $\mathcal{D}_{\tilde{d}}$ acts faithfully on $\mathcal{E}_{X \times \mathbb{C}}$.

**Note 2.13** The above considerations show that the element $\frac{\partial^\hbar}{\partial^\hbar z}$ can be interpreted as the differential operator $\hbar \frac{\partial}{\partial z}$ for $\hbar \neq 0$. The reason why we have chosen the special notation $\frac{\partial^\hbar}{\partial^\hbar z}$ is that we want to include the "classical limit" case $\hbar = 0$: then $\frac{\partial^\hbar}{\partial^\hbar z} \neq 0$, whereas $\hbar \frac{\partial}{\partial z} = 0$.

The representation of deformed observables on spaces of smooth functions enables us to write down easier formulas for quantized observables. Note that from now on we have to assume $X$ to be a RIEMANNIAN manifold.

**Proposition 2.14** *Let $a \in \mathcal{D}_0(U)$ be a polynomial symbol on a RIEMANNIAN manifold $X$, and $U \subset X$ open. Express $a$ locally for every $x \in U$ in the form*

$$a\Big|_{T_x^*U} = \sum_\alpha (a_{x,\alpha} \circ \pi) \xi_x^\alpha, \tag{68}$$

*where $a_{x,\alpha} \in \mathcal{C}^\infty(U_x)$ and $(z_x, \xi_x) : T^*U_x \to \mathbb{R}^{2n}$ is induced by a normal coordinate system $z_x : U_x \to \mathbb{R}^n$ on an open neighborhood $U_x \subset U$ of $x$. Then the action of $\mathfrak{q}_\hbar(a)$ on $\mathcal{C}^\infty(U)$ is given by*

$$\mathcal{C}^\infty(U) \ni f \mapsto \mathfrak{q}_\hbar(a)f = \left( U \ni x \mapsto \sum_\alpha (-i\hbar)^\alpha a_{x,\alpha}(x) \frac{\partial^{|\alpha|} f}{\partial z_x^\alpha}(x) \in \mathbb{C} \right) \in \mathcal{C}^\infty(U), \tag{69}$$

*and the action of $\mathfrak{q}(a)$ on $\mathcal{C}^\infty(U \times \mathbb{R})$ by*

$$\mathcal{C}^\infty(U \times \mathbb{R}) \ni r \mapsto \mathfrak{q}(a)r =$$
$$= \left( U \times \mathbb{R} \ni (x, \hbar) \mapsto \sum_\alpha (-i\hbar)^\alpha a_{x,\alpha}(x) \frac{\partial^{|\alpha|} r}{\partial z_x^\alpha}(x) \in \mathbb{C} \right) \in \mathcal{C}^\infty(U \times \mathbb{R}). \tag{70}$$

PROOF: By Eq. (44) and relation (45) we can write

$$a^{z_x}\Big|_{T_x^*X \times \mathbb{C}} = \sum_\alpha a_{x,\alpha}(x) \xi_x^\alpha, \tag{71}$$

hence the claim follows by (66) and (67). $\square$

**Note 2.15** Proposition 2.14 gives an easier description of the quantization operator $\mathfrak{q}_\hbar$ with $\hbar \neq 0$ than the original definition in section 2.2. The reason why we have chosen the more complicated definition as the basic one is that it also holds in the case $\hbar = 0$.

By the above considerations the quantized observables can be interpreted as differential operators over $X$. As $X$ is assumed to be a RIEMANNIAN manifold, we can form the HILBERT space $L^2(X)$ of square-integrable functions on $X$ in case $X$ is even orientable. The globally defined quantized observables $A \in \mathcal{D}_\hbar(X)$ then give rise to densely defined linear (in general unbounded) operators on $L^2(X)$. A common admissable dense domain for all quantized observables is the space $\mathcal{C}_0^\infty(X)$ of smooth functions on $X$ with compact support. The action of $\mathcal{D}_\hbar(X)$ on $L^2(X)$ is faithful for $\hbar \neq 0$.

It is an obvious fact that $X \to L^2(X)$ gives rise to a functor $L^2$ from the category of orientable RIEMANNIAN manifolds with isometric submersive embeddings to the category of HILBERT spaces. The following proposition now states that $L^2$ and the functor of normal order quantization fit together in a certain sense.

**Theorem 2.16** *Assume $\hbar \neq 0$. Then the functors of normal order quantization and of square integrable functions give rise to a monoidal functor $\mathcal{R}_\hbar$ going from the category* Riem$^{\text{or.}}$ *of orientable RIEMANNIAN manifolds with isometric submersive embeddings as morphisms to the category of HILBERT space representations* Rep *(with dense common admissable domain). In particular for every isometric and submersive embedding $f : X \to Y$ there is an isometric intertwining operator $\mathcal{R}_\hbar(f) = (f^*, f^*) : (\mathcal{D}_\hbar(Y), L^2(Y)) \to (\mathcal{D}_\hbar(X), L^2(X))$ given by the pull-back with $f$. Furthermore the equation*

$$f^*\left(\mathfrak{q}_\hbar(b)\, h\right) \;=\; \mathfrak{q}_\hbar(f^*b)\,(f^*h) \tag{72}$$

*holds for every $b \in \mathcal{D}_0(Y)$ and every $h \in \mathcal{C}_0^\infty(Y)$.*

Let us first give the proof of Theorem 2.10.

PROOF (of Theorem 2.10): In the same way like associating to a manifold its sheaf of differential operators is functorial, one can show that associating $\mathcal{D}_\hbar$ (resp. $\mathcal{D}_d$) to $X$ is functorial from Riem$^{\text{or.}}$ to the category of sheaves. Thus we have a functor from Riem$^{\text{or.}}$ to deformations. Now the construction of the exponential function exp and its corresponding normal coordinates is natural on the category Riem$^{\text{or.}}$, so even the process of normal order quantization is functorial. Let us show this in more detail, and assume $f : X \to Y$ to be an isometric embedding. Then we have

$$f \circ \exp = \exp \circ Tf. \tag{73}$$

To prove that quantization is natural, it now suffices to show that

$$f^*\left(\mathfrak{q}_\hbar(b)\, g\right) \;=\; \mathfrak{q}_\hbar(f^*b)\, f^*g \tag{74}$$

holds for every $b \in \mathcal{D}_0(U)$ and $g \in \mathcal{C}^\infty(U)$ with $U \subset Y$ open. So first choose $x \in f^{-1}(U)$ and expand locally $b = \sum_{\alpha \in \mathbb{N}^n} b_{f(x),\alpha}\, \xi_{f(x)}^\alpha$ with $b_{f(x),\alpha} \in \mathcal{C}_Y^\infty$ and $b \circ f(x) = \sum_{\alpha \in \mathbb{N}^n} (b \circ f)_{x,\alpha}\, \xi_x^\alpha$ with $(b \circ f)_{x,\alpha} \in \mathcal{C}_X^\infty$. Eq. (73) then entails $f_*\xi_x = \xi_{f(x)}$, so we have $b_{f(x),\alpha} = (b \circ f)_{x,\alpha}$.

But by Proposition 2.14 this just entails $\left(\mathfrak{q}_\hbar(b)\, g\right)(f(x)) = \left(\mathfrak{q}_\hbar(f^*b)\, f^*g\right)(x)$, so Eq. (74) holds. This proves the first part of claim. The second follows directly from the definition of $\underline{\mathsf{Qu}}_{\mathcal{C}^\infty}$ and its monoidal structure in section 1.3. $\square$

PROOF (of Theorem 2.16): As $f$ is an isometric submersive embedding, the pull-back $f^*$ induces an isometric embedding from $L^2(Y)$ to $L^2(X)$. The rest is now a direct consequence of the preceding proof, in particular of Eq. (74), and the fact that $L^2(X \times Y) = L^2(X) \hat{\otimes} L^2(Y)$, where $\hat{\otimes}$ is the tensor product in the category of HILBERT spaces. $\square$

## 2.4 Quantization of some examples from physics

The results of the last section give us the means to calculate the quantization of certain observables important for physics. Note that we still assume $X$ to be a RIEMANNIAN manifold. In the following we are going to calculate the quantization of observables which are of order 0,1 or 2 in the fibers of $T^*X$ or in other words which are of order 0, 1 or 2 in momentum.

($i$) Obviously the quantization of a smooth function $a = f \circ \pi \in \mathcal{C}^\infty_{T^*X}(0)(U)$ with $U \subset X$ open and $f \in \mathcal{C}^\infty(U)$ is trivial in the sense that

$$\mathfrak{q}(a) = f \circ q \quad \text{and} \quad \mathfrak{q}_\hbar(a) = f. \tag{75}$$

$\mathfrak{q}_\hbar(a)$ acts on $\mathcal{C}^\infty(U)$ as a multiplication operator:

$$\mathfrak{q}_\hbar(a) \cdot g = fg, \quad g \in \mathcal{C}^\infty(U). \tag{76}$$

($ii$) Now assume $\mathcal{V} : T^*U \to \mathbb{C}$ to be smooth and $\mathbb{R}$-linear in the fibers. Then we can regard $\mathcal{V}$ as a vector field over $U$ and write in local coordinates $\mathcal{V}\big|_V = \sum v_k \frac{\partial}{\partial z_k}$, where $V \subset U$ is open, $z : V \to \mathbb{R}^n$ a coordinate system and $v_k \in \mathcal{C}^\infty(V)$ for $k = 1, ..., n$. Hence the quantization of $\mathcal{V}$ is

$$\mathfrak{q}_{V \times \mathbb{C}}(\mathcal{V}) = -i \sum v_k \frac{\partial^d}{\partial^d z_k} = \mathcal{V}^d \quad \text{resp.} \quad (\mathfrak{q}_\hbar)_V(\mathcal{V}) = -i \sum v_k \frac{\partial^\hbar}{\partial^\hbar z_k} = \mathcal{V}^\hbar \tag{77}$$

and its action on smooth functions given by

$$(\mathfrak{q}_\hbar)_V(\mathcal{V}) \cdot g = \mathcal{V}^\hbar \cdot g = -i\hbar\,(\mathcal{V}g), \quad g \in \mathcal{C}^\infty(U). \tag{78}$$

($iii$) Finally choose a smooth $T : T^*U \to \mathbb{C}$ which is quadratic in the fibers. Then there exists a unique symmetric tensor field $\mathbf{T} : U \to T^2 X \otimes \mathbb{C}$ such that $T(\omega) = (\omega \otimes \omega) \circ \mathbf{T}_x$ for $\omega \in T^*X$, $x \in U$. Expand $\mathbf{T}$ in normal coordinates over $x$:

$$\mathbf{T} = \sum_{k,l} t_{x,kl} \frac{\partial}{\partial z_{x,k}} \frac{\partial}{\partial z_{x,l}} \tag{79}$$

with $t_{x,kl} \in \mathcal{C}^\infty(U)$. This gives $T = \sum\limits_{kl} (t_{x,kl} \circ \pi) \, \xi_{x,k} \, \xi_{x,l}$ and

$$T^\hbar \cdot f(x) = -\hbar^2 \sum t_{x,kl}(x) \left( \frac{\partial}{\partial z_{x,k}} \frac{\partial}{\partial z_{x,l}} f \right)(x). \tag{80}$$

Recalling that the HESSIAN of $f$ is the symmetric tensor field $f^{\mathrm{H}} : U \to T^{*2}X$ with $f^{\mathrm{H}}(x) = \sum\limits_{kl} \left( \frac{\partial}{\partial z_{x,k}} \frac{\partial}{\partial z_{x,l}} f \right)(x) \, dz_{x,k}\big|_{T_x X} \odot dz_{x,l}\big|_{T_x X}$, where $\odot$ denotes the symmetric tensor product, the relation

$$T^\hbar \cdot f = -\hbar^2 \left( \mathbf{T} \lrcorner f^{\mathrm{H}} \right) \tag{81}$$

now follows. In the same way

$$T^d \cdot f = -d^2 \left( \mathbf{T} \lrcorner f^{\mathrm{H}} \right). \tag{82}$$

Let us now calculate the quantization of some specific observables in three important mechanical systems. For further information on these classical systems see for example ABRAHAM, MARSDEN [1], SCHOTTENLOHER [48] or THIRRING [53, 54]. The last two references also treat their quantization.

**2.17 Harmonic oscillator.** The classical configuration space of the harmonic oscillator is given by $X = \mathbb{R}^n$ with its canonical EUCLIDEAN metric. Then the phase space $M = T^*X$ is isomorphic to $\mathbb{R}^{2n}$ with canonical coordinates $(x, \xi) = (x_1, ..., x_n, \xi_1, ..., \xi_n)$. Furthermore the HAMILTONIAN is given by $H = \frac{\|\xi\|^2}{2m} + \frac{1}{2} m \omega^2 \|x\|^2$, where $m > 0$ is the mass of the oscillator and $\omega$ its angular frequency. By the above considerations and the flatness of $X = \mathbb{R}^n$ we have $\mathfrak{q}_\hbar(\xi_k) = -i\hbar \frac{\partial}{\partial x_k}$ and $\mathfrak{q}_\hbar(x_k) = x_k$. Hence the the quantized HAMILTONIAN has the form

$$\mathfrak{q}_\hbar(H) = -\frac{\hbar^2}{2m} \sum_{k=1}^n \frac{\partial^2}{\partial x_k^2} + \frac{1}{2} m \omega^2 \|x\|^2, \tag{83}$$

which is just the quantized HAMILTONIAN of the harmonic oscillator in the SCHRÖDINGER picture. In geometric quantization (see WOODHOUSE [70]) this result can be derived only after a metaplectic correction.

**2.18 Free particle moving on a Riemannian manifold.** In the next example we let $X$ be a RIEMANNIAN manifold with metric $g$ and choose $M = T^*X$ with its canonical symplectic structure as the phase space of a particle moving on the RIEMANNIAN manifold $X$. The metric $g$ on $TX$ induces a metric $g^*$ and a "norm" $\| \ \|_*$ on $T^*X$ by the canonical isomorphism $TX \to T^*X$, $v \to g(-, v)$. The dynamics of a free particle with mass $m > 0$ is then described by the HAMILTONIAN $H = \frac{1}{2m} \| \ \|_*$. Now $H$ is an observable on $M$ which is quadratic in momentum, so paragraph $(iii)$ above entails $\mathfrak{q}_\hbar(H) \cdot f = -\frac{\hbar^2}{2m}\left(g^* \lrcorner f^{\mathrm{H}}\right)$ for any $f \in \mathcal{C}^\infty(M)$. Hence

$$\mathfrak{q}_\hbar(H) = -\frac{\hbar^2}{2m} \Delta_g \tag{84}$$

follows, where $\Delta_g$ is the metric LAPLACIAN on $X$. Compare this derivation again with the corresponding one in WOODHOUSE, *Geometric Quantization* [70].

**2.19 Kepler problem and Hydrogen atom.** The KEPLER problem describes a particle of mass $m$ moving in $\mathbb{R}^3 \setminus \{0\}$ under the influence of a central field force $F = -k\frac{x}{\|x\|^3}$, where $k > 0$ is the constant of force. Thus the phase space for the KEPLER problem is the cotangent bundle $T^*X$ with canonical coordinates $(x, \xi) \in (\mathbb{R}^3 \setminus \{0\}) \times \mathbb{R}^3$ of $T^*X$. The potential for the central field is $V(x) = -\frac{k}{x}$. This leads to the classical HAMILTONIAN

$$H = \frac{1}{2m}\|\xi\|^2 + V(x) = \frac{1}{2m}\|\xi\|^2 - \frac{k}{\|x\|}. \tag{85}$$

In case we consider an electron moving around a proton, i.e. we consider the Hydrogen atom, the constant $k$ is given by $k = e^2$, where $e$ is the elementary electric charge and $m$ is the mass $m_e$ of the electron.

The quantization of $H$ now obviously is

$$\mathfrak{q}_\hbar(H) = -\frac{\hbar^2}{2m}\sum_{k=1}^{3}\frac{\partial^2}{\partial x_k{}^2} + V(x). \tag{86}$$

We are not only interested in the quantization of the HAMILTONIAN but also of the angular momentum $L = (L_1, L_2, L_3)$ and the LENZ-RUNGE vector $A = (A_1, A_2, A_3)$. Recall that classically $L_j = x_k \xi_l - x_l \xi_k$ and $A_j = \frac{1}{m}(L_k \xi_l - L_l \xi_k) + k \frac{x}{|x|}$, where $(j, k, l)$ is a cyclic permutation of $(1, 2, 3)$. The normal order quantization of the angular momentum now produces the correct quantum observables, namely

$$\mathfrak{q}_\hbar(L_j) = -i\hbar\left(x_k \frac{\partial}{\partial x_l} - x_l \frac{\partial}{\partial x_k}\right). \tag{87}$$

But the normal order quantization of $A_1$ gives

$$\mathfrak{q}_\hbar(A_1) = -\frac{\hbar^2}{m}\left(x_3 \frac{\partial}{\partial x_1}\frac{\partial}{\partial x_3} - x_1 \frac{\partial}{\partial x_3}\frac{\partial}{\partial x_3} - x_1 \frac{\partial}{\partial x_2}\frac{\partial}{\partial x_2} + x_2 \frac{\partial}{\partial x_1}\frac{\partial}{\partial x_2}\right) + k\frac{x}{|x|}, \tag{88}$$

which is not a formally self-adjoint operator on $L^2(\mathbb{R}^3 \setminus \{0\})$ and thus not the "correct" quantization of $A_1$. An analog result holds for $A_2$ and $A_3$. The correct quantization of $A_k$ would be

$$\mathfrak{q}_\hbar(A_k) = \frac{1}{2m}\Big(\{\mathfrak{q}_\hbar(L_k), \mathfrak{q}_\hbar(\xi_l)\} - \{\mathfrak{q}_\hbar(L_l), \mathfrak{q}_\hbar(\xi_k)\}\Big) + k\frac{x}{|x|}, \tag{89}$$

where $\{\,,\,\}$ is the anticommutator in the algebra of quantized observables. See BOHM [8] and SCHOTTENLOHER [48] for more information on the quantization of the LENZ-RUNGE vector.

**Remark 2.20** As the quantization of the LENZ-RUNGE vector in the last example shows, normal order quantization does not always give the "correct" quantizations for higher order polynomials in momentum. This suggests that one should try to generalize normal order quantization to a quantization scheme which maps self-adjoint classical observables to (formally) self-adjoint quantized observables. A WEYL quantization scheme for arbitrary RIEMANNIAN manifolds or a mixed quantization between normal and antinormal quantizations would probably be an appropriate Ansatz for this.

Finally let us explain the importance of the HAMILTONIAN in physics or in other words the reason why we have quantized the HAMILTONIAN in all three examples above. To put it shortly, the HAMILTONIAN gives a classical or quantum mechanical system its dynamics. In the classical arena the HAMILTONIAN vector field $X_H$ corresponding to $H$ generates a one parameter family of local symplectomorphism of the phase space, or canonical transformations in physical terms, and thus induces the time evolution of the system. In the quantum setting a self-adjoint quantized HAMILTONIAN $\mathfrak{q}_\hbar(H)$ generates a one parameter family of unitary transformations of the HILBERT space of states. This family is responsible for the time evolution of the quantum mechanical system and is given by $\mathbb{R} \ni t \mapsto \exp\left(-i\mathfrak{q}_\hbar(H)t\right) \in U\left(L^2(X)\right)$.

## 2.5 Observables of infinite order in momentum

Let $X$ be a complex manifold. Then it is possible to generalize the deformation of the sheaf $\mathcal{D}_0$ over $X$ to a deformation of a larger sheaf $\mathcal{D}_0^\infty$. This sheaf consists of holomorphic functions over $T^*X$, whose TAYLOR coefficients fulfill a certain growth condition, and will be called the sheaf of **classical observables of infinite order in momentum**. In the following we will give its definition and a deformation.

Let $z : U \to \mathbb{C}^n$ be a local coordinate system of $X$ and define the linear space $\mathcal{D}^\infty(z)$ of holomorphic functions on $T^*U$ with **bounded growth** in the fibers by

$$\mathcal{D}^\infty(z) = \Bigg\{ F \in \mathcal{O}(T^*U) : F = \sum_{\nu \in \mathbb{N}} f_\nu \text{ where } f_\nu \in \mathcal{O}(\nu)(U) \text{ and}$$
$$\forall \varepsilon > 0 \, \forall K \subset T^*U \text{ compact } \exists C_{K,\varepsilon} > 0 \text{ such that } |f_\nu|_K \leq C_{K,\varepsilon}\frac{\varepsilon^\nu}{\nu!},\, \nu \in \mathbb{N} \Bigg\}. \tag{90}$$

Before proceeding let us first prove the little

**Lemma 2.21** *A formal sum $F = \sum_{\nu \in \mathbb{N}} f_\nu$ with $f_\nu = \sum_{|\alpha|=\nu} (f_\alpha \circ \pi)\, \xi^\alpha$ lies in $\mathcal{D}^\infty(z)$, iff for any $\varepsilon > 0$ and compact $K \subset U$ there exists a $D_{K,\varepsilon} > 0$ such that*

$$|f_\alpha|_K \leq D_{K,\varepsilon} \frac{\varepsilon^{|\alpha|}}{\alpha!} \tag{91}$$

*for all $\alpha \in \mathbb{N}^n$.*

PROOF: First let $F$ be an element of $\mathcal{D}^\infty(z)$, $K \subset U$ be compact and $\varepsilon > 0$. Then consider in local coordinates $(z,\xi): T^*U \to \mathbb{C}^{2n}$ the following Cauchy-integral:

$$f_\alpha(z) = \frac{1}{(2\pi i)^n} \int_{T_z^n} \frac{f_\nu(z,\xi)}{\xi^{\alpha+1}} d\xi, \quad |\alpha| = \nu, \tag{92}$$

where $z \in K$, $T_z^n$ is the $n$-torus in $T_z^*U \cong \mathbb{C}^n$ around 0 and $\alpha + 1 = (\alpha_1 + 1, ..., \alpha_n + 1)$. This integral entails

$$|f_\alpha|_K \leq |f_\nu|_{P(K,1)} \leq C_{P(K,1),\varepsilon} \frac{\varepsilon^\nu}{\alpha!}, \tag{93}$$

where $P(K,r)$ for $r > 0$ is the polycylinder $\{(z,\xi) \in T^*U : z \in K, |\xi_i| \leq r\}$. Hence Eq. (93) shows the condition being necessary.

The condition is also sufficient, because if $F$ is a formal sum fulfilling Eq. (91), then

$$|f_\nu|_{P(K,r)} \leq \sum_{|\alpha|=\nu} |f_\alpha|_K\, r^\nu \leq \frac{1}{\nu!} D_{K,\varepsilon} \sum_{|\alpha|=\nu} \frac{\nu!}{\alpha!} (\varepsilon r)^\nu = \frac{1}{\nu!} (n\varepsilon r)^\nu D_{K,\varepsilon} \tag{94}$$

is true. $\square$

Now let us furnish $\mathcal{D}^\infty(z)$ with parameter dependant products

$$m_\hbar : \mathcal{D}^\infty(z) \times \mathcal{D}^\infty(z) \to \mathcal{D}^\infty(z), \quad (F,G) = \left(\sum f_\nu, \sum g_\mu\right) \mapsto H = \sum h_\rho \tag{95}$$

by

$$h_\rho = \sum_{\nu+\mu-|\alpha|=\rho} \frac{(-i\hbar)^{|\alpha|}}{\alpha!} \frac{\partial^{|\alpha|}}{\partial \xi^\alpha} f_\nu \frac{\partial^{|\alpha|}}{\partial z^\alpha} g_\mu . \tag{96}$$

$H$ is indeed well-defined. To see that choose for $K \subset T^*U$ compact and $\varepsilon > 0$ constants $\tilde{\varepsilon} > 0$, $\delta > 0$ being small enough and constants $C_{K,\varepsilon} > 0$, $D_{K,\varepsilon} > 0$ being big enough such that

(i) $K_\delta := \{\omega \in T^*U : |(z,\xi)(\omega) - (z,\xi)(\tilde{\omega})| \leq \delta \text{ for all } \tilde{\omega} \in K\}$ is compact in $T^*U$,

(ii) $2\tilde{\varepsilon} < \varepsilon$ and $\frac{2\hbar\tilde{\varepsilon}}{\delta^2} < \frac{1}{2}$,

(iii) $|f_\nu|_{K_\delta} \leq \frac{1}{\nu!} C_{K,\varepsilon}\, \tilde{\varepsilon}^\nu$ and $|g_\mu|_{K_\delta} \leq \frac{1}{\mu!} D_{K,\varepsilon}\, \tilde{\varepsilon}^\mu$ for all $\nu, \mu \in \mathbb{N}$.

Applying the Cauchy inequalities then gives:

$$\begin{aligned}
|h_\rho|_K &\leq \sum_{\substack{\nu,\mu\in\mathbb{N},\,\alpha\in\mathbb{N}^n \\ \rho=\nu+\mu-|\alpha|}} \frac{\hbar^{|\alpha|}}{\alpha!} \left|\frac{\partial^{|\alpha|}}{\partial\xi^\alpha}f_\nu\right|_K \left|\frac{\partial^{|\alpha|}}{\partial z^\alpha}g_\mu\right|_K \\
&\leq \sum_{\substack{\nu,\mu\in\mathbb{N},\,\alpha\in\mathbb{N}^n \\ \rho=\nu+\mu-|\alpha|}} \left(\frac{\hbar}{\delta^2}\right)^{|\alpha|} \alpha!\, |f_\nu|_{K_\delta}\, |g_\mu|_{K_\delta} \\
&\leq \sum_{\substack{\nu,\mu\in\mathbb{N},\,\alpha\in\mathbb{N}^n \\ \rho=\nu+\mu-|\alpha|}} \left(\frac{\hbar}{\delta^2}\right)^{|\alpha|} \frac{\alpha!}{\nu!\,\mu!} C_{K,\varepsilon}\, \tilde{\varepsilon}^\nu\, D_{K,\varepsilon}\, \tilde{\varepsilon}^\mu \\
&\leq \frac{1}{\rho!} C_{K,\varepsilon}\, D_{K,\varepsilon} \sum_{\alpha\in\mathbb{N}^n} \left(\frac{\hbar}{\delta^2}\right)^{|\alpha|} \sum_{\substack{\nu,\mu\in\mathbb{N} \\ \rho=\nu+\mu-|\alpha|}} \frac{(\rho+|\alpha|)!}{\nu!\,\mu!} \tilde{\varepsilon}^\nu\,\tilde{\varepsilon}^\mu \\
&= \frac{1}{\rho!} C_{K,\varepsilon}\, D_{K,\varepsilon} \sum_{\alpha\in\mathbb{N}^n} \left(\frac{\hbar}{\delta^2}\right)^{|\alpha|} (2\tilde{\varepsilon})^{\rho+|\alpha|} \leq \frac{1}{\rho!} \left(C_{K,\varepsilon}\, D_{K,\varepsilon}\, 2^n\right) \varepsilon^\rho, \qquad (97)
\end{aligned}$$

so $H = \sum h_\rho$ is well-defined and lies in $\mathcal{D}^\infty(z)$. By a standard calculation one can prove that the bilinear map $m_\hbar$ associative. So we receive a scale of algebras $\mathcal{D}_\hbar^\infty(z)$, $\hbar \in \mathbb{C}$ such that $\mathcal{D}_\hbar$ is a subalgebra of $\mathcal{D}_\hbar^\infty(z)$.

By the above considerations it becomes also clear that $\mathcal{D}_\hbar^\infty(z)$ has the structure of a FRÉCHET space with seminorms

$$||F||_{K,\varepsilon} = \sup_{\nu\in\mathbb{N}} \left\{\frac{\nu!\,|f_\nu|_K}{\varepsilon^\nu}\right\},$$

and that the $m_\hbar$ are continuous with respect to that topology. Furthermore $\mathcal{O}(U)$ can be continuously embedded into $\mathcal{D}_\hbar^\infty(z)$ by $\mathcal{O}(U) \ni h \to H = h \circ \pi \in \mathcal{D}_\hbar^\infty(z)$.

Now we can define an action $\mathcal{D}_\hbar^\infty(z) \times \mathcal{O}(U) \to \mathcal{O}(U)$ on $\mathcal{O}(U)$ by

$$(F,h) \longmapsto F \cdot h = m_\hbar(F,H) \circ j|_U = \sum_{\alpha\in\mathbb{N}^n} (-i\hbar)^{|\alpha|} f_\alpha \frac{\partial^{|\alpha|} h}{\partial z^\alpha}, \qquad (98)$$

where $F = \sum_{\nu\in\mathbb{N}} f_\nu \in \mathcal{D}_\hbar^\infty(z)$, $f_\nu = \sum_{|\alpha|=\nu} (f_\alpha \circ \pi)\xi^\alpha$, $f_\alpha, h \in \mathcal{O}(U)$ and $j: X \to T^*X$ is the zero section. The action is continuous by definition and the fact that $m_\hbar$ and the pull-back $(j|_U)^* : \mathcal{D}_\hbar^\infty(z) \to \mathcal{O}(U)$ are continuous.

The question which now arises is, whether the algebras $\mathcal{D}_\hbar^\infty(z)$ can be patched together for different coordinate sytems. The answer is given in the following proposition.

**Proposition 2.22** *Let $\tilde{z} : U \to \mathbb{C}^n$ be another chart on $U$. Then there exists a unique algebra morphism $\mathcal{D}_\hbar^\infty(\tilde{z},z) : \mathcal{D}_\hbar^\infty(z) \to \mathcal{D}_\hbar^\infty(\tilde{z})$ such that*

$$
\begin{array}{ccc}
\mathcal{D}_\hbar^\infty(z) & \xrightarrow{\mathcal{D}_\hbar^\infty(\tilde{z},z)} & \mathcal{D}_\hbar^\infty(\tilde{z}) \\
\uparrow & & \uparrow \\
\mathcal{D}_\hbar(z) & \xrightarrow[\mathcal{D}_\hbar(\tilde{z},z)]{} & \mathcal{D}_\hbar(\tilde{z})
\end{array}
$$

*commutes. Furthermore*

$$\mathcal{D}_\hbar^\infty(\hat{z},z) \;=\; \mathcal{D}_\hbar^\infty(\hat{z},\tilde{z}) \circ \mathcal{D}_\hbar^\infty(\tilde{z},z) \tag{99}$$

*for any third chart $\hat{z}: U \to \mathbb{C}^n$.*

PROOF: This is a standard result of microlocal analysis. See for example BJÖRK [6] or PFLAUM [38]. □

By Proposition 2.22 it is possible to construct a sheaf $\mathcal{D}_\hbar^\infty$ of algebras over X such that the spaces $\mathcal{D}_\hbar^\infty(U)$ of sections over a coordinate patch $U$ with coordinates $z: U \to \mathbb{C}^n$ are naturally isomorphic to $\mathcal{D}_\hbar^\infty(z)$. Then $\mathcal{D}_\hbar$ is a subsheaf of $\mathcal{D}_\hbar^\infty$ called the sheaf of $\hbar$-**deformed observables of infinite order in momentum**. $\mathcal{D}_0^\infty$ is the above mentioned sheaf of **classical observables of infinite order in momentum**.

In the same way it is also possible to extend the sheaf $\mathcal{D}_d$ on $X \times \mathbb{C}$ to a sheaf $\mathcal{D}_d^\infty$ having the following properties:

(i) There exists an algebra monomorphism $\Delta^\infty: \mathcal{O}_\mathbb{C} \to \mathcal{D}_d^\infty$ of $\mathcal{O}_\mathbb{C}$ into the center of $\mathcal{D}_d^\infty$ such that the diagram

$$
\begin{array}{ccc}
& \mathcal{O}_\mathbb{C} & \\
{}^{\Delta}\swarrow & & \searrow^{\Delta^\infty} \\
\mathcal{D}_d & \longrightarrow & \mathcal{D}_d^\infty
\end{array}
$$

commutes.

(ii) $(\mathcal{D}_d^\infty)_\hbar = \mathcal{D}_d^\infty|_{X \times \{\hbar\}} / \mathfrak{m}_\hbar \mathcal{D}_d^\infty|_{X \times \{\hbar\}}$ is isomorphic to $\mathcal{D}_\hbar^\infty$ for all $\hbar \in \mathbb{C}$. Here $\mathfrak{m}_\hbar$ is the maximal ideal of $\mathcal{O}_\hbar$.

We do not work out the details, as they are analogous to the ones in section 2.1. Let us only point out that for the coordinate patch $U$ and an open $O \subset \mathbb{C}$ the space $\mathcal{D}_d^\infty(U \times O)$ of sections over $U \times O$ is isomorphic to

$$
\begin{aligned}
\mathcal{D}_d^\infty(z, O) \;=\; \Big\{ & P = \sum_{\nu \in \mathbb{N}} p_\nu \in \mathcal{O}(T^*U \times O) : p_\nu \in \mathcal{O}(\nu)(U \times O) \text{ and} \\
& \forall \varepsilon > 0 \, \forall K \subset (T^*U \times O) \text{ compact } \exists C_{K,\varepsilon} > 0 \text{ such that} \\
& |p_\nu|_K \le C_{K,\varepsilon} \frac{\varepsilon^\nu}{\nu!}, \; \nu \in \mathbb{N} \Big\}.
\end{aligned}
$$

In the following we refer to $\mathcal{D}_{\tilde{d}}^{\infty}$ as the sheaf of **universally deformed observables of infinite order in momentum** on the complex manifold $X$.

By the above conditions $(i)$ and $(ii)$ the morphism $D^{\infty} = (d, \Delta^{\infty}) : (X \times \mathbb{C}, \mathcal{D}_{\tilde{d}}^{\infty}) \to (\mathbb{C}, \mathcal{O}_{\mathbb{C}})$ over $d = pr_2 : X \times \mathbb{C} \to \mathbb{C}$ is fibered. It is even a deformation of $(X, \mathcal{D}_{\hbar}^{\infty})$ if we can yet show the flatness of $\mathcal{O}_{\hbar} \to (\mathcal{D}_{\tilde{d}}^{\infty})_{(x,\hbar)}$ for every $x \in X$.

To prove its flatness for an arbitrary $\hbar \in \mathbb{C}$ it is enough to show it for $\hbar = i$, because the diagram

$$\begin{array}{ccc} \mathcal{O}_i & \longrightarrow & \mathcal{O}_{\hbar} \\ \downarrow & & \downarrow \\ (\mathcal{D}_{\tilde{d}}^{\infty})_{(x,i)} & \longrightarrow & (\mathcal{D}_{\tilde{d}}^{\infty})_{(x,\hbar)} \end{array}$$

commutes. Here the isomorphism $\mathcal{O}_i \to \mathcal{O}_{\hbar}$ is given by $f_i \mapsto [f \circ (\_ + i - \hbar)]_{\hbar}$ and similarly for $(\mathcal{D}_{\tilde{d}}^{\infty})_{(x,i)} \to (\mathcal{D}_{\tilde{d}}^{\infty})_{(x,\hbar)}$. Now $\mathcal{D}_i$ (resp. $\mathcal{D}_i^{\infty}$) is isomorphic to the sheaf $\mathcal{D}_X$ (resp. $\mathcal{D}_X^{\infty}$) of differential operators of finite (resp. infinite) order over $X$. By results of microlocal analysis (see BJÖRK [6], Theorem 3.4.4) we already know that $\mathcal{D}_X \to \mathcal{D}_X^{\infty}$ is faithfully flat as well as $\mathcal{D}_{X \times \mathbb{C}} \to \mathcal{D}_{X \times \mathbb{C}}^{\infty}$. Furthermore we have the following commutative diagram:

$$\begin{array}{ccccc} \mathcal{O}_i & \xrightarrow{\Delta_i^{\infty}} & (\mathcal{D}_{\tilde{d}}^{\infty})_{(x,i)} & \xrightarrow{\kappa^{\infty}} & (\mathcal{D}_{X \times \mathbb{C}}^{\infty})_{(x,i)} \\ \uparrow & & \uparrow \beta & & \uparrow \\ \mathcal{O}_i & \xrightarrow{\Delta_i} & (\mathcal{D}_{\tilde{d}})_{(x,i)} & \xrightarrow{\kappa} & (\mathcal{D}_{X \times \mathbb{C}})_{(x,i)} \end{array}$$

where $\kappa$ (resp. $\kappa^{\infty}$) are the natural embeddings. We want to show $\Delta_i^{\infty}$ being faithfully flat. For that it suffices to prove the faithful flatness of $\beta$ because we already know $\Delta_i$ has this property. Now as $\mathcal{O}_i$-module $(\mathcal{D}_{X \times \mathbb{C}})_{(x,i)}$ is isomorphic to the direct sum $(\mathcal{D}_{\tilde{d}})_{(x,i)}^{\mathbb{N}}$, so $\kappa$ is faithfully flat. By the above result from microlocal analysis $\kappa^{\infty} \circ \beta$ then is faithfully flat. By MALGRANGE [33], Prop. 4.7 $\beta$ is faithfully flat iff for every ideal $J \subset (\mathcal{D}_{\tilde{d}})_{(x,i)}$ the identity

$$J (\mathcal{D}_{\tilde{d}}^{\infty})_{(x,i)} = J \mathcal{D}_{(x,i)}^{\infty} \cap (\mathcal{D}_{\tilde{d}}^{\infty})_{(x,i)} \tag{100}$$

holds. "$\subset$" is clear, so we prove "$\supset$". Let $d = \sum j_k d_k \in J \mathcal{D}_{(x,i)}^{\infty}$ such that $d \in (\mathcal{D}_{\tilde{d}}^{\infty})_{(x,i)}$, $d_k \in \mathcal{D}_{(x,i)}^{\infty}$, $j_k \in J$. Then $d_k$ can be written uniquely as a direct sum $d_k = d_{k0} + \tilde{d}_k$, where $d_{k0} \in (\mathcal{D}_{\tilde{d}}^{\infty})_{(x,i)}$ and $\tilde{d}_k = \sum_{\nu \geq 1} d_{k1} \left(\frac{\partial}{\partial \hbar}\right)^{\nu}$ with $d_{k\nu} \in (\mathcal{D}_{\tilde{d}}^{\infty})_{(x,i)}$. This gives $d = d_0 + \tilde{d}$, where $d_0 = \sum j_k d_{k0}$, $\tilde{d} = \sum j_k d_{k1} = \sum_{\nu \geq 1} \left(\sum d_{k\nu}\right) \left(\frac{\partial}{\partial \hbar}\right)^{\nu}$. As $d_0, d \in (\mathcal{D}_{\tilde{d}}^{\infty})_{(x,i)}$ and $\tilde{d} = 0$ or $\tilde{d} \notin (\mathcal{D}_{\tilde{d}}^{\infty})_{(x,i)}$, the expansion $d = d_0 + \tilde{d}$ gives $\tilde{d} = 0$, so $d = d_0 \in J (\mathcal{D}_{\tilde{d}}^{\infty})_{(x,i)}$. This proves the faithful flatness of $\beta$ and $\beta \circ \Delta_i$. Furthermore we receive the following theorem.

**Theorem 2.23** *The sheaf morphism $(d, \Delta^\infty) : (X \times \mathbb{C}, \mathcal{D}_d^\infty) \to (\mathbb{C}, \mathcal{O}_\mathbb{C})$ is fibered and a flat deformation of $(X, \mathcal{D}_\hbar^\infty)$ for every distinguished point $\hbar \in \mathbb{C}$.*

*Furthermore associating the deformation $(d, \Delta^\infty)$ of $(X, \mathcal{D}_\hbar^\infty)$ to a complex manifold $(X, \mathcal{O}_X)$ gives rise to a functor $\mathcal{D}^\infty$ from the category of complex manifolds to the category $\underline{\mathrm{Def}}_{0,(\mathbb{C},\mathcal{O})}^{\mathrm{flat}}$ of flat deformations over $(\mathbb{C}, \mathcal{O})$ with distinguished point $0$.*

PROOF: It only remains to show functoriality. But as the construction of the sheaf of infinite order differential operators on a complex manifold is natural, the claim follows.
□

**Corollary 2.24** *The sheaf $\mathcal{D}_0^\infty$ on a complex manifold $X$ of holomorphic functions of bounded growth on $T^*X$ can be be flatly deformed to the sheaf $\mathcal{D}^\infty = \mathcal{D}_i^\infty$ of infinite order differential operators on $X$.*

# 3 Quantization of Spaces of Symbols

In section 2.5 we found a deformation of the sheaf $\mathcal{D}_0^\infty$ of observables with infinite order in momentum on a complex cotangent bundle. Unfortunately it is not possible to carry over the methods used for the construction of $\mathcal{D}_0^\infty$ and its deformation to the real case. The reason is that we used complex techniques like the CAUCHY integral which are not applicable in the smooth and real arena. But, as ALAN WEINSTEIN [63] pointed out to me, the complete symbol calculus of Appendix B should be the tool to find a quantization of a rather general class of observables on a cotangent bundle. Roughly speaking this observable class will be a space of symbols on a smooth cotangent bundle. In the following these ideas will be worked out in detail. For necessary definitions and properties of symbol spaces the reader is refered to Appendix B.

## 3.1 Noncommutative algebra structures on symbol spaces

First let us define the POISSON space to be quantized. Let $X$ be a smooth RIEMANNIAN manifold and denote by $S_{\rho,\delta}^\infty$ the sheaf of symbols of type $(\rho, \delta)$ on $X$ according to Appendix B.1. We thus receive a commutativly ringed space $\left(X, S_{\rho,\delta}^\infty\right)$. The standard POISSON brackets $\{,\}$ over smooth function algebras on a cotangent bundle obviously give $\left(X, S_{\rho,\delta}^\infty\right)$ the structure of a POISSON space. Next let $a$ be an element of $S^{-\infty}(U)$ and $U$ a coordinate patch of $X$. Then take the derivatives $\frac{\partial^{|\alpha|}}{\partial z^\alpha}\frac{\partial^{|\beta|}}{\partial \xi^\beta}a$ with respect to local coordinates over $T^*U$ and check that $\frac{\partial^{|\alpha|}}{\partial z^\alpha}\frac{\partial^{|\beta|}}{\partial \xi^\beta}a$ again lies in $S^{-\infty}(U)$. Hence $S^{-\infty}$ is an ideal sheaf of $S_{\rho,\delta}^\infty$ with respect to the POISSON brackets. This gives rise to the POISSON space $\left(X, S_{\rho,\delta}^\infty/S^{-\infty}\right)$. By Example B.3 $(i)$ we also have a natural morphism $\left(X, S_{\rho,\delta}^\infty/S^{-\infty}\right) \to (X, \mathcal{D}_0)$ between POISSON spaces. Note that the arrow between the two POISSON spaces is inverse to the arrow between the corresponding sheaves of algebras.

Next let us introduce some linear mappings between symbol spaces. These mappings will be needed in the sequel to define the deformed products on $\left(X, S_{\rho,\delta}^\infty/S^{-\infty}\right)$. Denote by $\iota_\hbar : S_{\rho,\delta}^\infty(U, X) \to S_{\rho,\delta}^\infty(U, X)$ for $\hbar \in \mathbb{R}$ the function

$$a \;\mapsto\; \left(T^*U \ni \xi \mapsto a(\hbar\xi) \in \mathbb{C}\right). \tag{101}$$

In case $\hbar \neq 0$ $\iota_\hbar$ is a linear isomorphism with inverse $\iota_{\hbar^{-1}}$. By a slight abuse of language we sometimes also write $\iota_\hbar$ for the mapping

$$\begin{array}{rcl} S_{\rho,\delta}^\infty(U \times \mathbb{R}, pr_1^*(T^*X)) & \to & S_{\rho,\delta}^\infty(U, X), \\ b & \mapsto & \left(T^*U \ni \zeta \mapsto b(\hbar\zeta, \hbar) \in \mathbb{C}\right). \end{array} \tag{102}$$

Here $pr_1 : X \times \mathbb{R} \to X$ is the projection on the first coordinate and $pr_1^*(T^*X)$ the pull-back bundle of $T^*X$ via $pr_1$.

Now let $\hbar \in \mathbb{R} \setminus \{0\}$, $a, b \in S_{\rho,\delta}^{\infty}(U, X)$ be symbols and $A_{\hbar} = \Psi_{\psi}(\iota_{\hbar}(a))$ (resp. $B_{\hbar} = \Psi_{\psi}(\iota_{\hbar}(b))$ ) the so-called $\hbar$-**deformed** pseudo-differential operator corresponding to $a$ (resp. $b$). Then by Eq. (141) in Theorem B.13 the symbol $c_{\hbar} = \iota_{\hbar^{-1}}(\sigma_{A_{\hbar} B_{\hbar}})$ has asymptotic expansion

$$c_{\hbar}(\zeta) \sim \sum_{\substack{k \in \mathbb{N}}} \sum_{\substack{\alpha, \tilde{\alpha}, \alpha_1, \ldots, \alpha_k \in \mathbb{N}^n \\ \tilde{\alpha} + \alpha_1 + \ldots + \alpha_k = \alpha}} \sum_{\substack{\beta, \beta_1, \ldots, \beta_k \in \mathbb{N}^n \\ \beta_1 + \ldots + \beta_k = \beta \\ |\beta_1|, \ldots, |\beta_k| \geq 2}} \left(\frac{\hbar}{i}\right)^{|\alpha|+|\beta|-k} \frac{1}{k! \cdot \tilde{\alpha}! \cdot \alpha_1! \cdot \ldots \cdot \alpha_k! \beta_1! \cdot \ldots \cdot \beta_k!}$$
$$\left[\left.\frac{\partial^{|\alpha|}}{\partial \xi_{\pi(\zeta)}{}^{\alpha}}\right|_{\zeta} a\right] \left\{\left.\frac{\partial^{|\tilde{\alpha}|}}{\partial z_{\pi(\zeta)}{}^{\tilde{\alpha}}}\right|_{\pi(\zeta)} \left[\left.\frac{\partial^{|\beta|}}{\partial \xi_{(-)}{}^{\beta}}\right|_{d_{(-)}\varphi(\cdot,\zeta)} b\right]\right\} \cdot$$
$$\left\{\left.\frac{\partial^{|\alpha_1|}}{\partial z_{\pi(\zeta)}{}^{\alpha_1}}\right|_{\pi(\zeta)} \left[\left.\frac{\partial^{|\beta_1|}}{\partial z_{(-)}{}^{\beta_1}}\right|_{(-)} \varphi(\cdot, \zeta)\right]\right\} \cdot \ldots \cdot \left\{\left.\frac{\partial^{|\alpha_k|}}{\partial z_{\pi(\zeta)}{}^{\alpha_k}}\right|_{\pi(\zeta)} \left[\left.\frac{\partial^{|\beta_k|}}{\partial z_{(-)}{}^{\beta_k}}\right|_{(-)} \varphi(\cdot, \zeta)\right]\right\}.$$
(103)

In particular we have

$$c_{\hbar}(\zeta) =$$
$$a(\zeta) \cdot b(\zeta) + \left(\frac{\hbar}{i}\right) \sum_{l=1}^{n} \left[\frac{\partial a}{\partial \xi_{\pi(\zeta),l}}(\zeta)\right] \left\{\frac{\partial \left[b(d_{(-)}\varphi(\cdot, \zeta))\right]}{\partial z_{\pi(\zeta),l}}(\pi(\zeta))\right\} + \hbar^2 r(\zeta, \hbar), \quad (104)$$

where $r \in S_{\rho,\delta}^{\mu+\tilde{\mu}-\rho+\delta}(U \times \mathbb{R}, pr_1^*(T^*X))$, and $\mu$ (resp. $\tilde{\mu}$) is the order of the symbol $a$ (resp. $b$). Note that the bilinear mapping $S_{\rho,\delta}^{\infty}(U, X) \times S_{\rho,\delta}^{\infty}(U, X) \ni (a, b) \to a \cdot_{\hbar} b = c_{\hbar} \in S_{\rho,\delta}^{\infty}(U, X)$ is only associative modulo smoothing symbols. This follows easily from the fact that the symbol map $\sigma$ and the map $\Psi_{\psi}$ are inverse to each other up to smoothing symbols resp. smoothing pseudo-differential operators. Altogether we therefore receive a scale $\left(S_{\rho,\delta;\hbar}^{\infty}(\cdot, X)\right)_{\hbar \in \mathbb{R}}$ of sheaves of (noncommutative) algebras, such that the underlying linear structure of $S_{\rho,\delta;\hbar}^{\infty}(U, X)$ is that of $S_{\rho,\delta}^{\infty}/S^{-\infty}(U, X)$. The product of two equivalence classes $a + S^{-\infty}, b + S^{-\infty} \in S_{\rho,\delta;\hbar}^{\infty}(U, X)$ is given by the equivalence class of $a \cdot_{\hbar} b = c_{\hbar} = \iota_{\hbar^{-1}}(\sigma_{A_{\hbar} B_{\hbar}})$ in the case $\hbar \neq 0$ and by the equivalence class of the pointwise product $a \cdot b$ in the case $\hbar = 0$. The sheaf $S_{\rho,\delta;\hbar}^{\infty}(\cdot, X)$ is called the sheaf of $\hbar$-**deformed symbols of type** $(\rho, \delta)$ on a smooth RIEMANNIAN manifold $X$.

Let us now embed $\mathcal{D}_{\hbar}$ into $S_{\rho,\delta;\hbar}^{\infty}$ for $\hbar \neq 0$. Every $A \in \mathcal{D}_{\hbar}(U)$, $U \subset X$ open acts as a differential operator on $\mathcal{C}^{\infty}(U)$. Therefore the symbol $a = \iota_{\hbar^{-1}} \sigma_A$ defines a unique element $\sigma_{\hbar, A} = a + S^{-\infty} \in S_{\rho,\delta;\hbar}^{\infty}(U)$. If $B \in \mathcal{D}_{\hbar}$ is another $\hbar$-deformed polynomial symbol, we have for $b = \iota_{\hbar^{-1}} \sigma_B$

$$a \cdot_{\hbar} b = \iota_{\hbar^{-1}}(\sigma_{A_{\hbar} B_{\hbar}}) = \iota_{\hbar^{-1}}(\sigma_{AB}), \quad (105)$$

that means the mapping

$$\sigma_{\hbar} : \mathcal{D}_{\hbar}(U) \to S_{\rho,\delta;\hbar}^{\infty}(U), \ A \mapsto \sigma_{\hbar,A} = \iota_{\hbar^{-1}} \sigma_A + S^{-\infty} \quad (106)$$

is linear and multiplicative. It is also injective, as the symbol of a differential operator is smoothing if and only if the differential operator is zero. Let us also give a second definition for $a$. Assuming $U$ to be sufficiently small choose an orthonormal frame $e_1, ..., e_n$ of $TX$ over $U$ and for every $x \in U$ normal coordinates $z_x : U \to \mathbb{R}^n$ such that $\left.\frac{\partial}{\partial z_{x,k}}\right|_x = e_k$ for $k = 1, ..., n$. Then express $A \in \mathcal{D}(U)$ for every $x \in U$ in the form $A = a_x\left(-i\frac{\partial^\hbar}{\partial^\hbar z_x}\right)$, where $a_x$ is a polynomial symbol over $U$, and check that

$$a(\xi) = \left[A\left(\psi_{\pi(\xi)} e^{i\varphi(\cdot, \frac{\xi}{\hbar})}\right)\right](\pi(\xi)) = a_x(\xi). \tag{107}$$

In the same spirit like above we want to define a noncommutative product on the symbol sheaf $S^\infty_{\rho,\delta}/S^{-\infty}(\cdot, pr_1^*(T^*X))$. So let $a, b \in S^\infty_{\rho,\delta}(U \times O, pr_1^*(T^*X))$ be symbols, where $U \subset X$ and $O \subset \mathbb{R}$ are open with $O$ bounded, and denote for $\hbar \neq 0$ by $A_\hbar$ (resp. $B_\hbar$) the pseudo-differential operator $A_\hbar = \Psi_\psi(\iota_\hbar(a))$ (resp. $B_\hbar = \Psi_\psi(\iota_\hbar(b))$ ). Then define $c = a \cdot_d b$ by

$$c(\zeta, \hbar) = \begin{cases} a(\zeta, \hbar) \cdot b(\zeta, \hbar) & \text{for } \hbar = 0 \\ \sigma_{A_\hbar B_\hbar}\left(\frac{\zeta}{\hbar}\right) & \text{for } \hbar \neq 0 \end{cases} \tag{108}$$

We now want to show that $c$ is a symbol on $pr_1^*(T^*X)$ of order $\mu + \tilde{\mu}$, where $\mu$ (resp. $\tilde{\mu}$) is the order of $a$ (resp. $b$). Obviously $c$ is smooth on $U \times (O \setminus \{0\})$. According to the asymptotic expansion in Theorem B.13 we can find for every $N > 0$ a finite subset $I_N \subset J$ of the index set

$$J = \left\{ (k, \alpha, \tilde{\alpha}, \alpha_1, ..., \alpha_k, \beta, \beta_1, ..., \beta_k) \in \mathbb{N} \times (\mathbb{N}^n)^{2k+3} : \right.$$
$$\left. \tilde{\alpha} + \alpha_1 + ... + \alpha_k = \alpha, \ \beta_1, ..., \beta_k = \beta, \ |\beta_1|, ..., |\beta_k| \geq 2 \right\}$$

such that

$$\sigma_{A_\hbar B_\hbar}(\zeta) - \sum_{k,\alpha,\tilde{\alpha},\alpha_1,...,\alpha_k,\beta,\beta_1,...,\beta_k \in I} \hbar^{|\alpha|} \frac{i^{k-|\alpha|-|\beta|}}{k! \cdot \tilde{\alpha}! \cdot \alpha_1! \cdot ... \cdot \alpha_k! \beta_1! \cdot ... \cdot \beta_k!}$$
$$\left[\left.\frac{\partial^{|\alpha|}}{\partial \xi_{\pi(\zeta)}^\alpha}\right|_{(\zeta,\hbar)} a\right] \left\{ \left.\frac{\partial^{|\tilde{\alpha}|}}{\partial z_{\pi(\zeta)}^{\tilde{\alpha}}}\right|_{\pi(\zeta)} \left[\left.\frac{\partial^{|\beta|}}{\partial \xi_{(-)}^\beta}\right|_{(d_{(-)}\varphi(\cdot,\zeta),\hbar)} b\right] \right\} \cdot$$
$$\left\{ \left.\frac{\partial^{|\alpha_1|}}{\partial z_{\pi(\zeta)}^{\alpha_1}}\right|_{\pi(\zeta)} \left[\left.\frac{\partial^{|\beta_1|}}{\partial z_{(-)}^{\beta_1}}\right|_{(-)} \varphi(\cdot,\zeta)\right] \right\} \cdot ... \cdot \left\{ \left.\frac{\partial^{|\alpha_k|}}{\partial z_{\pi(\zeta)}^{\alpha_k}}\right|_{\pi(\zeta)} \left[\left.\frac{\partial^{|\beta_k|}}{\partial z_{(-)}^{\beta_k}}\right|_{(-)} \varphi(\cdot,\zeta)\right] \right\}$$

lies in $S^{-N}_{\rho,\delta}(U \times O, pr_1^*(T^*X))$. Hence we can find for $K \subset U$ compact a boundary $C_K > 0$ such that

$$\left| \sigma_{A_\hbar B_\hbar}\left(\frac{\zeta}{\hbar}\right) - \sum_{k,\alpha,\tilde{\alpha},\alpha_1,...,\alpha_k,\beta,\beta_1,...,\beta_k \in I} \left(\frac{\hbar}{i}\right)^{|\alpha|+|\beta|-k} \frac{1}{k! \cdot \tilde{\alpha}! \cdot \alpha_1! \cdot ... \cdot \alpha_k! \beta_1! \cdot ... \cdot \beta_k!} \right.$$

$$\left[\left.\frac{\partial^{|\alpha|}}{\partial \xi_{\pi(\zeta)}{}^{\alpha}}\right|_{(\zeta,\hbar)} a\right] \left\{\left.\frac{\partial^{|\tilde{\alpha}|}}{\partial z_{\pi(\zeta)}{}^{\tilde{\alpha}}}\right|_{\pi(\zeta)} \left[\left.\frac{\partial^{|\beta|}}{\partial \xi_{(-)}{}^{\beta}}\right|_{(d_{(-)}\varphi(\ ,\zeta),\hbar)} b\right]\right\} \cdot$$

$$\left\{\left.\frac{\partial^{|\alpha_1|}}{\partial z_{\pi(\zeta)}{}^{\alpha_1}}\right|_{\pi(\zeta)} \left[\left.\frac{\partial^{|\beta_1|}}{\partial z_{(-)}{}^{\beta_1}}\right|_{(-)} \varphi(\cdot,\zeta)\right]\right\} \cdot \ldots \cdot \left\{\left.\frac{\partial^{|\alpha_k|}}{\partial z_{\pi(\zeta)}{}^{\alpha_k}}\right|_{\pi(\zeta)} \left[\left.\frac{\partial^{|\beta_k|}}{\partial z_{(-)}{}^{\beta_k}}\right|_{(-)} \varphi(\cdot,\zeta)\right]\right\}\bigg|$$

$$\leq\ C_K \frac{|\hbar|^N}{(|\hbar|+|\zeta|)^N} \tag{109}$$

for all $(\pi(\zeta),\hbar) \in K \times O$. As the sum $\sum \ldots$ in (109) is a continuous function of $\hbar \in O$, (109) tells us that $c(\zeta,\hbar) = \sigma_{A_\hbar B_\hbar}\left(\frac{\zeta}{\hbar}\right)$ can be continuously expanded to $\hbar = 0$ such that the required growth condition is fulfilled. Similar considerations for all partial derivatives of $c$ now give the desired result, i. e. they show that $c$ has the asymptotic expansion

$$c(\zeta,\hbar) \sim \sum_{k,\alpha,\tilde{\alpha},\alpha_1,\ldots,\alpha_k,\beta,\beta_1,\ldots,\beta_k \in I} \left(\frac{\hbar}{i}\right)^{|\alpha|+|\beta|-k} \frac{1}{k! \cdot \tilde{\alpha}! \cdot \alpha_1! \cdot \ldots \cdot \alpha_k! \beta_1! \cdot \ldots \cdot \beta_k!}$$

$$\left[\left.\frac{\partial^{|\alpha|}}{\partial \xi_{\pi(\zeta)}{}^{\alpha}}\right|_{(\zeta,\hbar)} a\right] \left\{\left.\frac{\partial^{|\tilde{\alpha}|}}{\partial z_{\pi(\zeta)}{}^{\tilde{\alpha}}}\right|_{\pi(\zeta)} \left[\left.\frac{\partial^{|\beta|}}{\partial \xi_{(-)}{}^{\beta}}\right|_{(d_{(-)}\varphi(\ ,\zeta),\hbar)} b\right]\right\} \cdot$$

$$\left\{\left.\frac{\partial^{|\alpha_1|}}{\partial z_{\pi(\zeta)}{}^{\alpha_1}}\right|_{\pi(\zeta)} \left[\left.\frac{\partial^{|\beta_1|}}{\partial z_{(-)}{}^{\beta_1}}\right|_{(-)} \varphi(\cdot,\zeta)\right]\right\} \cdot \ldots \cdot \left\{\left.\frac{\partial^{|\alpha_k|}}{\partial z_{\pi(\zeta)}{}^{\alpha_k}}\right|_{\pi(\zeta)} \left[\left.\frac{\partial^{|\beta_k|}}{\partial z_{(-)}{}^{\beta_k}}\right|_{(-)} \varphi(\cdot,\zeta)\right]\right\} \tag{110}$$

and is an element of $S^{\mu+\tilde{\mu}}_{\rho,\delta}(U \times O, pr_1^*(T^*X))$. For later purposes let us write down the first order approximation of $c = a \cdot_d b$ in powers of $\hbar$:

$$c(\zeta,\hbar) = a(\zeta,\hbar) \cdot b(\zeta,\hbar)$$
$$+ \left(\frac{\hbar}{i}\right) \sum_{l=1}^n \left[\frac{\partial a}{\partial \xi_{\pi(\zeta),l}}(\zeta,\hbar)\right] \left\{\frac{\partial\left[b(d_{(-)}\varphi(\ ,\zeta),\hbar)\right]}{\partial z_{\pi(\zeta),l}}(\pi(\zeta))\right\} + \hbar^2 r(\zeta,\hbar), \tag{111}$$

where $r \in S^{\mu+\tilde{\mu}-\rho+\delta}_{\rho,\delta}(U \times \mathbb{R}, pr_1^*(T^*X))$.

The symbol space $S^{-\infty}(U \times O, pr_1^*(T^*X))$ forms an ideal with respect to the bilinear mapping $(a,b) \mapsto c = a \cdot_d b$. Furthermore the product $\cdot_d$ is associative up to smoothing symbols. This follows easily from the fact, that the maps $\Psi_\psi$ and $\sigma$ are inverse to each other modulo smoothing symbols resp. smoothing pseudo-differential operators. The quotient space

$$S^\infty_{\rho,\delta;d}(U \times O, pr_1^*(T^*X)) = S^\infty_{\rho,\delta}/S^{-\infty}(U \times O, pr_1^*(T^*X))$$

now becomes a noncommutative algebra and $S^\infty_{\rho,\delta;d}(\cdot, pr_1^*(T^*X))$ a sheaf of noncommutative algebras on $X \times \mathbb{R}$, the sheaf of **universally deformed symbols of type $(\rho,\delta)$** on a smooth RIEMANNIAN manifold $X$.

The embedding $\sigma_\hbar : \mathcal{D}_\hbar \to S^\infty_{\rho,\delta;\hbar}$ uniquely extends to an embedding $\sigma_d : \mathcal{D}_d \to S^\infty_{\rho,\delta;d}$ such that the diagram

$$\begin{array}{ccc} \mathcal{D}_d & \xrightarrow{\sigma_d} & S^\infty_{\rho,\delta;d} \\ \downarrow & & \downarrow \\ \mathcal{D}_\hbar & \xrightarrow{\sigma_\hbar} & S^\infty_{\rho,\delta;\hbar} \end{array}$$

commutes for every $\hbar \neq 0$. Just define $\sigma_d(P)$ for $P \in \mathcal{D}_d(U \times O)$, $U \subset X$ open and $O \subset \mathbb{R}$ open to be the function $p \in S^\infty_{\rho,\delta}(U \times O, pr_1^*(T^*X))$ such that $p(\xi,\hbar) = \left[P\left(\psi_{\pi(\xi)}(\cdot)e^{i\varphi(\cdot,\frac{\xi}{\hbar})}\right)\right](\pi(\xi))$ for $\hbar \in O \setminus \{0\}$ and $p(\xi,0) = \lim_{\substack{\hbar \to 0 \\ \hbar \in O \setminus \{0\}}} p(\xi,\hbar)$. To see that $p$ is well-defined indeed, write for every $x \in U$ on an open neighborhood $U_x$ of $x$

$$P|_{T^*U_x \times O} = p_x\left(-i\frac{\partial^d}{\partial^d z_x}\right) \tag{112}$$

with a smooth function $p_x$ on $T^*U_x \times O$, polynomial in the fibers. Then check that $p(\xi,\hbar) = p_{\pi(\xi)}(\xi,\hbar)$ for $\xi \in T^*U_x$ and $\hbar \in O$. As $p_x$ is also smooth with respect to $x$, $p$ is a smooth function on $T^*U \times O$ and polynomial in the fibers. The rest of the claim now follows easily.

## 3.2 Deformation and quantization

After having defined the sheaf $S^\infty_{\rho,\delta;d}(\cdot, pr_1^*(T^*X))$, we will now show that it gives rise to a deformation and quantization of the POISSON space $\left(X, S^\infty_{\rho,\delta}/S^{-\infty}\right)$.

Denote by $D_{\rho,\delta} = (d, \Delta_{\rho,\delta}) : (X \times \mathbb{R}, S^\infty_{\rho,\delta;d}) \to (\mathbb{R}, \mathcal{C}^\infty)$ the fibered morphism

$$\begin{array}{rccc} d = pr_2 : & X \times \mathbb{R} & \to & \mathbb{R}, \\ & (\zeta,\hbar) & \mapsto & \hbar, \end{array} \tag{113}$$

$$\begin{array}{rccc} (\Delta_{\rho,\delta})_O : & \mathcal{C}^\infty(O) & \to & S^\infty_{\rho,\delta;d}(X \times O, pr_1^*(T^*X)), & O \subset \mathbb{R} \text{ open}, \\ & f & \mapsto & f \circ pr_2 + S^{-\infty}(X \times O, pr_1^*(T^*X)). \end{array} \tag{114}$$

Furthermore let $(x,\hbar) \in (X \times \mathbb{R})$, and $\mathfrak{m}_\hbar \subset \mathcal{C}^\infty_\hbar$ be the maximal ideal in $\mathcal{C}^\infty_\hbar$. Then the stalk $\left(S^\infty_{\rho,\delta;d}\right)_{(x,\hbar)}$ is filtered by

$$\left(S^\infty_{\rho,\delta;d}\right)_{(x,\hbar)} = S^0_{(x,\hbar)} \supset S^1_{(x,\hbar)} \supset \ldots \supset S^m_{(x,\hbar)} \supset \ldots, \tag{115}$$

where $S^m_{(x,\hbar)}$ with $m \in \mathbb{N}$ consists of all (germs of) symbols $a \in \left(S^\infty_{\rho,\delta;d}\right)_{(x,\hbar)}$ such that $\left.\frac{\partial^m}{\partial \hbar^m}\right|_\hbar a = 0$. In other words $S^m_{(x,\hbar)} = \mathfrak{m}^m_\hbar \left(S^\infty_{\rho,\delta;d}\right)_{(x,\hbar)}$. As $\mathcal{C}^\infty/\mathfrak{m}_\hbar \cong \mathbb{C}$ is a field, $S^0_{(x,\hbar)}/S^1_{(x,\hbar)}$ is a flat $\mathcal{C}^\infty/\mathfrak{m}_\hbar$ module. Next define the graded algebras $gr\left(\mathcal{C}^\infty_\hbar\right)$ and $gr\left(S^0_{(x,\hbar)}\right)$ by

$$
\begin{aligned}
gr\left(\mathcal{C}^\infty_\hbar\right) &= \mathcal{C}^\infty_\hbar/\mathfrak{m}_\hbar \oplus \mathfrak{m}_\hbar/\mathfrak{m}^2_\hbar \oplus \ldots \oplus \mathfrak{m}^m_\hbar/\mathfrak{m}^{m-1}_\hbar \oplus \ldots \cong \mathbb{C}^\mathbb{N}, \\
gr\left(S^0_{(x,\hbar)}\right) &= S^0_{(x,\hbar)}/S^1_{(x,\hbar)} \oplus \ldots \oplus S^m_{(x,\hbar)}/S^{m-1}_{(x,\hbar)} \oplus \ldots \cong \left[\left(S^\infty_{\rho,\delta;\hbar}\right)_x\right]^\mathbb{N},
\end{aligned}
\tag{116}
$$

and check that the canonical morphism

$$
\gamma : gr\left(\mathcal{C}^\infty_\hbar\right) \otimes_{gr_0\left(\mathcal{C}^\infty_\hbar\right)} gr_0\left(S^0_{(x,\hbar)}\right) \to gr\left(S^0_{(x,\hbar)}\right) \tag{117}
$$

is bijective. By Theorem 1. in BOURBAKI [9], *Commutative Algebra*, Chapter III §5.2 this gives the flatness of the $\mathcal{C}^\infty_\hbar/\mathfrak{m}^m_\hbar$ module $S^0_{(x,\hbar)}/S^m_{(x,\hbar)}$. Furthermore the fiber $\left(S^\infty_{\rho,\delta;d}\right)_\hbar = S^\infty_{\rho,\delta;d}\big|_{X\times\{\hbar\}} / \mathfrak{m}_\hbar S^\infty_{\rho,\delta;d}\big|_{X\times\{\hbar\}}$ over $\hbar \in \mathbb{R}$ is isomorphic to $S^\infty_{\rho,\delta;\hbar}(\cdot, X)$, hence $D$ provides a flatly filtered deformation of $(X, S^\infty_{\rho,\delta;\hbar})$. Now we can formulate the main theorem.

**Theorem 3.1** *Let $X$ be a* RIEMANNIAN *manifold and $0 \leq \rho, \delta \leq 0$, $\rho + \delta \geq 1$, $\rho > \delta$. Then the fibered morphism $(d, \Delta_{\rho,\delta}) : (X \times \mathbb{R}, S^\infty_{\rho,\delta;d}) \to (\mathbb{R}, \mathcal{C}^\infty)$ is for every $\hbar \in \mathbb{R}$ a deformation of $(X, S^\infty_{\rho,\delta;\hbar})$ with distinguished point $\hbar$. In particular for $\hbar = 0$ it is a deformation of the ringed space $\left(X, S^\infty_{\rho,\delta}/S^{-\infty}\right)$.*

*Let $q = pr_1 : X \times \mathbb{R} \to X$ be the projection on the first coordinate, and $\mathfrak{q}_{\rho,\delta}$ the pullback. $pr_1^* : S^\infty_{\rho,\delta}/S^{-\infty}(\cdot, X) \to S^\infty_{\rho,\delta;d}(\cdot, X)$. Then the morphism $(q, \mathfrak{q}_{\rho,\delta}) : \left(X \times \mathbb{R}, S^\infty_{\rho,\delta;d}\right) \to \left(X, S^\infty_{\rho,\delta}/S^{-\infty}\right)$ together with the deformation $(d, \Delta_{\rho,\delta})$ give rise to a differentiable* DIRAC *quantization of the* POISSON *space $\left(X, S^\infty_{\rho,\delta}/S^{-\infty}\right)$, the so-called **normal order quantization** of $\left(X, S^\infty_{\rho,\delta}/S^{-\infty}\right)$ with respect to the* LEVI-CIVITA *connection.*

PROOF: It remains to show the DIRAC quantization condition

$$
[\mathfrak{q}(a), \mathfrak{q}(b)] = \frac{d}{i} \mathfrak{q}(\{a, b\}) + d^2 r, \tag{118}
$$

for $a, b \in S^\infty_{\rho,\delta}/S^{-\infty}(U, T^*X)$, $U \subset X$ open and a rest term $r \in S^\infty_{\rho,\delta}/S^{-\infty}(U, T^*X)$. By abuse of notation we also write $a$ (resp. $b$) for a representative of $a$ (resp. $b$) in $S^\infty_{\rho,\delta}(U, T^*X)$. Then we have by Eq. (111)

$$
\begin{aligned}
[\mathfrak{q}(a), \mathfrak{q}(b)](\zeta, \hbar) &= \frac{\hbar}{i} \sum_{k=1}^n \left[\frac{\partial a}{\partial \xi_{\pi(\zeta),k}}(\zeta, \hbar) \frac{\partial b}{\partial z_{\pi(\zeta),k}}(\zeta, \hbar) - \frac{\partial a}{\partial z_{\pi(\zeta),k}}(\zeta, \hbar) \frac{\partial b}{\partial \xi_{\pi(\zeta),k}}(\zeta, \hbar)\right] \\
&\quad + \hbar^2 r(\zeta, \hbar) \\
&= \frac{\hbar}{i} \mathfrak{q}(\{a, b\})(\zeta, \hbar) + \hbar^2 r(\zeta, \hbar),
\end{aligned}
\tag{119}
$$

where $\zeta \in T^*U$ and $\hbar \in \mathbb{R}$. Hence the claim holds. $\square$

**Note 3.2** The deformation in the preceding theorem can be interpreted as a convergent deformation with respect to the topology of asymptotic convergence on symbol spaces.

### 3.3 Functorial aspects

The normal order quantization on symbol spaces induces a certain functor $\mathcal{Q}^{no}_{symb}$ introduced in the following proposition.

**Proposition 3.3** *Associate to any* RIEMANNIAN *manifold $X$ its normal order quantization of the* POISSON *space $\left(X, S^\infty_{\rho,\delta}/S^{-\infty}\right)$. This gives rise to a functor $\mathcal{Q}^{no}_{symb}$ from the category* <u>Riem</u> *to the category* $\underline{\mathsf{Qu}}^{Dirac}_{0,(\mathbb{R},\mathcal{C}^\infty)}$ *of differentiable* DIRAC *quantizations.*

PROOF: It suffices to show that the symbol map is natural with respect to isometric embeddings between RIEMANNIAN manifolds. In other words we have to show that for an isometric embedding $f : X \to Y$ of RIEMANNIAN manifolds and every pseudo-differential operator $A \in \Psi^\infty_{\rho,\delta}(X)$ the relation

$$\sigma_{f_*A} = f_*(\sigma_A) \tag{120}$$

holds. So first choose a cut-off function $\psi : T^*Y \to \mathbb{R}$ like in Appendix B.3 and note that $f^*\psi : T^*X \to \mathbb{R}$ is a cut-off function on $X$. Now using the relation $f \circ \exp = \exp \circ TF$ and Eq. (139) we get for $\xi \in T^*Y\big|_{\operatorname{im} f}$

$$\begin{aligned}
\sigma_{f_*A}(\xi) &= f_*A\left(\psi_{\pi(\xi)}(\cdot) \, e^{i<\xi,\exp^{-1}_{\pi(\xi)}(\cdot)>}\right)(\pi(\xi)) \\
&= A\left((f^*\psi)_{f^{-1}(\pi(\xi))}(\cdot) \, e^{i<\xi,\exp^{-1}_{\pi(\xi)}(f(\cdot))>}\right)\left(f^{-1}(\pi(\xi))\right) \\
&= A\left((f^*\psi)_{f^{-1}(\pi(\xi))}(\cdot) \, e^{i<f^*(\xi),\exp^{-1}_{f^{-1}(\pi(\xi))}(\cdot)>}\right)\left(f^{-1}(\pi(\xi))\right) \\
&= A\left((f^*\psi)_{\pi(f^*(\xi))}(\cdot) \, e^{i<f^*(\xi),\exp^{-1}_{\pi(f^*(\xi))}(\cdot)>}\right)\left((\pi(f^*(\xi)))\right) \\
&= \sigma_A(f^*(\xi)) = f_*(\sigma_A)(\xi). \tag{121}
\end{aligned}$$

Hence the claim follows. □

By Theorem 3.1 and the preceeding proposition we know that the POISSON space $(X, S^\infty_{\rho,\delta}/S^{-\infty})$ can be quantized and that it gives rise to a functor $\mathcal{Q}^{no}_{symb}$ from <u>Riem</u> to $\underline{\mathsf{Qu}}^{0,(\mathbb{R},\mathcal{C}^\infty)}_{Dirac}$. But we do not yet know whether this normal order quantization quantization is consistent with the one of the POISSON space $(X, \mathcal{D}_0)$ of polynomial observables on $X$, resp. with the functor $\mathcal{Q}^{no}$. The following proposition shows that this is the case indeed.

**Proposition 3.4** *The canonical embeddings $\sigma_0 : \mathcal{D}_0 \to S^\infty_{\rho,\delta}/S^{-\infty}$ and $\sigma_d : \mathcal{D}_d \to S^\infty_{\rho,\delta;d}$ induce a morphism from the quantization map $\mathfrak{q} : \mathcal{D}_0 \to \mathcal{D}_d$ to the quantization map $\mathfrak{q}_{\rho,\delta} : S^\infty_{\rho,\delta}/S^{-\infty} \to S^\infty_{\rho,\delta;d}$. In other words the diagrams*

$$\begin{array}{ccc}
 & \mathcal{C}_{\mathbb{R}}^{\infty} & \\
\Delta \swarrow & & \searrow \Delta_{\rho,\delta} \\
\mathcal{D}_d & \xrightarrow{\sigma_d} & S_{\rho,\delta;d}^{\infty}
\end{array}
\qquad
\begin{array}{ccc}
\mathcal{D}_d & \xrightarrow{\sigma_d} & S_{\rho,\delta;d}^{\infty} \\
i^* \downarrow & & \downarrow i^* \\
\mathcal{D}_0 & \xrightarrow{\sigma_0} & S_{\rho,\delta}^{\infty}/S^{-\infty}
\end{array}$$

and

$$\begin{array}{ccc}
\mathcal{D}_0 & \xrightarrow{\sigma_0} & S_{\rho,\delta}^{\infty}/S^{-\infty} \\
\mathfrak{q} \downarrow & & \downarrow \mathfrak{q}_{\rho,\delta} \\
\mathcal{D}_d & \xrightarrow{\sigma_d} & S_{\rho,\delta;d}^{\infty}
\end{array}$$

*commute, where* $i : X \to X \times \mathbb{R}$ *is the canonical injection* $x \mapsto (x, 0)$. *Thus there exists a natural transformation from the functor* $\mathcal{Q}_{\text{symb}}^{\text{no}}$ *to the functor* $\mathcal{Q}^{\text{no}}$. *We say that these two functors resp. their quantization maps* $\mathfrak{q}$ *and* $\mathfrak{q}_{\rho,\delta}$ *are* **compatible**.

PROOF: To prove the proposition it suffices to show commutativity of the above diagrams. As the first two diagrams obviously commute, we show this only for the last one. Let $a \in \mathcal{D}_0(U)$. Then by Proposition 2.14 we have for $r \in \mathcal{C}^{\infty}(U \times \mathbb{R})$, $x \in U$ and $\hbar \in \mathbb{R}$

$$[\mathfrak{q}(f) \cdot r](x, \hbar) = \sum_{\alpha} (-i\hbar)^{|\alpha|} a_{x,\alpha}(x) \frac{\partial^{|\alpha|} r}{\partial z_x^{\alpha}}(x, \hbar), \tag{122}$$

where $a|_{T^*U_x} = \sum_{\alpha} (-i\hbar)^{|\alpha|} (a_{x,\alpha} \circ \pi) \xi_x^{\alpha}$, $a_{x,\alpha} \in \mathcal{C}^{\infty}(U_x)$, $U_x \subset U$ open neighborhood of $x$. Hence the equation

$$\begin{aligned}
\sigma_{d,\mathfrak{q}(f)}(\zeta, \hbar) &= \left[\mathfrak{q}(f) \left(\psi_{\pi(\zeta)}(\cdot) e^{i\varphi(\cdot, \frac{\xi}{\hbar})}\right)\right](\pi(\zeta), \hbar) = \sum_{\alpha} a_{\pi(\zeta),\alpha}(\pi(\zeta)) \xi_x^{\alpha}(\zeta) = \\
&= \mathfrak{q}_{\rho,\delta}(\sigma_0(a))(\zeta, \hbar)
\end{aligned} \tag{123}$$

holds for $\zeta \in T^*U$ and $\hbar \in \mathbb{R}$. But this gives the claim. $\square$

The above functorial properties of normal order quantization are not in contradiction to a theorem of VAN HOVE [60] saying that there is no plausible quantization functor going from the category **Symp** of symplectic manifolds and symplectomorphisms to the category of HILBERT spaces and isometric mappings. The reason why this is not the case lies in the fact that we have chosen different domain and range categories for our quantization functors. According to Theorem 2.16 the functors $\mathcal{Q}^{\text{no}}$ and $L^2$ induce a functor $\mathcal{R}_{\hbar}$ from **Riem** to HILBERT space representations. Now all these three functors have domain **Riem** which can be regarded as a subcategory of $\underline{\mathsf{Symp}}$, but only as a proper one. Thus there is no contradiction to VAN HOVES'S observation.

We hope to extend the quantization functors defined in this work to a larger (proper) subcategory of the symplectic category $\underline{\mathsf{Symp}}$ in a forthcoming project.

# Conclusion

By reasons of completeness let us mention at the end of this paper some other important quantization schemes besides the already explained ones of geometric quantization and formal deformation quantization and let us give some references:

(i) Probably the first mathematically rigorous quantization scheme is the WEYL quantization due to HERMANN WEYL [66]. Unfortunately it is well understood only in the case of quantization over EUCLIDEAN spaces. For attempts to generalize WEYL quantization to curved manifolds see UNDERHILL [55] or ZHANG, MIN [71]. MOYAL [36] used WEYL quantization to define a deformation of a certain commutative algebra of functions on EUCLIDEAN spaces (see Example 1.18).

(ii) Though it lacks of mathematical rigor, the path integral formalism by FEYNMAN [18] is very popular in theoretical physics and far-reaching for calculational and conceptual purposes. In the theory of functional integration one tries to give FEYNMAN'S ideas a precise mathematical meaning and wants to apply them for quantum field theory. See for example BARRY SIMON [50] or GLIMM, JAFFE [23].

(iii) The group quantization scheme together with a lot of topological and other aspects of quantization are extensively described in ISHAM [30].

(iv) BEREZIN'S quantization is in some respect similar to our approach, as BEREZIN regards a quantization as a deformation of a commutative algebra by a real parameter $\hbar$. But then it departs from our work, as BEREZIN'S quantization essentially uses complex methods and so-called coherent states. BEREZIN'S quantization is appropriate for the quantization of symmetric complex spaces or KÄHLER manifolds. See for example BEREZIN'S paper [4] and references therein or the more modern account by RAWNSLEY in [42, 43, 41]. RAWNSLEY does in fact construct a formal deformation for certain KÄHLER manifolds by means of BEREZIN'S quantization and claims that in some good cases those deformations probably come from converging ones.

(v) MARC RIEFFEL has established the notion of a strict deformation quantization, that is a framework for the continuous deformation of $C^*$-algebras. There is a close connection to the noncommutative geometry by ALAIN CONNES [11]. See the monograph [46] by RIEFFEL as well as his articles [44, 45, 47] for more information on the topic of strict deformation quantization.

The deformation theoretical quantization program we have introduced in this work can be regarded as the starting point for further developments in quantization theory, in particular if combined with one or the other of the above methods. As already mentioned in the introduction neither of the existing quantization schemes gives a completely satisfactory answer to the quantization problem. But each one has its own advantages, so it should be the goal to constructively combine different methods, if that is possible. This imperative holds for our program of deformation quantization as well, and I think that some generalizations and improvements in this direction can be done.

First it seems possible to extend the MOYAL product to arbitrary RIEMANNIAN manifolds. In other words we regard it as very promising to establish a complete WEYL symbol for the whole category of RIEMANNIAN manifolds and then to define a functor of WEYL quantizations on RIEMANNIAN manifolds according to the methods introduced in section 3. The papers UNDERHILL [55] and ZHANG, MIN [71] and the work by UNTERBERGER [56] and UPMEIER [57] could be very helpful for work in that direction.

The class of symplectic manifolds for which we have proven the existence of a concrete deformation quantization, namely the cotangent bundles with their canonial symplectic structure, should be enlarged. By using the general notions of conic manifolds and the theory of FOURIER integral operators we hope to extend our method of quantizing symbol spaces to conic symplectic manifolds. Similarly it should be possible to find concrete deformation quantizations of symplectic manifolds with a real polarization.

Up to now we have only regarded the quantization of bosonic systems. But it is well-known that there exists also a quantization condition for fermionic systems: POISSON brackets for classical fermionic variables or in other words for GRASSMANN variables go over under quantization to anticommutators (see for example HENNEAUX, TEITELBOIM [28]). Now on the mathematical side we have a notion of a fermionic symbol map going from a CLIFFORD algebra to an exterior algebra (cf. BERLINE, GETZLER, VERGNE [5]). It would then be very interesting to see whether it is possible to construct with such a fermionic symbol map a deformation quantization of GRASSMANN variables such that the above fermionic quantization condition is fulfilled.

Another important point which should be studied is the behaviour of deformation quantization under the action of symmetry groups. In this context one could think about the question whether first quantizing and then reducing by the symmetry group or first reducing and then quantizing gives the same result.

At the end I now hope to have convinced the reader that concrete deformation quantization gives new insights to the quantization process but also new questions for further research in this area. In forthcoming projects I intend to work on these.

# A  Some notions from algebraic geometry

For the convenience of the reader we briefly give some definitions from algebraic geometry used in this work. As a general reference and for more detailed explanations see HARTSHORNE [27]. In the following $k$ shall always denote a field.

First we want to recall the notion of flat modules. Let $R$ be a ring and $M$ a (left) $R$-module. Then $M$ is called **flat**, if for every injective morphism $A \to B$ of (right) $R$-modules the induced morphism $A \otimes M \to B \otimes M$ is injective, too. A homomorphism of $k$-algebras $A \to B$ is called flat, if $B$ regarded as an $A$-module is flat.

Now let $(f, \mathcal{F}) : (X, \mathcal{A}) \to (Y, \mathcal{B})$ be a morphism of $k$-ringed spaces. Then we call this morphism **flat over** $y \in Y$, if for every $x \in f^{-1}(y)$ the induced homorphism $\mathcal{F}_x : \mathcal{B}_y \to \mathcal{A}_x$ between the stalks $\mathcal{A}_x$ and $\mathcal{B}_y$ is flat. If $(f, \mathcal{F}) : (X, \mathcal{A}) \to (Y, \mathcal{B})$ is flat over every $y \in Y$, it is called a **flat morphism** of $k$-ringed spaces.

Finally we call a sheaf-morphism $(f, \mathcal{F}) : (X, \mathcal{A}) \to (Y, \mathcal{B})$ between $k$-ringed spaces $(X, \mathcal{A})$ and $(Y, \mathcal{B})$ linear, if for every $y \in Y$ and $x \in f^{-1}(y)$ the mapping $\mathcal{F}_x : \mathcal{B}_y \to \mathcal{A}_x$ is $k$-linear.

# B  A complete symbol calculus on manifolds

In this appendix we give the basic definitions and constructions for a complete symbol calculus on manifolds. To put it shortly, a symbol calculus associates to any differential or pseudo-differential operator $A$ on a RIEMANNIAN manifold $X$ a certain smooth function $\sigma_A$ on the cotangent bundle $T^*X$, such that $A$ is (almost) completely characterized by its symbol $\sigma_A$. In a certain way one can then regard the symbol $\sigma_A$ as the classical observable and the operator $A$ as its quantization. It is not too difficult to define a symbol calculus on EUCLIDEAN space $\mathbb{R}^n$, so one could try to set up a symbol calculus on a curved manifold by first defining it locally in charts. Unfortunately those locally defined symbols need not always fit together, so we have to find a global and chart independent definition of a complete symbol. Using and extending ideas of WIDOM [67, 68] we succeed in giving such a definition which also provides the essential tools for our further quantization purposes.

In B.1 we will give an introduction to symbols over vector bundles, and then define in B.2 the notion of pseudo-differential operators on manifolds. In the following section a new FOURIER transform on RIEMANNIAN manifolds is introduced. This and the symbols of B.1 will give us the means to construct a canonical integral representation for pseudo-differential operators on manifolds.

From B.4 on ideas of WIDOM [67, 68] come into play. We thus construct the complete symbol of a pseudo-differential operator by applying it to the exponential of a so-called phase function. But we use a different phase function than WIDOM [67, 68] does. The reason is that only our new phase function fits to the needs of deformation quantization and gives the means to explicitely construct an inverse (up to smoothing symbols resp. operators) of the symbol map by an integral representation (see Theorem B.8).

In B.5 we give an asymptotic expansion for the symbol of the product of two pseudo-differential operators. The results are analogous to the ones in WIDOM [67, 68].

All the missing proofs and details for the following considerations can be found in PFLAUM [38, 39].

## B.1 Symbols

Let $E \to X$ be a smooth vector bundle over the smooth manifold $X$. Denote by $\dot{E} \to X$ the bundle of all nonzero vectors in $E$. Then by the multiplication of vectors with scalars $\dot{E}$ carries in a natural way the structure of a conic manifold. In particular for any $\theta \in \mathbb{R}$ the bundle automorphism $\theta^* : \dot{E} \to \dot{E}$ is well-defined.

**Definition B.1** *Let $E \to X$ be a smooth vector bundle, $\mu \in \mathbb{R}$, $0 \leq \rho, \delta \leq 1$ and $U \subset X$ open. The space $S^\mu_{\rho,\delta}(U, \dot{E})$ of **symbols** on $\dot{E}$ over $U$ of **order** $\mu$ and **type** $(\rho, \delta)$ consists of all functions $a \in \mathcal{C}^\infty(\dot{E}|_U)$ with the following property:*

*Let $K \subset \dot{E}|_U$ compact, $L_1, ..., L_k$ $0$-homogeneous vector fields and $V_1, ..., V_l$ $(-1)$-homogeneous vertical vector fields on $\dot{E}|_U$. Then there exists $C = C_{K,L_1,...,L_k,V_1,...,V_l}(a) > 0$ such that*

$$[\theta^* (L_k \cdot ... \cdot L_1 \cdot V_l \cdot ... \cdot V_1 a)]|_K \leq C(1 + \theta)^{\mu+k\delta-l\rho}. \tag{124}$$

*for all $\theta \in \mathbb{R}^+$.*

By $S^{-\infty}(U, \dot{E}) = S^{-\infty}_{\rho,\delta}(U, \dot{E}) = \bigcap_{\mu \in \mathbb{R}} S^\mu_{\rho,\delta}(U, \dot{E})$ and $S^\infty_{\rho,\delta}(U, \dot{E}) = \bigcup_{\mu \in \mathbb{R}} S^\mu_{\rho,\delta}(U, \dot{E})$ we denote the space of smoothing symbols resp. the space of symbols of type $(\rho, \delta)$ over $U$.

The elements of $S^\mu_{\rho,\delta}(U, \dot{E})$ with $\mu \in \mathbb{R} \cup \{-\infty, \infty\}$ are the sections over $U$ of a certain sheaf $S^\mu_{\rho,\delta}(\cdot, \dot{E})$ of symbols on $E$. In case $E$ is the cotangent bundle $T^*X$ of a smooth manifold $X$ we just write $S^\mu_{\rho,\delta}(U, X)$ instead of $S^\mu_{\rho,\delta}(U, T^*X)$. Note that

$$\begin{array}{rcl} \mathcal{D}_0 & \to & S^\infty_{\rho,\delta}(\cdot, X) \\ \mathcal{D}_0(U) \ni f & \mapsto & f \in S^\infty_{\rho,\delta}(U, X) \end{array} \quad \text{for } U \subset X \text{ open} \tag{125}$$

gives rise to a monomorphism $\sigma_0 : \mathcal{D}_0 \to S^\infty_{\rho,\delta}/S^{-\infty}(\cdot, X)$ of sheaves of algebras. Here $\mathcal{D}_0$ is the sheaf of smooth fiberwise polynomial functions on $T^*X$ defined in Definition 2.1.

**Remark B.2** *The above definition of a symbol is an extension of the one given in DUISTERMAAT [15].*

To see whether a smooth function $a \in \mathcal{C}^\infty(E|_U)$ on a RIEMANNIAN vector bundle $E$ is a symbol of order $\mu$ and type $(\rho, \delta)$ it suffices to show that the following condition is satisfied: For every trivialization $(x, \xi) : E|_V \to \mathbb{R}^n \times \mathbb{R}^N$ with $V \subset U$ open, every $\alpha \in \mathbb{N}^n$, $\beta \in \mathbb{N}^N$ and every $K \subset V$ compact there exists $C = C_{K,\alpha,\beta} > 0$ such that the inequality

$$\left| \frac{\partial^{|\alpha|}}{\partial x^\alpha} \frac{\partial^{|\beta|}}{\partial \xi^\beta} a(e) \right| = C (1 + |e|)^{\mu+\delta|\alpha|-\rho|\beta|} \tag{126}$$

holds for every $e \in E|_K$.

We can attach seminorms $p = p_{K, L_1, ..., L_k, V_1, ..., V_l} : \mathrm{S}^\mu_{\rho,\delta}(U, \dot{E}) \to \mathbb{R}^+ \cup \{0\}$ to the symbol spaces $\mathrm{S}^\mu_{\rho,\delta}(U, \dot{E})$ by

$$p(a) = \sup \left\{ \frac{|\theta^*(L_k \cdot ... \cdot L_1 \cdot V_l \cdot ... \cdot V_1 \, a)(c)|}{(1+\theta)^{\mu+k\delta-l\rho}} : \theta \in \mathbb{R}^+, \, c \in K \right\}, \qquad (127)$$

where $K$ and the vector fields $L_1, ..., L_k, V_1, ..., V_l$ are like in Definition B.1. Then $\mathrm{S}^\mu_{\rho,\delta}(U, \dot{E})$ becomes a FRÉCHET space, such that the restriction morphisms $\mathrm{S}^\mu_{\rho,\delta}(U, \dot{E}) \to \mathrm{S}^\mu_{\rho,\delta}(V, \dot{E})$ for $V \subset U$ open are continuous. Additionally we have natural and continuous inclusions $\mathrm{S}^\mu_{\rho,\delta}(U, \dot{E}) \subset \mathrm{S}^{\tilde{\mu}}_{\tilde{\rho},\tilde{\delta}}(U, \dot{E})$ for $\tilde{\mu} \geq \mu$, $\tilde{\rho} \leq \rho$ and $\tilde{\delta} \geq \delta$.

An easy but somewhat technical consideration shows that pointwise multiplication of $a \in \mathrm{S}^\mu_{\rho,\delta}(U, \dot{E})$ and $b \in \mathrm{S}^{\tilde{\mu}}_{\rho,\delta}(U, \dot{E})$ defines an element $ab \in \mathrm{S}^{\mu+\tilde{\mu}}_{\rho,\delta}(U, \dot{E})$. This map is bilinear and gives a continuous, commutative and associative product on $\mathrm{S}^\infty_{\rho,\delta}(\cdot, \dot{E})$ which even commutes with restrictions. So $\mathrm{S}^\infty_{\rho,\delta}(\cdot, \dot{E})$ becomes a sheaf of commutative algebras on $X$, and $\mathrm{S}^{-\infty}(\cdot, \dot{E})$ an ideal sheaf in $\mathrm{S}^\infty_{\rho,\delta}(\cdot, \dot{E})$. We denote by $\mathrm{S}^\infty_{\rho,\delta}/\mathrm{S}^{-\infty}(\cdot, \dot{E})$ the quotient $\mathrm{S}^\infty_{\rho,\delta}(\cdot, \dot{E})/\mathrm{S}^{-\infty}(\cdot, \dot{E})$.

**Example B.3** (i) Let $X$ be a smooth manifold, and denote by $\mathcal{D}_0(U)$ with $U \subset X$ open the space of smooth functions $a : T^*U \to \mathbb{C}$ which are polynomials on the fibers of $T^*U$. Then an element $a \in \mathcal{D}_0$ lies in $\mathrm{S}^\mu_{\rho,\delta}(U, T^*X)$, iff it is a polynomial of order $\leq \mu$ on every fiber. In particular we have a monomorphism of sheaves $\mathcal{D}_0 \to \mathrm{S}^\infty_{\rho,\delta}(\cdot, T^*X)$.

(ii) Let $X = \mathbb{R}^n$, $n \in \mathbb{N}$ with its canonical EUCLIDEAN structure. Then the mapping $l : T^*X = \mathbb{R}^n \times \mathbb{R}^n \to \mathbb{C}$, $(x, \xi) \mapsto |\xi|^2 = \sum_{k=1}^n \xi_k^2$ is a symbol of order 2 and type $(1,0)$ on $\mathbb{R}^n$, but not one of order $\mu < 2$. Next regard the function $a : T^*X = \mathbb{R}^n \times \mathbb{R}^n \to \mathbb{C}$, $(x, \xi) \mapsto \frac{1}{1+|\xi|^2} = \sum_{k=1}^n \frac{1}{1+\xi_k^2}$. This function is a symbol of order $-2$ and type $(1,0)$ on $\mathbb{R}^n$, but not one of order $\mu < -2$.

See HÖRMANDER [29] or GRIGIS, SJÖSTRAND [24] for more material about and examples of symbols.

The following theorem is an essential tool for the use of symbols in the theory of partial differential equations. We will not give the proof, but refer the interested reader to HÖRMANDER [29], GRIGIS, SJÖSTRAND [24], SHUBIN [49] or any other good book on pseudo-differential operators.

**Theorem B.4** *Let $a_j \in S_{\rho,\delta}^{\mu m_j}(U, E)$ be symbols such that $m_j \in \mathbb{N}$ for all $j \in \mathbb{N}$ and $\lim_{j \mapsto \infty} m_j = -\infty$. Then there exists a symbol $a \in S_{\rho,\delta}^{\mu m_j}(U, E)$ unique up to smoothing symbols, such that $a - \sum a_j \in S_{\rho,\delta}^{\mu m_j}(U, E)$ for all $j \in \mathbb{N}$.*

*This induces a locally convex HAUSDORFF topology on the vector space $S_{\rho,\delta}^\infty / S^{-\infty}(U, E)$, which is called the* **topology of asymptotic convergence**.

Now composing the morphism $\mathcal{D}_0 \to S_{\rho,\delta}^\infty(\cdot, X)$ with the projection $S_{\rho,\delta}^\infty(\cdot, X) \to S_{\rho,\delta}^\infty / S^{-\infty}(\cdot, X)$ we receive a monomorphism $\sigma_0 : \mathcal{D}_0 \to S_{\rho,\delta}^\infty / S^{-\infty}(\cdot, X)$ of sheaves of algebras.

## B.2 Pseudo-differential operators on manifolds

For the convenience of the reader we will briefly state the definition of a pseudo-differential operator on $\mathbb{R}^n$. For more information on this topic and the theory of pseudo-differential operators on manifolds see HÖRMANDER [29], GRIGIS, SJÖSTRAND [24] and SHUBIN [49].

**Definition B.5** *Let $U \subset \mathbb{R}^n$ be an open set, $\mu \in \mathbb{R}$ and $0 < \rho \leq 1$, $0 \leq \delta < 1$. Then a* **pseudodifferential operator** *on $\mathbb{R}^n$ (over $U$) of* **order** *$\mu$ and* **type** *$(\rho, \delta)$ is a continuous operator $A : \mathcal{C}_0^\infty(U) \to \mathcal{C}^\infty(U)$ which can be written as an oscillatory integral of the form*

$$A(u)(x) = \frac{1}{(2\pi)^n} \int_{\mathbb{R}^n} e^{2\pi i <x-y,\theta>} a(x,y,\theta) u(y) \, dy \, d\theta, \quad u \in \mathcal{C}_0^\infty(U), \ x \in U \qquad (128)$$

*such that $a \in S_{\rho,\delta}^\mu(U \times U, U \times U \times \mathbb{R}^n)$ is a symbol of order $\mu$ and type $(\rho, \delta)$ on $U \times U \times \mathbb{R}^n$. The space of these pseudo-differential operators is denoted by $\Psi_{\rho,\delta}^\mu(U, \mathbb{R}^n)$.*

In the following proposition we will give a more convenient integral representation of pseudo-differential operators on $\mathbb{R}^n$.

**Proposition B.6** *Let $A \in \Psi_{\rho,\delta}^\mu(U, \mathbb{R}^n)$ be a properly supported pseudo-differential operator. Then the function $b(x, \xi) = e^{-i<x,\xi>} A\left(e^{i<(\cdot),\xi>}\right)$ belongs to $S_{\rho,\delta}^\mu(U, \mathbb{R}^n)$ and has an asymptotic expansion*

$$b(x, \xi) \sim \sum_{\alpha \in \mathbb{N}^n} \frac{i^{-|\alpha|}}{\alpha!} \left( \frac{\partial^{|\alpha|}}{\partial \xi^\alpha} \frac{\partial^{|\alpha|}}{\partial y^\alpha} a(x, y, \xi) \right) \bigg|_{y=x}. \qquad (129)$$

*Moreover $A$ can be written in the form*

$$Au(x) = \frac{1}{(2\pi)^n} \int e^{i<x,\xi>} b(x, \xi) \hat{u}(\xi) \, d\xi, \quad u \in \mathcal{C}_0^\infty(U), \ x \in U, \qquad (130)$$

*where $\hat{u}$ denotes the FOURIER transform of $u$. $b$ is called the* **complete symbol** *of $A$.*

Now we generalize the above definition to pseudo-differential operators on manifolds.

**Definition B.7** *Let $X$ be a smooth manifold of dimension $n$, $U \subset X$ be open, $\mu \in \mathbb{R}$ and $0 \leq \rho, \delta \leq 1$ such that $\rho + \delta \geq 1$ and $\rho > \delta$. A continuous operator $A : \mathcal{C}_0^\infty(U) \to \mathcal{C}^\infty(U)$ is called a **pseudo-differential operator** of order $\mu$ and type $(\rho, \delta)$ on $X$ over $U$, if $A$ is pseudo-local and if there exist a covering $\{U_j\}_{j \in J}$ of $U$ together with coordinates $z_j : U_j \to V_j \subset \mathbb{R}^n$ such that for every $j \in J$ the operator $A_j : \mathcal{C}_0^\infty(V_j) \to \mathcal{C}^\infty(V_j)$, $u \mapsto (A(u \circ z_j)) \circ z_j^{-1}$ is a pseudo-differential operator of order $\mu$ and type $(\rho, \delta)$ on $\mathbb{R}^n$. We denote the space of such pseudo-differential operators by $\Psi_{\rho,\delta}^\mu(U, X)$.*

The complete symbol calculus introduced in this appendix has the purpose to generalize Proposition B.6 to arbitrary manifolds and find a natural and coordinate independant integral representation for pseudo-differential operators on manifolds.

## B.3   Fourier transform on manifolds

Assume $X$ to be a RIEMANNIAN manifold and consider the exponential function exp with respect to the LEVI-CIVITA connection on $X$. Choose a neighborhood $W \subset TX$ of the zero section in $TX$ such that $(\pi, \exp) : W \to X \times X$ maps $W$ diffeomorphically onto an open neighborhood of the diagonal $\Delta$ of $X \times X$. Then there exists a smooth function $\psi : TX \to [0, 1]$ called a **cut-off function** such that $\psi|_{\tilde{W}} = 1$ and $\operatorname{supp}(\psi) \subset W$ for an open neighborhood $\tilde{W} \subset W$ of the zero section in $TX$. Now the following microlocal lift $\mathcal{M}_\psi$ is well-defined:

$$\mathcal{M}_\psi : \mathcal{D}(X) \to \mathrm{S}^{-\infty}(X, TX), \quad f \mapsto \left(v \mapsto {}^\psi\!f(v) = \psi(v) f(\exp v)\right). \tag{131}$$

$\mathcal{M}_\psi$ is a linear but not multiplicative map between function spaces.

Over the tangent bundle $TX$ we can define the **Fourier transform** as the following sheaf morphism:

$$\mathcal{F} : \mathrm{S}^{-\infty}(U, TX) \to \mathrm{S}^{-\infty}(U, T^*X), \quad a \mapsto \hat{a}(\xi) = \frac{1}{(2\pi)^{n/2}} \int_{T_{\pi(\xi)}X} e^{-i\langle \xi, v \rangle} a(v) \, dv. \tag{132}$$

We also have a **reverse Fourier transform**:

$$\mathcal{F}^{-1} : \mathrm{S}^{-\infty}(U, T^*X) \to \mathrm{S}^{-\infty}(U, TX), \quad b \mapsto \check{b}(v) = \frac{1}{(2\pi)^{n/2}} \int_{T_{\pi(v)}X} e^{i\langle \xi, v \rangle} b(\xi) \, d\xi. \tag{133}$$

It is easy to check that $\mathcal{F}^{-1}$ and $\mathcal{F}$ are inverse to each other, indeed.

Composing $\mathcal{M}_\psi$ and $\mathcal{F}$ gives rise to the **Fourier transform** $\mathcal{F}_\psi = \mathcal{F} \circ \mathcal{M}_\psi$ on the RIEMANNIAN manifold $X$. $\mathcal{F}_\psi$ has the left inverse

$$\mathrm{S}^{-\infty}(U, T^*X) \to \mathcal{C}^\infty(U), \quad a \mapsto \mathcal{F}^{-1}(a)|_{0_U}, \tag{134}$$

where $|_{0_U}$ means the restriction to the natural embedding of $U$ into $T^*X$ as zero section.

By definition $\mathcal{M}_\psi$ and $\mathcal{F}_\psi$ do depend on the smooth cut-off function $\psi$, but this arbitrariness will only have minor effects on the study of pseudo-differential operators defined through $\mathcal{F}_\psi$. This will become clear in the following.

Let $a$ be a smooth function defined on $T^*U$ and polynomial in the fibers. Then locally $a = \sum_\alpha (a_{x,\alpha} \circ \pi) \, \xi_x$ with respect to a normal coordinate system at $x \in U$ and functions $a_{x,\alpha} \in \mathcal{C}^\infty(U)$. Define the operator $A : \mathcal{D}(U) \to \mathcal{C}^\infty(U)$ by

$$f \mapsto Af = \left( U \ni x \mapsto \frac{1}{(2\pi)^{n/2}} \int_{T_x^*X} a(\xi) \, \mathcal{F}\left(^\psi f\right)(\xi) \, d\xi \in \mathbb{C} \right) \tag{135}$$

and check that

$$Af(x) = \sum_\alpha a_{x,\alpha}(x) \frac{\partial^{|\alpha|}}{\partial \xi_x^\alpha} [\psi (f \circ \exp)](0_x) = \sum_\alpha a_{x,\alpha}(x) \frac{\partial^{|\alpha|}}{\partial z_x^\alpha} f(x) \tag{136}$$

for all $x \in U$. Therefore $Af$ is a differential operator independent of the choice of $\psi$.

However, if $a$ is an arbitrary element of the symbol space $S^\infty_{\rho,\delta}(U)$, Eq. (135) defines a continuous operator $A = A_\psi : \mathcal{C}^\infty(U) \to \mathcal{C}^\infty(U)$, which is a pseudo-differential operator and independent of the cut-off function $\psi$ only up to smoothing operators. Let us give the precise statement.

**Theorem B.8** *Let $X$ be a* RIEMANNIAN *manifold and* $\exp$ *the exponential function corresponding to the* LEVI-CIVITA *connection on $X$. Further assume that $0 \leq \rho, \delta \leq 1$, $\rho + \delta \geq 1$ and $\rho > \delta$. Then any smooth cut-off function $\psi : TX \to [0,1]$ gives rise to a linear sheaf morphism $\Psi_\psi : S^\infty_{\rho,\delta}(\cdot, X) \to \Psi^\infty_{\rho,\delta}(\cdot, X)$, $a \mapsto A_\psi$ defined by*

$$A_\psi f(x) = \frac{1}{(2\pi)^{n/2}} \int_{T_x^*X} a(\xi) \, \hat{^\psi f}(\xi) \, d\xi, \tag{137}$$

*where $a \in S^\infty_{\rho,\delta}(U, X)$, $f \in \mathcal{D}(U)$, $x \in U$ and $U \subset X$ open. This morphism preserves the natural filtrations of $S^\infty_{\rho,\delta}(\cdot, X)$ and $\Psi^\infty_{\rho,\delta}(\cdot, X)$. In particular it maps the subsheaf $S^{-\infty}(\cdot, X)$ of smoothing symbols to the subsheaf $\Psi^{-\infty}(\cdot, X) \subset \Psi^\infty_{\rho,\delta}(\cdot, X)$ of smoothing pseudo-differential operators.*

*The quotient morphism $\Psi : (S^\infty_{\rho,\delta}/S^{-\infty})(\cdot, X) \to (\Psi^\infty_{\rho,\delta}/\Psi^{-\infty})(\cdot, X)$ is an isomorphism and independent of the choice of the cut-off function $\psi$.*

## B.4 The symbol map

In the sequel denote by $\varphi : (id \times \pi)^{-1}(V) \to \mathbb{C}$ the smooth phase function defined by

$$(id \times \pi)^{-1}(V) \ni (x, \xi) \mapsto <\xi, \exp^{-1}_{\pi(\xi)}(x)> \,=\, <\xi, z_{\pi(\xi)}(x)> \,\in \mathbb{C}, \tag{138}$$

where $V \subset X \times X$ is an open neighborhood of the diagonal $\Delta = \{(x,y) \in X \times X : x = y\}$ such that $\exp^{-1}_y(x)$ is well-defined for every $(x,y) \in V$.

**Theorem and Definition B.9** *Let $A \in \Psi_{\rho,\delta}^\mu (U, X)$ be a pseudo-differential operator on a RIEMANNIAN manifold $X$, $0 \leq \rho, \delta \leq 1$, $\rho + \delta \geq 1$, $\rho > \delta$ and $\psi : T^*X \to [0,1]$ a cut-off function. Then the $\psi$-cutted symbol of $A$ with respect to the LEVI-CIVITA connection on $X$ is the function*

$$\sigma_{\psi,A} : T^*U \to \mathbb{C}, \quad \xi \mapsto \left[ A \left( \psi_{\pi(\xi)}(\cdot) \, e^{i\varphi(\cdot, \xi)} \right) \right] (\pi(\xi)), \tag{139}$$

*where $\psi_x = \psi \circ \exp_x^{-1} = \psi \circ z_x$ for all $x \in X$. $\sigma_{\psi,A}$ is an element of $S_{\rho,\delta}^\mu(U, X)$. The corresponding element $\sigma_A$ in the quotient $(S_{\rho,\delta}^\mu / S^{-\infty})(U, X)$ is called the **complete symbol** of $A$. It is independent of the choice of $\psi$.*

**Remark B.10** In the above theorem and definition we depart from the consideration of WIDOM in [68] insofar, as we have chosen a different (notion of a) phase function. Therefore our symbols are not the same as the ones in WIDOM'S article.

After having defined the notion of a complete symbol, we will now give its essential properties in the following theorem.

**Theorem B.11** *Let $0 \leq \rho, \delta \leq 1$, $\rho + \delta \geq 1$ and $\rho > \delta$. Furthermore let $A \in \Psi_{\rho,\delta}^\mu(U)$ be a pseudo-differential operator on a RIEMANNIAN manifold $X$ and $a = \sigma_{\psi,A} \in S_{\rho,\delta}^\mu(U)$ its $\psi$-cutted symbol with respect to the LEVI-CIVITA connection. Then $A$ and the pseudo-differential operator $A_\psi$ defined by Eq. (137) are equal modulo smoothing operators. Moreover the sheaf morphisms $\sigma : (\Psi_{\rho,\delta}^\infty / \Psi^{-\infty})(\cdot, X) \to (S_{\rho,\delta}^\infty / S^{-\infty})(\cdot, X)$ and $\Psi : (S_{\rho,\delta}^\infty / S^{-\infty})(\cdot, X) \to (\Psi_{\rho,\delta}^\infty / \Psi^{-\infty})(\cdot, X)$ are inverse to each other.*

**Example B.12** Consider the EUCLIDEAN space $X = \mathbb{R}^n$ and the symbols $l : T^*X \to \mathbb{C}$, $(x, \xi) \mapsto |\xi|^2$ and $a : T^*X \to \mathbb{C}$, $(x, \xi) \mapsto \frac{1}{1+|\xi|^2}$ of Example B.3 *(ii)*. Then the pseudo-differential operator corresponding to $l$ is minus the LAPLACIAN: $\Psi(l) = -\Delta$. As one can now calculate directly, $\Psi(a)$ is an inverse in $\Psi_{1,0}^\infty(X)$ of the differential operator $1 - \Delta$ having symbol $T^*X \ni (x, \xi) \mapsto 1 + |\xi|^2 \in \mathbb{C}$.

So with the help of a symbol calculus one can construct inverses to certain differential operators in the algebra of pseudo-differential operators. This is the main reason why a symbol calculus is so important for the theory of partial differential equations.

## B.5 Product expansions

Most considerations of elliptic partial differential operators as well as our quantization program for symbol spaces require to know how to express the symbol of the product of two pseudo-differential operators $A, B$ in terms of the symbols of its components. In the flat case on $\mathbb{R}^n$ it is a well known fact (see for example SHUBIN [49] or GRIGIS, SJÖSTRAND [24]) that $\sigma_{AB}$ has an asymptotic expansion of the form

$$\sigma_{AB}(x, \xi) \sim \sum_\alpha \frac{1}{i^{|\alpha|} \alpha!} \frac{\partial^{|\alpha|}}{\partial \xi^\alpha} \sigma_a(x, \xi) \frac{\partial^{|\alpha|}}{\partial x^\alpha} \sigma_B(x, \xi). \tag{140}$$

Note that precisely speaking, the product $AB$ of pseudo-differential operators is only well-defined, if at least one of them is properly supported. But this is only a minor set-back, as any pseudo-differential operator is properly supported modulo a smoothing operator and the above asymptotic expansion also describes a symbol only up to smoothing symbols.

In the following we want to derive a similar but more complicated formula for the case of pseudo-differential operators on manifolds. To achieve this let us first state a proposition.

**Theorem B.13** *Let $A, B \in \Psi_{\rho,\delta}^{\infty}(U, X)$ with $\rho+\delta \geq 1$ and $\rho > \delta$ be two pseudo-differential operators on the RIEMANNIAN manifold $X$, one of them properly supported. Then the symbol $\sigma_{AB}$ of the product $AB$ has the following asymptotic expansion.*

$$\sigma_{AB}(\zeta) \sim \sum_{k\in\mathbb{N}} \sum_{\substack{\alpha,\tilde{\alpha},\alpha_1,\ldots,\alpha_k \in \mathbb{N}^n \\ \tilde{\alpha}+\alpha_1+\ldots+\alpha_k=\alpha}} \sum_{\substack{\beta,\beta_1,\ldots,\beta_k \in \mathbb{N}^n \\ \beta_1+\ldots+\beta_k=\beta \\ |\beta_1|,\ldots,|\beta_k|\geq 2}} \frac{i^{k-|\alpha|-|\beta|}}{k!\cdot\tilde{\alpha}!\cdot\alpha_1!\cdot\ldots\cdot\alpha_k!\beta_1!\cdot\ldots\cdot\beta_k!}$$
$$\left[\left.\frac{\partial^{|\alpha|}}{\partial\xi_{\pi(\zeta)}{}^{\alpha}}\right|_\zeta \sigma_A\right] \left\{\left.\frac{\partial^{|\tilde{\alpha}|}}{\partial z_{\pi(\zeta)}{}^{\tilde{\alpha}}}\right|_{\pi(\zeta)} \left[\left.\frac{\partial^{|\beta|}}{\partial\xi_{(-)}{}^{\beta}}\right|_{d_{(-)}\varphi(\,,\zeta)} \sigma_B\right]\right\} \cdot$$
$$\left\{\left.\frac{\partial^{|\alpha_1|}}{\partial z_{\pi(\zeta)}{}^{\alpha_1}}\right|_{\pi(\zeta)} \left[\left.\frac{\partial^{|\beta_1|}}{\partial z_{(-)}{}^{\beta_1}}\right|_{(-)} \varphi(\cdot,\zeta)\right]\right\} \cdot \ldots \cdot \left\{\left.\frac{\partial^{|\alpha_k|}}{\partial z_{\pi(\zeta)}{}^{\alpha_k}}\right|_{\pi(\zeta)} \left[\left.\frac{\partial^{|\beta_k|}}{\partial z_{(-)}{}^{\beta_k}}\right|_{(-)} \varphi(\cdot,\zeta)\right]\right\} \cdot$$
(141)

**Note B.14** The differential operators of the form $\frac{\partial^{|\alpha|}}{\partial z_{\pi(\zeta)}{}^{\alpha}}$ act on the variables denoted by $(-)$, the differential operators of the form $\frac{\partial^{|\beta|}}{\partial z_{(-)}{}^{\beta}}$ on the variables $(\cdot)$.